\documentclass[11pt]{article}
\pdfoutput=1

\usepackage{jheppub}
\usepackage[utf8]{inputenc}
\usepackage{booktabs}
\usepackage{graphicx}
\usepackage{bbold}
\usepackage{subcaption}
\usepackage{mathtools}
\usepackage{mathrsfs}
\usepackage{amsmath,amssymb}
\usepackage{kantlipsum}
\usepackage{comment}
\allowdisplaybreaks
\usepackage{autobreak}
\usepackage{color}
\usepackage[dvipsnames]{xcolor}

\numberwithin{equation}{section}

\newcommand{\nn}{\nonumber}

\newcommand\beq{\begin{equation}}
\newcommand\eeq{\end{equation}}
\newcommand\beal{\begin{aligned}}
\newcommand\eeal{\end{aligned}}
\newcommand\bea{\begin{eqnarray}}
\newcommand\eea{\end{eqnarray}}

\newcommand\dd{{\mathrm d}}
\usepackage{xcolor}

\newcommand{\bb}{{\boldsymbol b}}
\newcommand{\bk}{{\boldsymbol k}}

\newcommand{\bn}{{\boldsymbol n}}
\newcommand{\bp}{{\boldsymbol p}}
\newcommand{\bP}{{\boldsymbol P}}
\newcommand{\bL}{{\boldsymbol L}}

\newcommand{\bq}{{\boldsymbol q}}
\newcommand{\br}{{\boldsymbol r}}

\newcommand{\bx}{{\boldsymbol x}}
\newcommand{\bv}{{\boldsymbol v}}

\newcommand{\bz}{{\boldsymbol z}}
\newcommand{\bS}{{\boldsymbol S}}
\newcommand{\ba}{{\boldsymbol a}}

\newcommand{\bF}{{\boldsymbol F}}

\newcommand{\PV}{\mathscr{P}}
\newcommand{\sg}{\textrm{Sign}}

\newcommand{\eps}{{\varepsilon}}
\newcommand{\ord}[1]{^{(#1)}}

\def\ddl{\delta\!\!\!{}^-}

\makeatletter
\newcommand{\Biggg}{\bBigg@{3.5}}
\makeatother

\begin{document}
\preprint{DESY\, 26-050\\\phantom{~}}
\title{Black Hole Dynamics at Fifth Post-Newtonian Order} 
\author[a]{\center \large Rafael A. Porto}
\author[b]{\large and Massimiliano M. Riva}
\affiliation[a]{Deutsches Elektronen-Synchrotron DESY, Platanenallee 6, 15738 Zeuthen, Germany.}
\affiliation[b]{Department of Physics and Center for Theoretical Physics,\\ Columbia University, New York, 538 West 120th Street, NY 10027, USA.}


\abstract{Using the worldline in-in effective action obtained in~\cite{Porto:2024cwd}, we derive the total even-in-velocity {\it relative} impulse, scattering angle, and time delay at fifth post-Newtonian (5PN) order, including radiation-reaction and hereditary contributions at ${\cal O}(G^5\nu^2)$ and ${\cal O}(G^6\nu^2)$. We introduce an isotropic-like description which, together with the associated losses of energy and angular momentum, fixes the evolution of the system entirely from scattering data. This framework opens the door to an unambiguous characterization of the underlying two-body dynamics solely in terms of scattering observables. Following \cite{Porto:2024cwd}, we isolate a {\it conservative} component using Feynman's $i0^+$ prescription. This sector contains both ``tail-like'' and ``memory-like'' contributions, the latter being intrinsically nonlocal in time and described by a double Principal-Value integral. Owing to the local-in-time character of the corresponding (in-in) action, we establish a systematic procedure that is consistent with Feynman's prescription while preserving the complete local dynamics. This provides a universal contribution to the conservative (isotropic) Hamiltonian at 5PN order and, as a byproduct, also fixes the value of the Effective One Body coefficients $\{{\bar d}_{5{\rm loc}}, a_{6{\rm loc}}\}$ consistently with the {\it Tutti-Frutti} framework. For completeness, we analyse the ``$\gamma\text{-}3$" prescription introduced in recent post-Minkowskian computations. When implemented in our formalism, we find exact agreement over the overlapping regime of validity. In contrast, Feynman's prescription yields a (local) memory-like contribution with the {\it opposite} sign at ${\cal O}(G^5\nu^2)$. We also find that an analogous  $\gamma\text{-}3$ rerouting at ${\cal O}(G^6\nu^2)$ would be incompatible with the conjecture that all $\pi^2$ terms arise solely from the potential region, while Feynman's formulation preserves this expectation.}
\maketitle

\section{Introduction}

The maturity of gravitational-wave (GW) astronomy \cite{LIGOScientific:2025slb} combined with the  scientific potential of next-generation detectors \cite{ET:2025xjr,Caprini:2025mfr}, has turned precision modeling of the two-body problem in general relativity into a top priority \cite{Buonanno:2014aza,Porto:2016zng,Porto:2017lrn,AlvesBatista:2021eeu,Bernitt:2022aoa}.  For compact binaries, Post-Newtonian (PN) theory \cite{Blanchet:2013haa,Rothstein:2014sra,Schafer:2018jfw,Porto:2016pyg,Goldberger:2022ebt} remains the natural language for the slow-motion, weak-field regime relevant to the inspiral phase, while Post-Minkowskian (PM) methods endowed with modern amplitude techniques provide complementary access to the high-velocity unbound~regime~\cite{Neill:2013wsa,Bjerrum-Bohr:2018xdl,Cheung:2018wkq, Bern:2019crd, Kalin:2020mvi,Kalin:2020fhe,Liu:2021zxr,Kalin:2022hph,Mougiakakos:2021ckm,Dlapa:2021npj,Dlapa:2021vgp,Dlapa:2022lmu,Jakobsen:2022psy,Goldberger:2022ebt,Dlapa:2023hsl, Brandhuber:2021eyq,Bern:2021dqo,Bern:2021yeh,Manohar:2022dea,DiVecchia:2023frv,Driesse:2024xad,Driesse:2024feo,Bern:2024adl,Cheung:2024byb,Bini:2024icd,Bini:2025vuk,Heissenberg:2025fcr,Heissenberg:2025ocy,Bern:2025wyd,Driesse:2026qiz}. In~principle, scattering observables can also be {\it analytically continued} to the bound problem through the Boundary-to-Bound (B2B) correspondence \cite{Kalin:2019rwq,Kalin:2019inp,Cho:2021arx} to describe elliptic motion. In recent years it has therefore become increasingly clear that these expansions are not merely parallel developments: gauge-invariant observables such as the scattering angle furnish a sharp bridge between unbound and bound dynamics, offer powerful internal consistency checks, and provide an efficient route to construct local-in-time (isotropic-like) universal descriptions once hereditary pieces are properly identified \cite{Dlapa:2024cje,Bini:2024tft,Dlapa:2025biy}.\vskip 4pt

The near-zone two-body dynamics is governed not only by instantaneous (``potential") interactions, but also nonlinear radiation-reaction effects that include hereditary phenomena, such as tails, tail-of-tails, and memory-type effects, which intertwine the near-zone dynamics with radiative degrees of freedom \cite{Blanchet:2013haa,Rothstein:2014sra,Schafer:2018jfw,Porto:2016pyg,Goldberger:2022ebt}.  Capturing these corrections requires a  formalism that incorporates all phenomena in a unified framework that can extend to high PN orders. This is precisely where the (``in-in") worldline effective field theory (WEFT) approach has become a particularly powerful and systematic tool to tackle the two-body problem in gravity \cite{Goldberger:2004jt,Goldberger:2005cd,Porto:2005ac,Porto:2006bt,Porto:2008tb,Porto:2008jj,Goldberger:2009qd,Galley:2009px,Galley:2010es,Galley:2012qs,Galley:2012hx,Goldberger:2012kf,Galley:2013eba,Galley:2014wla,Galley:2015kus,Maia:2017yok,Maia:2017gxn,Porto:2017dgs,Foffa:2019rdf,Foffa:2019yfl,Foffa:2019hrb,Goldberger:2020fot,Goldberger:2020wbx,Blumlein:2020pyo,Almeida:2021xwn,Blumlein:2021txe,Almeida:2023yia,Cho:2022syn,Leibovich:2023xpg,Amalberti:2023ohj,Amalberti:2024jaa,Almeida:2024lbv,Brunello:2025gpf,Brunello:2026anu}. Using the WEFT formalism, we have recently derived in~\cite{Porto:2024cwd} the contribution to the \emph{relative} effective action from nonlinear radiative effects at 5PN order. Moreover, following \cite{Galley:2009px, Kalin:2022hph}, we also identified a conservative sector via Feynman's $i0^+$ prescription---leading naturally to Principal-Value structures---and highlighted additional, qualitatively distinct sources of conservative-like contributions, {\it i)} terms \emph{quadratic} in the leading radiation-reaction force, and {\it ii)} nonlocal-in-time memory effects, which had not been included in previous treatments.\vskip 4pt 

In our previous work \cite{Porto:2024cwd}, we derived the relative (even-in-velocity) impulse and deflection angle at ${\cal O}(G^4\nu^2)$, finding agreement in the overlap with the 4PM value in~\cite{Dlapa:2022lmu}.  This resolved the apparent tension with earlier WEFT computations in~\cite{Blumlein:2021txe,Almeida:2023yia},\footnote{A rederivation by the authors of \cite{Almeida:2023yia}, with a waveform reconstruction shown to agree with the Multipolar-Post-Minkowskian (MPM) result~\cite{Blanchet:1985sp,Blanchet:2013haa}, appeared in~\cite{Almeida:2024lbv}. It~was also explicitly shown that the expression inferred from our stress-energy tensor in~\cite{Porto:2024cwd} is equivalent to the MPM waveform. Unfortunately, the scattering angle in~\cite{Almeida:2024lbv} remains in tension with the values in~\cite{Porto:2024cwd,Dlapa:2021npj,Dlapa:2021vgp,Dlapa:2022lmu} as well as the results presented here (and in \cite{Driesse:2026qiz}).} while emphasizing the bigger picture of how nonlinear radiation-reaction effects organize themselves. For instance, the existence of tail-like effects including not only the standard mass, current, and angular-momentum couplings, but also quadrupole moments; as well as memory-like contributions which translate to products of Principal-Value integrals in the conservative~sector. \vskip 4pt  

In the present work, we build directly on the effective action obtained in \cite{Porto:2024cwd} to compute the remaining radiation-reaction and hereditary ${\cal O}(G^5)$ and ${\cal O}(G^6)$ corrections to black hole scattering---thus completing the knowledge of the even-in-velocity binary dynamics at 5PN order.\footnote{In contrast, in the odd-in-velocity sector, several results for the relative impulse and the radiated energy and angular momentum are already available, some of which have been derived using linear response theory~\cite{Bini:2021gat,Bini:2022enm}. A comparison between high-order PM \cite{Dlapa:2022lmu,Driesse:2024feo} against PN results \cite{Bini:2020wpo}, including both odd- and even-in-velocity contributions, can also be found in~\cite{Bini:2025vuk}.}  
In particular, we compute the impulse, $\Delta \bp$, scattering angle, $\chi$, and time delay, $\tau$, as well as the losses of energy, $\Delta E$, and angular momentum, $\Delta \bL_z$, perpendicular to the plane. While all of the nonlinear effects contribute to the impulses and time delay, only terms at second order in the leading radiation-reaction force dissipate energy/angular-momentum. Although energy-conserving, it can be shown that nonlinear gravitational forces do not necessarily derive from a Hamiltonian/Lagrangian description. Nonetheless, we show that, once the isotropic gauge inherited from scattering computation is at hand, the evolution of the system is unambiguously fixed by the knowledge of the set $\{\chi,\tau,\Delta E,\Delta \bL_z\}$ of boundary quantities. The resulting picture thus provides a compact, self-contained framework that can be used to reconstruct the full  (dissipative) binary dynamics entirely from scattering data.\vskip 4pt

Following the analysis of \cite{Porto:2024cwd}, we isolate a conservative sector by means of Feynman's prescription. Beginning at ${\cal O}(G^5)$ the Feynman decomposition generically entails both local and  nonlocal-in-time memory-like contributions. Exploiting the lack of nonlocality in the corresponding in-in effective action, we establish a local-in-time separation of the full dynamics into conservative and dissipative sectors in a way that both {\it i)} remains consistent with the {\it on-shell} Feynman prescription and {\it ii)} preserves the complete local dynamics.~This is achieved by computing the on-shell (radial) action and identifying a local-in-time component through the Poincar\'e-Bertrand (PB) decomposition of the Principal-Value integrals.~The resulting PB-inspired procedure provides a universal contribution to the conservative sector at 5PN order, while preserving locality in time throughout. As a side product, the PB prescription determines the universal value of the Effective One Body (EOB) coefficients ${\bar d}_{5{\rm loc}}$ and $a_{6{\rm loc}}$, in a manner that is consistent with the ``Tutti-Frutti" framework~\cite{Bini:2020wpo}, as well as the conjecture that all $\pi^2$ contributions originate exclusively from the potential region to 5PN order \cite{Bini:2025vuk}.\vskip 4pt

We examine other possible ways of disentangling conservative and dissipative contributions within the effective theory. For instance, we show that performing an {\it off-shell} split of the Feynman action leads to an explicit cancellation among the nonlocal terms, thus allowing for a formulation that remains entirely local and consistent with a variational principle. However, because this construction eliminates the information from the Principal Value, its straightforward universal implementation---across all integration regions---would lead to a scattering dynamics that  deviates from the previously established conservative sector already at ${\cal O}(G^4)$. We also show that an appropriate {\it rerouting} of the in-in action can be exploited to define a conservative-like sector, in close analogy with the $\gamma\text{-}3$ prescription recently proposed in the context of PM scattering in \cite{Driesse:2026qiz}. When implemented within our WEFT formalism in the PN context, we find perfect agreement throughout their common regime of validity. This provides nontrivial support for an equivalence between the two approaches at the level of the integrand construction. However, we find that the $\gamma\text{-3}$ rerouting  produces a memory-like contribution at ${\cal O}(G^5\nu^2)$ with precisely the {\it opposite} sign relative to Feynman's prescription. We~further observe that the analogous $\gamma\text{-}3$ prescription, formulated in terms of oriented retarded propagators, would be difficult to reconcile with the $\pi^2$ conjecture. On the other hand, the PB prescription preserves this expectation in a particularly natural manner.  We elaborate on the broader implications of these constructions in \S\ref{outlook}.\vskip 4pt  The rest of this paper is organized as follows:\vskip 4pt  

In~\S\ref{eft} we briefly review the WEFT framework and revisit the conservative/dissipative split introduced in \cite{Porto:2024cwd}. We further distinguish two separate classes of second-order radiation-reaction contributions. The first is generated by feeding the radiation-reaction acceleration back into itself; following \cite{Porto:2024cwd}, we refer to these as RR$^2$ effects. The second arises from evaluating the linear force along radiation-reacted trajectories. We denote these by RR--RR, and show that they can likewise induce conservative-type contributions. For the conservative sector, we will often refer to the combination of both such terms as 2RR effects. \vskip 4pt 

In \S\ref{data} we summarize the resulting expressions for the total even-in-velocity relative impulse and scattering angle, together with the local conservative sectors. In \S\ref{isotropic} we introduce an isotropic-like representation of the full dynamics, and show that it is fixed by the set of boundary observables $\{\chi,\tau,\Delta E,\Delta \bL^z\}$. We also demonstrate the  equivalence to the forces obtained in the original (background) harmonic gauge. This representation makes it possible to disentangle conservative and dissipative effects in a transparent manner, while also revealing non-Hamiltonian conservative-like terms. Building on the analysis of \S\ref{eft}, in conjunction with the B2B dictionary, we isolate a (universal) local-in-time Hamiltonian sector through the corresponding scattering angle. We use this information to determine the missing EOB coefficients in the Tutti-Frutti formalism at 5PN order. \vskip 4pt

In \S\ref{alternative}  we explore alternative ways of separating conservative and dissipative contributions. We discuss both an off-shell split of the Feynman action and possible reroutings of the in-in action that can define a local-in-time conservative sector. We show how these prescriptions are incompatible with the conservative results established through the on-shell Feynman approach at 4PM order. Finally, we discuss the $\gamma\text{-}3$ prescription introduced in \cite{Driesse:2026qiz}.  We explicitly show how it effectively removes the dependence on the time routing provided by the Principal Value, such that its application leads to various modifications relative to the PB prescription. We conclude with a summary and outlook in~\S\ref{outlook}. Several details are relegated to various appendices, and a collection of the main results can be found in the ancillary file.   
\newpage
\section*{List of conventions}

\begin{itemize}
	\item We use the mostly minus signature $\eta_{\mu\nu} = \textrm{diag}(+, -, -, -)$ for the Minkowski metric.
	\item $\hbar = c = 1$, $ \kappa = \sqrt{32 \pi G} = m_{ \textrm{pl}}^{-1}$.
	\item  $\ddl{}^n(x) = (2\pi)^n\delta^n(x)$.
	\item We use Einstein's conventions for summations over repeated indices. To avoid confusion with the choice of metric signature, we use the Euclidean ${\bf 3}$-metric whenever results are written with space-like indices irrespectively of their (up or down) position.  
	\item We use (square) round brackets to identify a group of totally (anti-)simmetrized indices, e.g.
		\[
			A_{(\mu|} C_{\rho} B_{|\nu)} = \frac{1}{2}\big(A_\mu C_\rho B_\nu + A_\nu C_\rho B_\mu\big) \,,
			\qquad 
			A_{[\mu|} C_{\rho} B_{|\nu]} = \frac{1}{2}\big(A_\mu C_\rho B_\nu - A_\nu C_\rho B_\mu\big) \, .
		\]
	\item We work with dim. reg. in $d= 3-2\epsilon$ dimensions, and use the following shorthand for the $d+1$ and $d$ dimensional integrals,
		\[
		\int_{k, q, \cdots} \equiv \int \frac{\dd^{d+1} k}{(2\pi)^{d+1}}\frac{\dd^{d+1} q}{(2\pi)^{d+1}}\cdots \, ,
		\qquad
		\int_{\bk, \bq, \cdots} \equiv \int \frac{\dd^{d} \bk}{(2\pi)^{d}}\frac{\dd^{d} \bq}{(2\pi)^{d}}\cdots
		\]

	\item We use the convention
\begin{equation}
	I\ord{n}_{ab}(t) \equiv \frac{\dd ^n I_{ab}(t)}{\dd t^n}
\end{equation}
for the time derivatives of the quadrupole moment(s).
	\item We use the convention
	\beq
	f(x) = \int_k f(k) e^{-ik \cdot x}\,,
	\eeq
	for the Fourier transform.
	
\end{itemize}

\newpage

\section{Worldline effective theory}\label{eft} 

\subsection{Brief introduction}
The effective action describing the binary takes the form%
\begin{equation}
	S = S_{\textrm{EH}} + S_{\textrm{source}} \, ,
	\label{eq:Seff}
\end{equation}
where $S_{\textrm{EH}}$ is the standard Einstein-Hilbert term, 
\beq
S_{\rm EH} = -\frac{2}{\kappa^2} \int d^{d+1} x \sqrt{-g} R\,,
\eeq
and the source part given by \cite{Goldberger:2009qd}
\begin{align}
		S_{\textrm{source}} & = - \int \dd \lambda \,  \bigg\{ \sqrt{g_{\mu\nu} V^\mu(\lambda) V^\nu(\lambda)} M(\lambda) + \frac{1}{2} \omega^{ab}_\mu L_{ab}(\lambda)V^\mu(\lambda)  - \frac{1}{2} I^{ab}(\lambda) \frac{E_{ab}}{\sqrt{g_{\mu\nu}V^\mu V^\nu} }+ \cdots \bigg\}\, ,
		\label{eq:StartingAction}
\end{align}
with $\big\{M(\lambda), L^{ab}(\lambda),I^{ab}(\lambda),\cdots \big\}$, the total mass/energy, angular momentum, symmetric-trace-free (STF) quadrupole moment, and so on, with the ellipses denoting higher multipoles. Greek indices $\mu,\nu,\ldots$ label spacetime components, whereas latin indices $a,b,\ldots$ refer to local-frame tensors obtained by projection with a tetrad $e^\mu_a$, with $e^\mu_0 = V^\mu$, satisfying
$g^{\mu\nu} = e_0^\mu e_0^\nu - \delta^{ab} e_a^\mu e_b^\nu \,$. The parameter $\lambda$ is an arbitrary affine parameter along the center-of-mass worldline $X^\mu(\lambda)$ describing the binary, and $V^\mu \equiv \dd X^\mu/\dd\lambda$ is its four-velocity. We choose $\lambda \equiv X^0=t$ and focus solely on the relative dynamics, so that $X^\mu(t)=(t,0,0,0)$ and $V^\mu=(1,0,0,0)$.  The binary's rotation is encoded in the angular-momentum tensor, which couples to gravity through the spin connection $\omega^{ab}_\mu$, defined in the standard way as
\begin{equation}
	\omega^{ab}_\mu \equiv g^{\rho\sigma}e^b_\sigma  e^a_{\rho;\mu} \, .
\end{equation}
Similarly, the quadrupole couples to the electric part of the Weyl tensor $\mathcal{C}_{\mu\rho\nu\sigma}$, projected onto the local frame,
\begin{align}
	E_{ab} \equiv e_a^\mu e^\nu_b \mathcal{C}_{\mu\rho\nu\sigma} V^{\rho} V^{\sigma} \, .
\end{align}
We impose a supplementary condition for the rotational sector, which in the local frame becomes $L^{\mu\nu}e_\mu^{0}=L^{\mu0}=0$. An analogous restriction holds for the quadrupole, $I^{a0}=0$, since $E_{\mu\nu}V^\mu=0$. We also introduce the angular-momentum three-vector $\bL^i=(1/2)\varepsilon^{ijk}L^{jk}$, with indices contracted using the Euclidean metric $\delta^{ij}$, and where $\varepsilon^{ijk}$ is the flat Levi-Civita symbol normalized by $\varepsilon^{123}=+1$.\vskip 4pt In the weak-field regime, the metric is decomposed into a background plus fluctuations,
\begin{equation}
	g_{\mu\nu} = \bar{g}_{\mu\nu} + \kappa h_{\mu\nu} \, , \qquad \qquad 
	\bar{g}_{\mu\nu} =  \eta_{\mu\nu} + \kappa \bar{h}_{\mu\nu} \, ,
\end{equation}
and likewise for the tetrad.\vskip 4pt 

Dissipative effects are included via the in-in formalism \cite{Keldysh:1964ud,Calzetta:1986cq}, which requires doubling all fields and leads to a closed-time-path action \cite{Galley:2009px,Galley:2012qs,Goldberger:2012kf,Galley:2013eba,Galley:2014wla,Galley:2015kus,Maia:2017yok,Maia:2017gxn,Kalin:2022hph,Dlapa:2022lmu},
\begin{equation}
S^{c} \equiv S_1 - S_2 = S_{\textrm{EH}, 1} + S_{\textrm{source}, 1} - S_{\textrm{EH}, 2} + S_{\textrm{source}, 2} \, .
\end{equation}
We adopt Keldysh's $\pm$ parametrization, i.e. for any field $\Phi$,
\begin{equation}
	\Phi_+ = \frac{\Phi_1+\Phi_2}{2}  \, , \qquad \qquad \Phi_- = \Phi_1 - \Phi_2 \, .
\end{equation}
The effective action is obtained by integrating out the $h_{\pm,\mu\nu}$ fluctuations. To the order that is relevant here, the result  depends on the following (STF) multipole moments 
\beq
\begin{aligned}	
M_+ &= \sum_{a=1,2} m_a \left[
1+\frac{v^2_{a, +}}{2} - \frac{1}{2}\sum_{b\neq a}\frac{G m_b}{r_+}
\right]\,,\\
M_- &= \sum_{a=1,2} m_a \left[
1+\bv_{a, +}\cdot \bv_{a, -} + \frac{1}{2}\sum_{b\neq a}G m_b\frac{\br_{+}\cdot \br_-}{r_+^3}
\right]\,,\\
L_-^{ij} &= 2\sum_{a=1,2} m_a\bigg ( \bx_{a,-}^{[i}\bv_{a,+}^{j]} + \bx_{a,+}^{[i}\bv_{a,-}^{j]} \bigg)\,,\\
L_+^{ij} &= 2\sum_{a=1,2} m_a \bx_{a,+}^{[i}\bv_{a,+}^{j]}\,,\\
I_-^{ij} &= \sum_{a=1,2} m_a \bigg( 2 \bx_{a, -}^{(i}\bx_{a, +}^{j)}- \frac{2}{3}\delta^{ij}\bx_{a, -}\cdot \bx_{a, +} \bigg) \,,\\
I_+^{ij} &= \sum_{a=1,2} m_a \bigg(  \bx_{a, +}^{i}\bx_{a, +}^{j}- \frac{1}{3}\delta^{ij}\bx_{a, +}\cdot \bx_{a, +} \bigg)\,,
\end{aligned}
\eeq
with $m_a$, $\bx^i_a$, and $\bv^j_a$, the masses, positions, and velocities, and $r\equiv |\bx_1-\bx_2|$. It is useful to introduce a non-TF part of the quadrupole, $Q_{ij} \equiv \sum_a m_a \bx^i_a\bx^j_a$, which enters in the conservative sector through nonlinear radiation-reaction effects. See \cite{Porto:2024cwd} for more details.

\subsection{Radiative action} 
After integrating out the gravitational field, we arrived at the result \cite{Porto:2024cwd},\footnote{
We ignore here the boundary term in the soft limit discussed in \cite{Porto:2024cwd}, which does not affect the total impulse.}
\begin{align}	
\label{eq:fin} 
	&S_{\rm 5PN} =  S_{\rm (P)}+S_{
	\rm (RR)} + S_{\rm (T)}  -\frac{G^2}{15}\int \! \dd t \, L_{+}^{kl}I\ord{4}_{-,kj}I\ord{3}_{+, jl}
		 + \frac{G^2}{30}\int \! \dd t \, L_{-}^{kl} I\ord{4}_{+,kj}I\ord{3}_{+, jl}  \\
		& + \frac{G^2}{5} \int \dd t \left(
	\frac{1}{2} I_{-, ij} I\ord{4}_{+, jk} I\ord{4}_{+, ki} 
	- I\ord{4}_{-, ij} I\ord{4}_{+, jk} I_{+, ki} 
	+ \frac{1}{7} I\ord{2}_{-, ij} I\ord{3}_{+, jk} I\ord{3}_{+, ki}
	+ \frac{2}{7} I\ord{3}_{-, ij} I\ord{3}_{+, jk} I\ord{2}_{+, ki} 
	\right) \,,\nn
	\end{align}
where $S_{\rm (P)}$ is the potential-only part \cite{Blumlein:2020pyo}, $S_{\rm (RR)}$ is the leading radiation-reaction contribution, 
\beq
\label{eq:srr}
S_{\rm (RR)} = -\frac{G}{5} \int \dd t \, I^{ij}_{-}(t) I^{ij(5)}_+(t)\,,
\eeq
and $S_{\rm (T)}$ encompasses all the relevant (mass/energy) tail corrections \cite{Galley:2015kus,Almeida:2021xwn},  
\begin{align}\label{tails}
S_{I_2\rm (T)} &= \frac{2M_+G^2}{5}\int \frac{\dd\omega}{2\pi} \omega^6  I_-^{ij}(-\omega)I_+^{ij}(\omega)  \left[
	\frac{1}{2\epsilon} +\frac{41}{30} + i \pi\, \textrm{sign}(\omega) - \log \left(\frac{\omega^2e^{\gamma_{\textrm{E}}}}{\pi\mu^2}\right)
	\right] \,   \\
	&
	-\frac{G^2M_-}{5}\int \frac{\dd\omega}{2\pi} \omega^6  I_+^{ij}(-\omega)I_+^{ij}(\omega)  \left[
	\frac{1}{2\epsilon} +\frac{41}{30}  - \log \left(\frac{\omega^2e^{\gamma_{\textrm{E}}}}{\pi\mu^2}\right)
	\right]\,,\nn
	\\ 
S_{I_3\rm (T)} &= \frac{2G^2 M_+}{189}\int \frac{\dd\omega}{2\pi} \omega^8  I_-^{ijk}(-\omega)I_+^{ijk}(\omega)  \left[
	\frac{1}{2\epsilon} + \frac{82}{35} + i \pi \,\textrm{sign}(\omega) - \log \left(\frac{\omega^2e^{\gamma_{\textrm{E}}}}{\pi\mu^2}\right)
	\right] \, , \nn\\
S_{J_2\rm(T)} &= \frac{32G^2 M_+}{90}\int \frac{\dd\omega}{2\pi} \omega^6  J_-^{a|ij}(-\omega)J_+^{a|ij}(\omega) \left[
	\frac{1}{2\epsilon} + \frac{49}{20} + i \pi \,\textrm{sign}(\omega) - \log \left(\frac{\omega^2 e^{\gamma_{\textrm{E}}} }{\pi\mu^2}\right)
	\right] \,. \nn 
\end{align}
The remaining pieces in \eqref{eq:fin}, computed in \cite{Porto:2024cwd}, include the so-called `failed-tail' and `memory' contributions, which involve the angular-momentum and quadrupole moments, respectively.\vskip 4pt   

The equations of motion follow from (with $L_{\rm(P)}$ the potential-only Lagrangian) \cite{Galley:2015kus,Porto:2024cwd}, 
\begin{equation}
\label{eq:eom}
\left({\delta \over \delta \bx^i_{a,-}} \int S[\bx_\pm] dt \right)\bigg|_{\textrm{PL}}=0 \to 
		\frac{\dd}{\dd t}\frac{\partial L_{\rm (P)}[\bx_a]}{\partial \bv^i_{a}}-\frac{\partial L_{\rm (P)}[\bx_a]}{\partial \bx^i_{a}}  = \bigg[\frac{\partial R[\bx_\pm]}{\partial \bx^i_{a,-}} 
		-\frac{\dd}{\dd t}\frac{\partial R[\bx_\pm]}{\partial \bv^i_{a,-}} \bigg]\bigg|_{\textrm{PL}} \,,
\end{equation}
in the  ``Physical Limit" (PL), i.e., $\bx_{a, +} \to \bx_a$, $\bx_{a, -} \to 0$, and we obtained~\cite{Porto:2024cwd}
\begin{align}
	\ba_{(\textrm{FT})} & = \frac{G^3 M^3 \nu^2}{r^4}\bigg[
	\bigg(
	36 v^6 - 288 v^4 (\bv\cdot \bn)^2 + 112 v^2 (\bv\cdot \bn)^4 + 252 (\bv\cdot \bn)^6
	\bigg) \bn \notag \\
	& \qquad\qquad - \bigg(
	84 v^4 (\bv\cdot \bn) - 588 v^2 (\bv\cdot \bn)^3 + 616 (\bv\cdot \bn)^5 
	\bigg)
	\bv
	\bigg] \notag \\
	& -
	\frac{4G^4 M^4 \nu^2}{5r^5} \bigg[
	\bigg(
	47 v^4  + 163 v^2 (\bv\cdot \bn)^2 - 630 (\bv\cdot \bn)^4
	\bigg) \bn  - \bigg(
	350 v^2 (\bv\cdot \bn) - 770 (\bv\cdot \bn)^3 
	\bigg)
	\bv
	\bigg] \notag \\
	& +\frac{8G^5 M^5 \nu^2}{15r^6} \bigg[
	\bigg(
	11 v^2  + 199 (\bv\cdot \bn)^2 \bigg) \bn 
	- 210 (\bv\cdot \bn) \bv
	\bigg]  \, , \label{eq:FTrela} \\
		\ba_{(\textrm{M})} & = -\frac{2 G^3 M^3 \nu^2}{35 r^4}\bigg[
	\bigg(
	2484 v^6 - 38325 v^4 (\bv\cdot \bn)^2 + 91770 v^2 (\bv\cdot \bn)^4 - 56385 (\bv\cdot \bn)^6
	\bigg) \bn \notag \\
	& \qquad\qquad + \bigg(
	9633 v^4 (\bv\cdot \bn) - 33870 v^2 (\bv\cdot \bn)^3 + 24885 (\bv\cdot \bn)^5 
	\bigg)
	\bv
	\bigg] \notag \\
	& +
	\frac{4G^4 M^4 \nu^2}{315r^5} \bigg[
	\bigg(
	13900 v^4  - 95892 v^2 (\bv\cdot \bn)^2 + 76680 (\bv\cdot \bn)^4
	\bigg) \bn +   \notag \\
	& \hspace{2cm} \bigg(
	32163 v^2 (\bv\cdot \bn) - 35955 (\bv\cdot \bn)^3 
	\bigg)
	\bv
	\bigg] \notag \\
	& -\frac{16G^5 M^5 \nu^2}{1260 r^6} \bigg[
	\bigg(
	2028 v^2  + 9421 (\bv\cdot \bn)^2 \bigg) \bn 
	- 393 (\bv\cdot \bn) \bv
	\bigg]  
	-\frac{824G^6 M^6 \nu^2}{63 r^7} \bn\, , \label{eq:Mrela} 
\end{align}
where $\ba \equiv \ba_1-\ba_2$, $M=m_1+m_2$, $\nu = \frac{m_1m_2}{M^2}$, $\br \equiv \bx_1-\bx_2$, $\bn \equiv \br/r$, and $\bv \equiv \bv_1-\bv_2$.\vskip 4pt The leading radiation-reaction $S_{\rm (RR)}$ induces both, the standard (Burke-Thorne)  force,\begin{align}
	\ba_{\rm (RR)} & =  \frac{8 G^2 M^2 \nu}{5 r^3}\bigg[
	\bigg(18 v^2-25 (\bv\cdot \bn)^2 \bigg) (\bv\cdot \bn) \bn
	-\bigg(6 v^2 - 15(\bv\cdot \bn)^2
	\bigg) \bv
	\bigg]   + \frac{16 G^3 M^3 \nu}{5 r^4} \bigg[
	\bv +\frac{\bv\cdot \bn}{3} \bn
	\bigg] \, ,
	\label{eq:BTaLO} 
\end{align}
plus a second-order term,
\begin{align}
\label{eq:BTaNLO}
\ba_{\rm (RR^2)}& =  \frac{16 G^3 M^3 \nu^2}{5 r^4}\bigg[
	\bigg(
	168 v^6 -2496 v^4 (\bv \cdot \bn)^2+5789 v^2 (\bv \cdot \bn)^4
	-3465 (\bv \cdot \bn)^6
	\bigg)  \bn \\
	& \hspace{2cm} +\bigg(
	738 v^4-2449 v^2 (\bv \cdot \bn)^2
	+1715 (\bv \cdot \bn)^4
	\bigg) (\bv \cdot \bn)\bv
	\bigg]  \notag \\
	& - \frac{32 G^4 M^4\nu^2}{225 r^5}\bigg[
	\bigg(
	3898 v^4-28131 v^2 (\bv \cdot \bn)^2+22755 (\bv \cdot \bn)^4
	\bigg)\bn \notag \\
	& \hspace{2cm} +6 \bigg(
	1921 v^2-1830 (\bv \cdot \bn)^2
	\bigg)(\bv \cdot \bn) \bv
	\bigg] \notag \\
	& + \frac{64 G^5 M^5 \nu^2}{225 r^6}\bigg[
	\bigg(
	311 v^2 + 689(\bv \cdot \bn)^2
	\bigg)\bn
	-234(\bv \cdot \bn) \bv
	\bigg]
	+\frac{1792 G^6 M^6 \nu^2}{75 r^7} \bn\, , \notag
\end{align}
after plugging the Newtonian acceleration, \beq \label{eq:N} \ba_{\rm (N)} = -GM\frac{\br}{r^3}\,,\eeq
and the radiation-reaction acceleration onto itself, respectively. 

\subsection{Deriving the impulse}\label{sec:impulse}

 We derive the coefficients of the relative impulse in a PM expansion,
\begin{equation}
	\Delta \bp
	= M \nu \int_{-\infty}^{+\infty} \dd t \, \ba(t) = \sum_n G^n \Delta^{(n)} \bp\,,
	\label{eq:relpgen_imp}
\end{equation}
 up to ${\cal O}(G^6\nu^2)$, arising from contributions involving all of the above accelerations,\footnote{The effects due to $S_{\rm (P+T)}$ are already known in the literature, see e.g. \cite{Bini:2021gat,Bini:2022enm}, and will be added later on.} 
\begin{equation}
	\ba(t) = \ba_{(\rm N)}(t) + \ba_{(\rm RR)}(t) + \ba_{(\rm RR^2)}(t)
	+ \ba_{(\rm FT)}(t) + \ba_{(\rm M)}(t)\,.
	\label{eq:accsplit}
\end{equation}
We solve the equations of motion iteratively, and expand the trajectory around straight motion,
\begin{equation}
	\br(t) = \br_0(t) + \delta \br(t)\,,
	\qquad
	\bv(t) = \bv_0 + \delta \bv(t)\,,
	\end{equation}
where	
	\begin{equation}
	\br_0(t) = \bb + \bv_\infty t\,,
	\qquad
	\bv_0 = \bv_\infty\,,
\end{equation}
with $(\bb$, $\bv_\infty)$ the impact parameter and incoming (relative) velocity at infinity, respectively,  
and the associated deflections given by
	\begin{equation}
	\delta \br(t) = \sum_X \delta_{(\textrm{X})} \br(t)\,, \quad \delta \bv(t) =  \sum_X \delta_{(\textrm{X})} \bv(t)\,.
\end{equation}
For each $X\in\{\rm N,RR,RR^2,FT,M\}$ the perturbation satisfies
\begin{equation}
	\delta_{(\textrm{X})} \ddot{\br}(t) = \ba_{(\textrm{X})}[\br(t)]\,,
	\label{eq:deltax}
\end{equation}
with the understanding that the right-hand side is consistently expanded to the desired order.\vskip 4pt

For the derivation of effects at second order in the radiation-reaction force it is convenient to separate the computation into two types. In addition to the impulse generated by $\ba_{\rm (RR^2)}$, we also have the effects due to the linear acceleration, $\ba_{\rm (RR)}$, on the radiation-reacted deflection, schematically,
\beq
 \delta_{\rm (RR\text{-}RR)} \ddot\br = \ba_{(\rm RR)}[\delta_{(\rm RR)}\br]\,,\label{eq:rrrr}
\eeq
which we denote as RR--RR effects. As it turns out, restricted to the 5PN order, the former conserves energy and angular-momentum, whereas the latter is the only source of dissipation from the novel nonlinear effects derived in \cite{Porto:2024cwd}. Nevertheless, as we shall see, a subset of RR--RR terms may as well be incorporated into a conservative sector.

\subsubsection*{N \& RR deflections}

The derivation of the impulse involves the solution for the trajectories.  In order to derive the impulse through ${\cal O}(G^6)$, the Newtonian deflection up to ${\cal O}(G^2)$, %
\begin{equation}
\label{eqbrN}
	\delta_{(\textrm{N})}\br
	= G \, \delta_{(\textrm{N})}\ord{1}\br
	+ G^2 \, \delta_{(\textrm{N})}\ord{2}\br\,,
\end{equation}
and similarly for $\delta_{(\textrm{N})}\bv$, turn out to be sufficient. At first order we have\footnote{There is a subtlety in the derivation of the trajectory in $d=3$ dimensions, where the expression for the deflection is formally logarithmically divergent,
\begin{equation}
	\delta^{(1)}\br(t) \xrightarrow[t\to -\infty]{} -G M\frac{\bv}{v_\infty^3}\log\bigg(\frac{2v_\infty t}{b}\bigg) \, .
\end{equation}
This can be remediated by dim. reg., such that  $\delta_d^{(1)}\br(t) \sim t^{2\epsilon}$, and regularizing the infrared divergence by taking $d>3\, (\varepsilon<0)$ while sending $t\to-\infty$ prior to $\epsilon \to 0$.}

\begin{equation}
	 \, \delta_{(\textrm{N})}\ord{1}\bv(t)
	= - M \int_{-\infty}^{t} \dd t'\,
	\frac{\bb + \bv_\infty t'}{\big(b^2 + v_\infty^2 t'^2\big)^{3/2}}\,,
	\qquad
	 \, \delta_{(\textrm{N})}\ord{1}\br(t)
	=  \int_{-\infty}^{t} \dd t'\, \delta_{(\textrm{N})}\ord{1}\bv(t')\,.
\end{equation}
The second-order Newtonian deflection then follows from
\begin{equation}
	 \, \delta_{(\textrm{N})}\ord{2}\bv(t)
	= G^{-2}\int_{-\infty}^{t} \dd t'\; \ba_{\rm (N)}\Big|_{\delta^{(1)}_{\rm N}}(t')\,,
	\qquad
 \, \delta_{(\textrm{N})}\ord{2}\br(t)
	=  \int_{-\infty}^{t} \dd t'\, \delta_{(\textrm{N})}\ord{2}\bv(t')\,,
\end{equation}
where we use the notation
\begin{equation}
	\ba_{\rm (N)}\Big|_{\delta^{(1)}_{\rm N}}
	= -G^2 M\bigg[
	\frac{\delta_{(\textrm{N})}\ord{1}\br}{r_0^3}
	-3 \frac{(\br_0 \cdot \delta_{(\textrm{N})}\ord{1}\br)\,\br_0}{r_0^5}
	\bigg]\,,
	\end{equation}
for the acceleration expanded to leading order in the deflection, and so on and so forth.\vskip 4pt For the radiation-reaction corrections, we organize the trajectory as follows (and similarly for $\delta \bv$) 
\begin{align}
	\delta \br(t) & \supset
	G^2 \delta_{\rm (RR)}\ord{1} \br
	+ G^3 \delta_{\rm (RR)}\ord{2} \br
	+ G^4 \delta_{\rm (RR)}\ord{3} \br \, .
	\end{align}
The first- and second-order terms can be solved analytically via,
\begin{align}
	\delta_{\rm (RR)}\ord{1} \bv(t)
	& = G^{-2}\int_{-\infty}^{t} \dd t'\; \ba_{\rm (RR)}\Big|_{\br_0}\,,
	\qquad
	\delta_{\rm (RR)}\ord{1} \br(t)
	= \int_{-\infty}^{t} \dd t'\, \delta_{\rm (RR)}\ord{1}\bv(t')\,,
	\label{eq:RRLOdefl}
	\\
\label{eq:RRNLOdefl}	\delta_{\rm (RR)}\ord{2} \bv(t)
	& = G^{-3} \int_{-\infty}^{t} \dd t'\, \bigg[
	\ba_{\rm (N)}\Big|_{\delta\ord{1}_{\rm RR}}(t')
	+ \ba_{\rm (RR)}\Big|_{\delta\ord{1}_{\rm N}}(t')
	\bigg]\,,
	\qquad
 \delta_{\rm (RR)}\ord{2} \br(t)
	=  \int_{-\infty}^{t} \dd t'\, \delta_{\rm (RR)}\ord{2}\bv(t')\,.
\end{align}
Notice that $\delta_{\rm (RR)}\ord{2}\bv$ receives two contributions, namely the Newtonian acceleration evaluated on the first-order RR deflection, as well as the RR acceleration evaluated on the first-order Newtonian deflection.\vskip 4pt 

At third-order, the RR deflection can be written as
\begin{equation}
	\delta_{\rm (RR)}\ord{3} \bv(t)
	= G^{-4} \int_{-\infty}^{t} \dd t'\, \bigg[
	\ba_{\rm (N)}\Big|_{\delta\ord{1}_{\rm RR}+\delta\ord{1}_{\rm N},\,\delta\ord{2}_{\rm RR}}(t')
	+ \ba_{\rm (RR)}\Big|_{(\delta\ord{1}_{\rm N})^2,\,\delta\ord{2}_{\rm N}}(t')
	\bigg]\,.
\end{equation}
A closed-form expression for $\delta_{\rm (RR)}\ord{3}\br$ will not be necessary. Instead, we keep it as
\begin{equation}
	\delta_{\rm (RR)}\ord{3} \br(t)
	=  \int_{-\infty}^{t} \dd t'\, \delta_{\rm (RR)}\ord{3} \bv(t')\,,
	\label{eq:RRnoExpr}
\end{equation}
which, as discussed in \S\ref{app:integration}, suffices for the computation of the impulse.

\subsubsection*{RR--RR deflections}

For the RR--RR terms the velocity deflections can be written as follows
\begin{align}
	 \delta_{\textrm{(RR\text{-}RR)}}\ord{2} \bv(t)
	& =G^{-4} \int_{-\infty}^{t} \dd t'\;
	\ba_{\rm (RR)}\Big|_{\delta\ord{1}_{\rm RR}}(t') \, ,\\
	\delta_{\textrm{(RR\text{-}RR)}}\ord{3} \bv(t)
	& = G^{-5} \int_{-\infty}^{t} \dd t' \bigg[
	\ba_{\rm (N)}\Big|_{(\delta\ord{1}_{\rm RR})^2,\,\delta\ord{2}_{\textrm{RR\text{-}RR}}}(t')
	+ \ba_{\rm (RR)}\Big|_{\delta\ord{1}_{\rm N} + \delta\ord{1}_{\rm RR},\,\delta\ord{2}_{\rm RR}}(t')
	\bigg] \, ,
\end{align}
while we keep the associated $\delta_{\rm (RR^2)}\ord{3}\br$ and $\delta_{\textrm{(RR--RR)}}\ord{3}\br$ in integral form. The computation of the impulse then takes the form,
\begin{align}
	\Delta^{(4)} \bp_{\rm (RR\text{-}RR)} & = M \nu \int \dd t \, \ba_{\rm (RR)}\Big|_{\delta\ord{1}_{\rm RR}} \, , \\
	\Delta^{(5)} \bp_{\rm (RR\text{-}RR)} & = M \nu \int \dd t \bigg[
	\ba_{\rm (N)}\Big|_{(\delta\ord{1}_{\rm RR})^2, \delta\ord{2}_{\rm RR-RR}}
	+ \ba_{\rm (RR)}\Big|_{\delta\ord{2}_{\rm RR}, \delta\ord{1}_{\rm RR}+\delta\ord{1}_{\rm N}}
	\bigg] \, , \\
	\Delta^{(6)} \bp_{\rm (RR\text{-}RR)} & = M \nu \int \dd t \bigg[
	\ba_{\rm (N)}\Big|_{(\delta\ord{1}_{\rm RR})^2+\delta\ord{1}_{\rm N}, \delta\ord{2}_{\rm RR-RR}+\delta\ord{1}_{\rm N}, \delta\ord{3}_{\rm RR\text{-}RR}}
	+ \ba_{\rm (RR)}\Big|_{\delta\ord{3}_{\rm RR}, \delta\ord{2}_{\rm RR}+\delta\ord{1}_{\rm N}, \delta\ord{1}_{\rm RR}+\delta\ord{2}_{\rm N}}
	\bigg] \, .
\end{align}

\subsubsection*{RR$^2$ FT \& M deflections}

Finally, for the $X \in \{\rm RR^2,FT,M\}$ deflections we have
\begin{align}
	 \delta_{(\textrm{X})}\ord{1} \bv(t)
	& = G^{-3} \int_{-\infty}^{t} \dd t'\; \ba_{(\textrm{X})}\Big|_{\br_0}(t') \, , \\
	 \delta_{(\textrm{X})}\ord{2} \bv(t)
	& = G^{-4} \int_{-\infty}^{t} \dd t' \bigg[
	\ba_{\rm (N)}\Big|_{\delta\ord{1}_{\rm X}}(t')
	+ \ba_{(\textrm{X})}\Big|_{\delta\ord{1}_{\rm N}}(t')
	\bigg] \, , \\
	\delta_{(\textrm{X})}\ord{3} \bv(t)
	& = G^{-5} \int_{-\infty}^{t} \dd t' \bigg[
	\ba_{\rm (N)}\Big|_{\delta\ord{1}_{\rm X}+\delta\ord{1}_{\rm N},\,\delta\ord{2}_{\rm X}}(t')
	+ \ba_{(\textrm{X})}\Big|_{(\delta\ord{1}_{\rm N})^2,\,\delta\ord{2}_{\rm N}}(t')
	\bigg] \, .
\end{align}
As in the previous cases, we also keep the position deflections $\delta_{(\textrm{X})}\ord{3}\br$  in integral form.\vskip 4pt

The computation  can be further simplified by isolating {\it Schott}-like terms from the accelerations. In particular, using the identity%
\begin{equation}
	\frac{\bv \cdot \br}{r^\alpha} = \frac{1}{2-\alpha}\frac{\dd}{\dd t}\bigg(\frac{1}{r^{2-\alpha}}\bigg) \,,
\end{equation}
together with integration by parts,
\begin{equation}
	\mathcal{F}(\br, \bv)\frac{(\bv \cdot \br)^\beta}{r^\alpha}
	= \frac{1}{2-\alpha}\frac{1}{r^{2-\alpha}} \frac{\dd}{\dd t}\bigg(\mathcal{F}(\br, \bv)(\bv \cdot \br)^{\beta-1}\bigg)
	+ \frac{\dd}{\dd t}\bigg(\mathcal{F}(\br, \bv)\frac{(\bv \cdot \br)^{\beta-1}}{r^{\alpha-2}}\bigg) \, ,
	\label{eq:IntParts}
\end{equation}
we can split each contribution to the acceleration into an irreducible piece plus a Schott term as follows:
\begin{equation}
	\ba_{(\textrm{X})}(t) = \ba_{(X),\rm \not{S}}(t) + \frac{\dd}{\dd t}\, \bS_{(\textrm{X})}(t)\,,
\end{equation}
such that
\begin{equation}
	\Delta \bp_{(\textrm{X})} = M \nu \int_{-\infty}^{+\infty} \dd t \, \ba_{(X),\rm \not{S}}(t)
	+ M \nu \, \bS_{(\textrm{X})}\bigg|_{-\infty}^{+\infty}\,,
\end{equation}
with the boundary term dropping out of the impulse, provided $\bS(t \to \pm \infty) \to 0$, as expected. Since $\ba_{(X),\rm \not{S}}$ is nontrivial starting at ${\cal O}(G^4)$, the computation is then reduced to 
\begin{align}
	\Delta^{(4)} \bp_{(\textrm{X})} & = M \nu \int \dd t \bigg[
	 \ba_{\rm (N)}\Big|_{\delta\ord{1}_{(\textrm{X})}}
	+\ba_{(X), \rm \not{S}}\Big|_{\br_0}
	\bigg] \, , \\
	\Delta^{(5)} \bp_{(\textrm{X})} & = M \nu \int \dd t \bigg[
	 \ba_{\rm (N)}\Big|_{\delta\ord{1}_{(\textrm{X})}+\delta\ord{1}_{\rm N}, \delta\ord{2}_{(\textrm{X})}}
	+\ba_{(X), \rm \not{S}}\Big|_{\delta\ord{1}_{\rm N}}
	\bigg] \, , \\
	\Delta^{(6)} \bp_{(\textrm{X})} & = M \nu \int \dd t \bigg[
	 \ba_{\rm (N)}\Big|_{\delta\ord{1}_{(\textrm{X})}+\delta\ord{2}_{\rm N}, \delta\ord{2}_{(\textrm{X})}+\delta\ord{1}_{\rm N}, \delta\ord{3}_{(\textrm{X})}}
	+\ba_{(X), \rm \not{S}}\Big|_{(\delta\ord{1}_{\rm N})^2, \delta\ord{2}_{\rm N}}
	\bigg] \, ,
\end{align}
which avoids the need of introducing the third-order Newtonian deflection. We will return to the explicit derivation of the impulse in \S\ref{data}.

\subsection{Conservative sector}

As discussed in \cite{Kalin:2022hph,Porto:2024cwd}, a conservative effective action can be defined by implementing Feynman's $i0^+$-prescription. Besides the familiar potential and tail contributions in \eqref{tails}, we also generate the terms 
\begin{align}
\label{SFT}	S^{\rm cons}_{\text{(FT)}} & = -\frac{G^2}{30} \int  \frac{\dd \omega_1 \dd \omega_2}{(2\pi)^2} \, (i\omega_1^4\omega_2^3) L_{ki}I_{kj}(\omega_1)I_{ij}(\omega_2)\ddl(\omega_1+\omega_2) \\ &= \frac{G^2}{30} L_{kl}\int  \frac{\dd \omega}{2\pi} \, (-i\omega^7) I_{i k}(-\omega)I_{il}(\omega) \; ,\nn \\
S^{\rm cons}_{\text{(M)}} & = \frac{G^2}{70} \int \frac{\dd \omega_1 \dd \omega_2\dd \omega_3}{(2\pi)^3}
	I_{ij}(\omega_1)I_{jk}(\omega_3)I_{ki}(\omega_2) \sg(\omega_3)\sg(\omega_1)\ddl(\omega_1+\omega_2+\omega_3)\nn  \\
	&\hspace{3cm}  \times\left(7 \omega_1 ^4 \omega_3^4-2 \omega_1 ^3 \omega_3^3 \omega_2^2+2 \omega_1 ^2 \omega_3^2 \omega_2^4\right)\nn  \, , \\
	& = -\frac{G^2}{70} \int \frac{\dd \omega \dd \omega_1}{(2\pi)^2}
	I_{ij}(-\omega)I_{jk}(\omega_1)I_{ki}(\omega-\omega_1) \sg(\omega)\sg(\omega_1)\label{Smem1}  \\
	&\hspace{3cm} \times\omega^2 \omega_1^2 \left(2 \omega ^4 -6 \omega ^3 \omega_1+15 \omega ^2 \omega_1^2 - 6 \omega  \omega_1^3+2 \omega_1^4\right)\,,\nn
\end{align}
corresponding to failed-tail and memory effects, respectively.\vskip 4pt  In principle, the effective actions in \eqref{SFT}--\eqref{Smem1} constitute a well-defined (and systematic) conservative sector.  Nevertheless, owing to the intrinsically nonlinear character of gravitational radiation-reaction, in~\cite{Porto:2024cwd} we identified further terms that may be consistently incorporated into the conservative dynamics. Following closely the derivation in the PM regime, a conservative part at second order in the radiation-reaction may be derived from the dynamical equations that follow from the (complex) Feynman action, 
\beq
S_{(\rm RR)}^{\rm cons} = -\frac{2\pi G}{5} \int \frac{\dd\omega}{2\pi} \, I_{ij} (-\omega) I_{ij}(\omega)\Delta_{F}(\omega) \omega^4 \,, \qquad \Delta_{F}(\omega) = \int_\bk \frac{1}{\omega^2-\bk^2+i0^+}\,.\label{eq:srr2}\eeq

A particular example arises from plugging the linear RR acceleration back into the action, which we have denoted as RR$^2$ effects.  As we have shown~\cite{Porto:2024cwd}, these naturally encompass the needed components that yield perfect consistency with the previous 4PM results \cite{Dlapa:2021npj,Dlapa:2021vgp,Bern:2021dqo,Bern:2021yeh,Dlapa:2022lmu}. As we demonstrate, there are in principle additional corrections---arising from evaluating the leading radiation-reaction force along the leading radiation-reacted trajectory (that we termed RR--RR effects) [cf.~\eqref{eq:rrrr}]---that may also contribute the conservative sector. At first sight this may appear counterintuitive. Indeed,  RR--RR effects are responsible for the loss of energy and angular momentum (see \eqref{chi4red}-\eqref{chi6red} below). Moreover, by construction, they are obtained by inserting the trajectory into an (a priori) dissipative force, and thus they do not obviously admit a Hamiltonian-type characterization. Despite these issues, as we show in App.~\ref{app:split} (see also \S\ref{isotropic}), we may isolate a combined conservative-like (2RR) contribution to the scattering angle which, in turn, can be absorbed into a conservative Hamiltonian.

\subsubsection*{Tail-like}

As argued in \cite{Porto:2024cwd}, despite the intrinsic nonlocality, the effective action can be decomposed into two sharply distinct contributions. This separation is naturally organized within a PM approximation, where the multipole moments admit an expansion of the form \beq \label{I0N} I^{ij}(\omega) = I_{(0)}^{ij}(\omega) + I_{\rm (N)}^{ij}(\omega) +\cdots,\eeq 
with the leading term determined by straight-line motion,
\begin{equation}
\label{i02d}
	I_{(0)}^{ ij}(\omega) = \bb^{\langle i} \bb^{j \rangle} \ddl(\omega)
	- i 2 \bb^{\langle i} \bv_\infty^{j \rangle} \ddl{}^\prime(\omega)
	-v_\infty^{\langle i} \bv_\infty^{j \rangle} \ddl{}^{\prime\prime}(\omega) \; ,
\end{equation}
with $\langle ij \rangle$ denoting an STF projection, primes are derivatives with respect to $\omega$, and  $I_{\rm (N)}^{ij}(\omega)$ is the quadrupole evaluated on the Newtonian trajectory given by \eqref{eqbrN}. In particular, at zeroth order we have
\begin{equation}
\label{w3I}
\omega_1^2 I_{(0)}^{ij}(\omega_1)  = -2 \bv_\infty^{\langle i} \bv_\infty^{j \rangle} \ddl(\omega_1) \,, \quad	\omega^n I_{(0)}^{ij}(\omega) = 0 \,\,\, (\text{ for } n\geq 3). 
\end{equation}
Together with the distributional identity
\begin{equation}
\label{w2sign}	\int \frac{\dd\omega}{2\pi} \sg(\omega)\ddl(\omega) f(\omega) = 0  \,,
\end{equation}
these relations allow us to isolate a `tail-like' contribution, described by the (on-shell) action
\beq
\begin{aligned}
	S^{\rm cons}_{\rm(T\text{-}like)} \label{Tlike}& = +
	\frac{17G^2}{150} \int \frac{\dd\omega}{2\pi}
	(-i\omega^7)L^{kl}I^{ki}(-\omega)I^{il}(\omega)\\
	& +\frac{G^2}{5}\int\!\!\frac{\dd \omega_1\dd \omega_2\dd\omega_3}{(2\pi)^3}\ddl(\omega_1+\omega_2+\omega_3)
	\bigg\{
	\bigg(
	\frac{\omega_1 ^3 \omega_3^3 \omega_2^2}{7}
	-\frac{\omega_1 ^4 \omega_3^4}{2} 
	\bigg) I^{ij}_{\rm (N)}(\omega_1)I^{jk}_{\rm (N)}(\omega_3)I_{(0)}^{ki}(\omega_2) \notag \\
	& \hspace{3cm}
	+ \bigg(
	\frac{2\omega_1 ^4 \omega_3^4}{5}
	+\frac{2\omega_1 ^3 \omega_3^3 \omega_2^2}{5}
	\bigg) I^{ij}_{\rm (N)}(\omega_1)I^{jk}_{\rm (N)}(\omega_3)Q_{(0)}^{ki}(\omega_2)
	\bigg\}\; ,
\end{aligned}
\eeq
where we have added the tail-like corrections induced by 2RR effects---that are dominated by the RR$^2$ contributions \cite{Porto:2024cwd} (see App~\ref{app:split}). In coordinate space, it becomes (see Eq. (7.2) in \cite{Porto:2024cwd}),
\begin{align}
\label{7.2}
	S^{\rm cons}_{\rm(T\text{-}like)} &
	=   \frac{G^2}{5}\int \dd t
	\bigg\{
	-\frac{17}{30}L_{kl}I^{(4)}_{ki}I^{(3)}_{il}
	 \notag \\
	&
	+\frac{1}{7}I\ord{2}_{(0),ij}I\ord{3}_{jk}I\ord{3}_{ki}
	-\frac{1}{2}I_{(0),ij}I\ord{4}_{jk}I\ord{4}_{ki}
	  +\frac{2}{5}\bigg(
	Q\ord{2}_{(0),ij}I\ord{3}_{jk}I\ord{3}_{ki}
	+Q_{(0),ij}I\ord{4}_{jk}I\ord{4}_{ki}
	\bigg)
	\bigg\} \;.
\end{align}
 As it was argued in \cite{Porto:2024cwd}, this identification precisely matches with the tail-like (2rad) region of integration encountered in PM scattering calculations. This includes not only the standard (mass/energy) tail and failed-tail terms, but also captures contributions arising from quadrupole moments for which one of the insertions is evaluated on a {\it static} configuration. 
\subsubsection*{Memory-like} 

Upon implementing \eqref{I0N} into the action \eqref{Smem1}, we find a contribution of the sort,
\begin{align}
	S^{\rm cons}_{\rm(M\text{-}like)} \supset	& -\frac{G^2}{70}\int \frac{\dd \omega \dd \omega_1}{(2\pi)^2}
	I_{(\rm N)}^{ij}(-\omega)I_{(\rm N)}^{jk}(\omega_1)I_{(\rm N)}^{ki}(\omega-\omega_1) \sg(\omega)\sg(\omega_1) \label{eq:SmemFourier2} \\
	&\hspace{3cm} \times\omega^2 \omega_1^2 \left(2 \omega ^4 -6 \omega ^3 \omega_1+15 \omega ^2 \omega_1^2 - 6 \omega  \omega_1^3+2 \omega_1^4\right)\, .\nn

\end{align}
Simple power counting then makes it manifest that genuine time nonlocality can only arise starting at ${\cal O}(G^5)$. Since the relevant kernels involve two independent frequencies, $\omega_1$ and $\omega_2$, we naturally interpret these contributions as memory-like (2rad) effects. As we show in App~\ref{app:split}, memory-like contributions also arise in the full 2RR sector. In the PM language, these can be mapped directly onto the corresponding region of the 5PM integration in \cite{Driesse:2026qiz}.\vskip 4pt

Nonlocal-in-time effects are already known to pollute the two-body dynamics \cite{Galley:2015kus}. As in the case of the standard tail terms, it is also advantageous to isolate a local-in-time correction from memory-like effects, which will enable a universal implementation across both (unbound) scattering and (bound) inspiral dynamics \cite{Dlapa:2024cje,Bini:2024tft,Dlapa:2025biy}.  Noticing that the above memory-like correction has the structure
\begin{align}
\label{localnloc}
	S^{\rm cons}_{\rm(M\text{-}like)}  &= \int\frac{\dd \omega_1\dd\omega_2 }{(2\pi)^2}\sg(\omega_1)\sg(\omega_2) A(-\omega_1)B(\omega_1-\omega_2)C(\omega_2)  \\
	& \qquad = \int \dd t_1 \dd t_2 \dd t_3\bigg(-\frac{i}{\pi}\frac{\PV}{t_1-t_2} \bigg)
	\bigg(-\frac{i}{\pi}\frac{\PV}{t_2-t_3} \bigg) A(t_1) B(t_2) C(t_3)\nn \,,
\end{align}
we can then apply the PB theorem to extract a local-in-time contribution, which (for the above routing) becomes	
\beq
	S^{\rm cons (PBloc)}_{\rm(M\text{-}like)}  =\int \dd t_1 \dd t_2 \dd t_3 \delta (t_1-t_2) \delta (t_2-t_3) A(t_1)B(t_2) C(t_3) \; .
\eeq
Equivalently, in the routing of \eqref{eq:SmemFourier2}, we can arrive at the local correction via the Fourier-space relation (see App.~\ref{app:PBFourier}) 
\begin{equation}
  \sg(\omega) \sg( \omega_1)
  = 1 -2\theta(\omega)\theta(-\omega_1) - 2\theta(-\omega)\theta(\omega_1) \, .  \label{eq:PBf2}
\end{equation}
Hence, the memory-like contribution form \eqref{eq:SmemFourier2} decomposes into a local part, 
\begin{align}
	S^{\rm cons (PBloc)}_{\rm(M\text{-}like)} \supset  \frac{G^2}{5}\int \dd t
	\bigg\{
	\frac{1}{7}I\ord{2}_{{\rm (N)},ij}I\ord{3}_{jk}I\ord{3}_{ki}
	-\frac{1}{2}I_{{\rm (N)},ij}I\ord{4}_{jk}I\ord{4}_{ki}
	-\frac{1}{7}I\ord{2}_{{\rm (N)},ij}I\ord{2}_{{\rm (N)}jk}I\ord{4}_{ki} \bigg\}\,,
\end{align}
truncated to 5PN order, and a nonlocal remainder 
\begin{align}
&  S^{\rm cons (PBnloc)}_{\rm(M\text{-}like)}  \supset \frac{4G^2}{70}\int \frac{\dd \omega \dd \omega_1}{(2\pi)^2}
	I_{(\rm N)}^{ij}(-\omega)I_{(\rm N)}^{jk}(\omega_1)I_{(\rm N)}^{ki}(\omega-\omega_1) \theta(\omega)\theta(-\omega_1)\\
	&\hspace{3cm} \times\omega^2 \omega_1^2 \left(2 \omega ^4 -6 \omega ^3 \omega_1+15 \omega ^2 \omega_1^2 - 6 \omega  \omega_1^3+2 \omega_1^4\right)\nn
\;.
\end{align}
Somewhat expectedly, the form of the nonlocal-in-time contributions resembles the convolution of `cut'  (Wightman) propagators which is the hallmark distinction between retarded and Feynman Green's functions. A similar decomposition, although slightly more intricate, applies to the full 2RR sector. See  \S\ref{alternative} and App.~\ref{app:split} for more on this point. 

\subsubsection*{Combined local terms}

Upon applying the reduction to a local-in-time contribution via the PB prescription, the combined (on-shell) action may be written as follows
\begin{align}
\label{7.2n}
	&S^{\rm cons(PB)}_{\rm  (tot)}
	  
	=\frac{G^2}{5}\int \dd t \bigg\{ {\color{ForestGreen}

	-\frac{17}{30}L_{kl}I^{(4)}_{ki}I^{(3)}_{il}
	}  \\
	& {\color{ForestGreen}
	+\frac{1}{7}I\ord{2}_{(0)ij}I\ord{3}_{jk}I\ord{3}_{ki}
	-\frac{1}{2}I_{(0)ij}I\ord{4}_{jk}I\ord{4}_{ki}
	  +\frac{2}{5}\bigg(
	Q\ord{2}_{(0)ij}I\ord{3}_{jk}I\ord{3}_{ki}
	+Q_{(0)ij}I\ord{4}_{jk}I\ord{4}_{ki}
	\bigg)
	}\nn
	  \,  \\
&{\color{RoyalBlue}
	+
	\frac{1}{7}\check I\ord{2}_{ij}I\ord{3}_{jk}I\ord{3}_{ki}
	-\frac{1}{2}\check I_{ij}I\ord{4}_{jk}I\ord{4}_{ki}
	}
			{\color{RoyalBlue}	
	+\frac{2}{5}\left(	 \check Q\ord{2}_{ij}I\ord{3}_{jk}I\ord{3}_{ki}	+ \check Q_{ij}I\ord{4}_{jk}I\ord{4}_{ki}	\right)} {\color{Violet}
	-\frac{1}{7}\check I\ord{2}_{ij}\check I\ord{2}_{jk}I\ord{4}_{ki}}  \bigg\}{\color{red}+ S^{\rm cons(PB)}_{\rm RR\text{-}RR}}\nn \,,
\end{align}
where $\check I^{ij} \equiv I^{ij} - I_{(0)}^{ij}$ and similarly for $\check Q^{ij}$. We highlight in {\color{ForestGreen}green} the tail-like contribution and decompose the remaining (local) memory-like corrections into three distinct parts. The {\color{RoyalBlue}blue} terms include memory and RR$^2$ effects that complete the tail-like action to full multipole moments. The {\color{Violet}violet} part corresponds to a residual memory contribution that does not participate in the tail region. Finally, the remaining RR--RR corrections [highlighted in {\color{red}{red}}] depend on the evaluation of the radiation-reacted trajectory on the leading radiation-reaction effective action. See App.~\ref{app:split} for the explicit expression. \vskip 4pt Starting from the known potential and tail contributions, $S^{\rm cons}_{\rm (P+T)}$, and combining them with \eqref{7.2n}, we may then compute the scattering angle via $\chi^{\rm cons}=- \partial_J S^{\rm cons}$ (see App.~\ref{app:split}). We will present the corresponding results through 5PN order momentarily. The derivation of the nonlocal-in-time contribution to the conservative scattering angle through Feynman's prescription is deferred to future work. Instead, we return shortly to explain how \eqref{7.2n} can be used to introduce a fully local-in-time decomposition of the complete dynamics into conservative and dissipative sectors.

\section{Scattering data at 5PN}\label{data}

In what follows we report the value of the impulse and scattering angle up to ${\cal O}(G^6)$, obtained from the procedure outlined in \S\ref{sec:impulse}.

\subsection{Impulse}\label{sec:ResFullImpRel}
We start by quoting the value of the impulse from the linear radiation-reaction force, 
\begin{align}
\label{prr1}	\Delta^{(3)} \bp_{\rm RR} & = -\frac{G^3 M^4 \nu^2}{b^3}\bigg( \frac{16}{5}\frac{\bb}{b} + \frac{37}{15}\pi\frac{\bv_\infty}{v_\infty} \bigg) \, , \\
	\Delta^{(4)} \bp_{\rm RR} & = -\frac{G^4 M^5 \nu^2}{b^4}\bigg( \frac{47}{15}\pi\frac{\bb}{b} + \frac{1856}{35}\frac{\bv_\infty}{v_\infty} \bigg)\frac{1}{v_\infty^2} \, , \\
	\label{prr3} \Delta^{(5)} \bp_{\rm RR} & = -\frac{G^5 M^6 \nu^2}{b^5}\bigg( -\frac{416}{45}\frac{\bb}{b} + \frac{178}{5}\pi\frac{\bv_\infty}{v_\infty} \bigg)\frac{1}{v_\infty^4} \, ,
\end{align}
which is in agreement with the results in \cite{Bini:2021gat,Bini:2022enm}.\vskip 4pt

Moving on, for the contribution we dubbed RR--RR we find
\begin{align}
\label{prr21}	\Delta^{(4)} \bp_{\rm RR\text{-}RR} & = -\frac{G^4 M^5 \nu^3}{b^4}\bigg( \frac{479}{25}\pi\frac{\bb}{b}  \bigg)v_\infty^3 \, , \\
	\Delta^{(5)} \bp_{\rm RR\text{-}RR} & = -\frac{G^5 M^6 \nu^3}{b^5}\bigg( \frac{740864}{1575}\frac{\bb}{b} + \frac{1254}{25}\pi\frac{\bv_\infty}{v_\infty} \bigg) v_\infty \, , \\
\label{prr22}	\Delta^{(6)} \bp_{\rm RR\text{-}RR} & = -\frac{G^6 M^7 \nu^3}{b^6}\bigg( \frac{468389}{900}\pi \frac{\bb}{b} + \bigg[\frac{1841024}{1575} + \frac{1739}{150} \pi^2 \bigg]\frac{\bv_\infty}{v_\infty} \bigg)\frac{1}{v_\infty} \, ,
\end{align}
whereas for the remaining RR$^2$, FT, and M parts we have 
\begin{align}
	\Delta^{(4)} \bp_{\rm RR^2} & = -\frac{G^4 M^5 \nu^3}{b^4}\bigg( \frac{97}{50}\pi\frac{\bb}{b}  \bigg)v_\infty^3 \, , \\
	\Delta^{(5)} \bp_{\rm RR^2} & = -\frac{G^5 M^6 \nu^3}{b^5}\bigg( \frac{49664}{675}\frac{\bb}{b} + \frac{97}{25}\pi\frac{\bv_\infty}{v_\infty} \bigg) v_\infty \, , \\
	\Delta^{(6)} \bp_{\rm RR^2} & = -\frac{G^6 M^7 \nu^3}{b^6}\bigg( \frac{16003}{150}\pi\frac{\bb}{b} + \frac{99328}{675} \frac{\bv_\infty}{v_\infty} \bigg)\frac{1}{v_\infty} \, , \\
	\Delta^{(4)} \bp_{\rm FT} & = -\frac{G^4 M^5 \nu^3}{b^4}\bigg( \frac{23}{8}\pi\frac{\bb}{b}  \bigg)v_\infty^3 \, , \\
	\Delta^{(5)} \bp_{\rm FT} & = -\frac{G^5 M^6 \nu^3}{b^5}\bigg( \frac{1088}{15}\frac{\bb}{b} + \frac{23}{4}\pi\frac{\bv_\infty}{v_\infty} \bigg) v_\infty \, , \\
	\Delta^{(6)} \bp_{\rm FT} & = -\frac{G^6 M^7 \nu^3}{b^6}\bigg( \frac{661}{8}\pi\frac{\bb}{b} + \frac{2176}{15} \frac{\bv_\infty}{v_\infty} \bigg)\frac{1}{v_\infty} \, , \\
	\Delta^{(4)} \bp_{\rm M} & = \frac{G^4 M^5 \nu^3}{b^4}\bigg( \frac{529}{140}\pi\frac{\bb}{b}  \bigg)v_\infty^3 \, , \\
	\Delta^{(5)} \bp_{\rm M} & = \frac{G^5 M^6 \nu^3}{b^5}\bigg( \frac{30272}{315}\frac{\bb}{b} + \frac{529}{70}\pi\frac{\bv_\infty}{v_\infty} \bigg) v_\infty \, , \\
	\Delta^{(6)} \bp_{\rm M} & = \frac{G^6 M^7 \nu^3}{b^6}\bigg( \frac{34847}{315}\pi\frac{\bb}{b} + \frac{60544}{315} \frac{\bv_\infty}{v_\infty} \bigg)\frac{1}{v_\infty} \, .
\end{align}
As a nontrivial consistency check, the resulting impulses---including the full set of nonlinear radiation-reaction and hereditary effects---can be shown to satisfy the on-shell condition
\beq
(p^\mu + \Delta p^\mu)^2 = M^2\nu^2 + {\cal O}(G^6 v_\infty^4),\label{eqDp1}
\eeq
with $p^\mu = (M\nu+ \frac{\bp_\infty^2}{2M\nu}+\cdots, \bp_\infty)$ the incoming spacetime momentum at infinity, and $\Delta p^\mu = (E_{\rm rad},\Delta \bp)$, up to higher PN orders.\footnote{This follows from the on-shell condition for the full spacetime momenta of each particle, which is related to the relative momentum via $\Delta \bp = \Delta\bp_1 + \frac{E_1}{E} \bP_{\rm rad} + {\cal O}\left(\bP_{\rm rad}^2\right)$, with $\bP_{\rm rad} \equiv -(\Delta \bp_1+\Delta \bp_2)$ the radiated $\bf 3$-momentum, which starts at ${\cal O}(G^3)$ but first contributes at 3.5PN order \cite{Bini:2021gat}.}
Moreover, restricting to the $X \in \{\rm RR^2,FT,M\}$ sectors, one finds that the corresponding individual relative impulses already obey
\begin{equation}
\bp_\infty \cdot \Big(\Delta^{(4)} \bp_{\rm X} + \Delta^{(5)} \bp_{\rm X}+ \Delta^{(6)} \bp_{\rm X}\Big) +\Delta \bp_{\rm N}\cdot \Big(\Delta^{(4)} \bp_{\rm X} + \Delta^{(5)} \bp_{\rm X}+ \Delta^{(6)} \bp_{\rm X}\Big) = 0 + {\cal O}(G^7),
\end{equation}
and hence $|\bp_\infty + \Delta \bp_{\rm N}+ \Delta \bp_{\rm X}| = |\bp_\infty|$. Together with \eqref{eqDp1}, this shows that only RR--RR effects can generate a nonzero contribution to $\Delta p^0$. This provides a direct confirmation of the conservative-like character of the RR$^2$, failed-tail, and memory contributions.\vskip 4pt

In summary, we find for the total impulse the structure:
\begin{align}
	\Delta \bp = \Delta \bp_{\rm P}+\Delta \bp_{\rm RR} + \Delta \bp_{\rm T} & +
	\frac{G^4 M^{5}}{b^4}\nu^2
	\bigg[
	\frac{\bb}{b}\Big(c^{(4, \nu)}_b + \nu c^{(4, \nu^2)}_b
	\Big)
	+\frac{\bv}{v_\infty}\Big(
	c^{(4, \nu)}_v + \nu c^{(4, \nu^2)}_v
	\Big)
	\bigg] \notag \\
	& + \frac{G^5 M^{6}}{b^5}\frac{\nu^2}{v_\infty^2}
	\bigg[
	\frac{\bb}{b}\Big(c^{(5, \nu)}_b + \nu c^{(5, \nu^2)}_b
	\Big)
	+\frac{\bv}{v_\infty}\Big(
	c^{(5, \nu)}_v + \nu c^{(5, \nu^2)}_v
	\Big)
	\bigg] \notag \\
	& + \frac{G^6 M^{7}}{b^6}\frac{\nu^2}{v_\infty^4}
	\bigg[
	\frac{\bb}{b}\Big(c^{(6, \nu)}_b + \nu c^{(6, \nu^2)}_b
	\Big)
	+\frac{\bv}{v_\infty}\Big(
	c^{(6, \nu)}_v + \nu c^{(6, \nu^2)}_v
	\Big)
	\bigg] \, ,
\end{align}
with the value of the $c^{(n, \nu^2)}_{b,v}$ coefficients due to the nonlinear radiation-reaction and hereditary effects adding up to:
\begin{align}
	c^{(4, \nu^2)}_b & = \bigg(-\frac{1451}{1400} {\color{red} -\frac{479}{25} }\bigg) 
	\pi  v_\infty^3 \, ,\\
	 c^{(4, \nu^2)}_v & =0 \, ,\\
	 c^{(5, \nu^2)}_b & = \bigg(-\frac{236288}{4725} {\color{red} -{\frac{740864}{1575}}}\bigg) 
	  v_\infty^3 \, ,\\
	 c^{(5, \nu^2)}_v & = \bigg(-\frac{1451}{700} {\color{red} -{\frac{1254}{25}}}\bigg)\pi v_\infty^3 \, , \\
	 c^{(6, \nu^2)}_b & = \bigg(-\frac{991447}{12600} {\color{red} -{\frac{468389}{900}}}\bigg)\pi  v_\infty^3 \, , \\
	 c^{(6, \nu^2)}_v & = \bigg(-\frac{472576}{4725} {\color{red} -{\frac{1841024}{1575}}-\frac{1739}{150}\pi^2}\bigg)  v_\infty^3 \,,
\end{align}
and the remaining $c^{(n, \nu)}_{b,v}$ vanishing at ${\cal O}(\nu^1)$. We highlight [in {\color{red} red}] the full RR--RR contributions, which are the only ones from this set that contains dissipate terms at this order.
\subsection{Scattering angle}

We can now extract the contribution to the relative scattering angle. This is achieved via 
\begin{equation}
	\cos\chi_{\rm rel} = \bigg(\frac{\bp_- \cdot \bp_+}{|\bp_-||\bp_+|}\bigg) \, , \qquad
	\sin\chi_{\rm rel} = \bigg(\frac{|\bp_- \times \bp_+|}{|\bp_-||\bp_+|}\bigg) \; ,
\end{equation}
with $\bp_+ = \bp_\infty +\Delta \bp$;
or equivalently,
\begin{equation}
	\sin\chi_{\rm rel} = \frac{|\Delta p_b|}{|p_+|} \; ,
\end{equation}
where $\Delta p_b$ is the component along the impact parameter of the relative impulse. We decompose the angle in the usual PM-expanded fashion, 
\begin{equation}
	\frac{\chi_{\rm rel}}{2} = \sum_n \bigg(\frac{G M}{b}\bigg)^n\chi_{b,\,{\rm rel}}^{(n)} 
	= \sum_n \frac{\tilde{\chi}^{(n)}_{j,\,{\rm rel}}}{j^n \Gamma^{n-1} } \; , \label{chirel}
\end{equation}
with $j \equiv J/(G M^2\nu)$ the reduced angular momentum, and $\Gamma \equiv \sqrt{1+2\nu(\gamma-1)}$, with $\gamma$ the center-of-mass Lorentz factor. For simplicity, we remove the `rel' label in what follows.\vskip 4pt 

Focusing on the ${\cal O}(\nu^2)$ contributions, and in terms of the previous coefficients, we find
\begin{align}
	\chi^{ (4, \nu^2)}_b & = -\frac{c_b^{(4, \nu^2)}}{2}\frac{1}{v_\infty} 
	= \bigg(\frac{1451}{2800} + {\color{red} \frac{479}{50}}\bigg)\pi v_\infty^2 \, \, , \\
\chi^{ (5, \nu^2)}_b &  = -  \frac{1}{v_\infty^3} \left(\frac{c_b^{(5, \nu^2)}}{2}+\frac{c_v^{(4, \nu^2)}c_{(\rm N)}}{2}\right)
	=\frac{118144}{4725} + {\color{red} {\frac{370432}{1575}}}\, , \\
\chi^{ (6, \nu^2)}_b &  = -\left(  \frac{c_b^{(6, \nu^2)}}{2} + \frac{c_v^{(5, \nu^2)}c_{(\rm N)}}{2} -\frac{c_b^{(4, \nu^2)}c_{(\rm N)}(c_{(\rm N)}-1)}{2}\right)\frac{1}{v_\infty^5} +
	{\color{Orange}\frac{296}{75}\pi}\frac{1}{v_\infty^2} \\
	&= \bigg(\frac{203513 }{5040} + {\color{red}
	{\frac{524189 }{1800}}} +{\color{Orange}\frac{296}{75}}\bigg)\frac{\pi}{v_\infty^2}\nn  \, .
\end{align}
where $c_{\rm (N)}$ is defined analogously to the other coefficients, but for the leading Newtonian trajectory:
\beq
	\Delta {\boldsymbol p}_{\rm N} =-c_{\rm (N)}
	\left[
	\frac{GM^2\nu}{b v_\infty}\frac{{\boldsymbol b}}{b}
	+
	\frac{G^2M^3\nu}{b^2 v_\infty^3}
	\frac{{\boldsymbol v}_\infty}{v_\infty}
	+ \dots
	\right],
	\eeq
with the value $c_{\rm (N)}=2$. We highlight [in {\color{Orange} orange}] the contribution coming from the iteration of the linear radiation-reaction impulse in the scattering angle. Equivalently, using the parameterization in \eqref{chirel}, we have %
\begin{align}
\label{chi4red}	 \tilde{\chi}^{ (4, \nu^2)}_j & = \bigg(\frac{1451}{2800} {\color{red} +\frac{479}{50}}\bigg)\pi v_\infty^6 \, \, , \\
\label{chi5red}	 \tilde{\chi}^{ (5, \nu^2)}_j &  = 
	\bigg(
	\frac{118144}{4725} + {\color{red} {\frac{370432}{1575}}}
	\bigg)v_\infty^5\, , \\
\label{chi6red}	 \tilde{\chi}^{ (6, \nu^2)}_j &  = 
	\bigg(\frac{203513 }{5040} + {\color{red}
	{\frac{524189 }{1800}}} + {\color{Orange}\frac{296}{75}}\bigg)\pi v_\infty^4  \, .
\end{align}
The above result incorporates the FT, M, RR$^2$ and  RR--RR [in {\color{red} red}/{\color{orange}orange}] contributions. For the remaining P and T corrections we resort to the derivations in \cite{Bini:2021gat,Bini:2022enm}, obtained via the results in \cite{Blumlein:2020pyo,Almeida:2021xwn}. The values reported in \cite{Bini:2021gat,Bini:2022enm} are already conveniently split into a conservative and dissipative parts. In particular, we have
\begin{align}
	\tilde{\chi}^{(4, \nu^2)\rm cons}_{j,(\rm P+T)} & = -\frac{8919}{1400}\pi v_\infty^6 \, , \label{eq:chiTpPcons4}\\
	\tilde{\chi}^{(5, \nu^2)\rm cons}_{j,(\rm P+T)}
	& = 
	\bigg(\frac{217866437}{151200} - \frac{224057}{1440}\pi^2 + \frac{1408}{45}\log(2v_\infty)\bigg)v_\infty^5 \, , \label{eq:chiTpPcons5}\\
	\tilde{\chi}^{(6, \nu^2)\rm cons}_{j,(\rm P+T)}
	& = 
	\bigg(
	\frac{7572253}{2240} - \frac{10812865}{32768}\pi^2 + \frac{201}{2}\log\left(\frac{v_\infty}{2}\right) + \frac{2817}{16}\zeta(3) \bigg) \pi v_\infty^4 \, , \label{eq:chiTpPcons6}
\end{align}
whereas for the dissipative tails,\footnote{In an earlier version, we quoted the values in Eq.~(C12) of~\cite{Bini:2022enm}. However, despite being labeled as ``from tail,'' we have since realized they include additional contributions at ${\cal O}(\nu^2)$. The values reported below retain only the dissipative tail terms. We thank G. Brunello for sharing a draft of his paper, whose results agree with all our findings leading to \eqref{chi5tot1}.}  
\begin{align}
	\tilde{\chi}^{(4, \nu^2)\rm diss}_{j,(\rm T)} & = 0 \; ,\\
	\tilde{\chi}^{(5, \nu^2)\rm diss}_{j,(\rm T)}& = -\frac{1216}{21}v_\infty^5  \; ,\\
\label{chi62dis}	\tilde{\chi}^{(6, \nu^2)\rm diss}_{j,(\rm T)} & = -\pi^3\frac{11013}{560} v_\infty^4  \; .
\end{align}
Combining all terms we arrive at the following final result for the total (even-in-velocity) scattering angle to 5PN order:
\begin{align}
\label{chi4tot1}	 \tilde{\chi}^{ (4, \nu^2)\rm tot}_{j({\rm even})} & = \frac{1491}{400} \pi v_\infty^6 \, \, , \\
\label{chi5tot1}	 \tilde{\chi}^{ (5, \nu^2)\rm tot}_{j({\rm even})} &  = 
	\bigg({\frac{35493331}{21600}} - \frac{224057}{1440}\pi^2 + \frac{1408}{45}\log(2v_\infty)\bigg)v_\infty^5 \, , \\
\label{chi6tot1}	 \tilde{\chi}^{ (6, \nu^2)\rm tot}_{j({\rm even})} &  = 
	\bigg(
	{\frac{53510579}{14400} - \frac{401004899}{1146880}\pi^2} + \frac{201}{2}\log\left(\frac{v_\infty}{2}\right) + \frac{2817}{16}\zeta(3) \bigg) \pi v_\infty^4  \, .
\end{align}
The contribution at ${\cal O}(G^4)$ was already reported in \cite{Porto:2024cwd}, and is in perfect agreement with the 4PM results in \cite{Dlapa:2022lmu}. The remaining values constitute new results that will serve as benchmarks for future 5PM and 6PM computations.

\subsection{Conservative terms}\label{consA}

Garnering all the pieces together, and keeping only the local-in-time part of the memory-like region---obtained through the PB prescription---the total conservative contribution to the scattering angle at 5PN order becomes 
\begin{align}
	\tilde\chi_{j (\rm tot)}^{(4,\nu^2)\rm cons} &  =  \pi \bigg(
	-\frac{8919}{1400}
	+ 
	{\color{ForestGreen}
	\frac{8919}{1400}} 
	\bigg) v_\infty^6 = 0 \; , \label{eq:chi4nu2const}\\
	\tilde\chi_{j(\rm tot)PB}^{(5, \nu^2)\rm cons} &  =  \bigg(
	\frac{217866437}{151200} - \frac{224057}{1440}\pi^2 + \frac{1408}{45}\log(2v_\infty) 
	+ 
	{\color{ForestGreen}
	\frac{38144}{225}} 
	-{\color{RoyalBlue}
	\frac {3584} {1125}}
  + {\color {Violet}  \frac{1024}{675}}  
+{\color{red}\frac {127232 } {3375}}	\bigg) v_\infty^5 \nn \\
	& =  \bigg(
	\frac{48699841}{30240} - \frac{224057}{1440}\pi^2 + \frac{1408}{45}\log(2v_\infty) 	
	{\color{cyan}+\frac{9728}{270}  }
	\bigg) v_\infty^5  \; , 
	\label{eq:chi5nu2const}\\
	\tilde\chi_{j \rm (tot)PB}^{(6, \nu^2)\rm cons} &  = \pi \bigg(
	\frac{7572253}{2240} - \frac{10812865}{32768}\pi^2 + \frac{201}{2}\log\left(\frac{v_\infty}{2}\right) + \frac{2817}{16}\zeta(3) 
	+ 
	{\color{ForestGreen}
	\frac{5864113}{26880}}
	+{\color{ForestGreen} \pi^2 \frac{2721}{5120}}\notag \\ 
	&\hspace{2cm} -{\color{RoyalBlue}
	\frac{63005}{5376}}
	- {\color{RoyalBlue} \pi^2 \frac{2721}{5120}}
	+{\color{Violet} \frac{149}{42}} 
	+{\color{red} \frac{9427}{120}}\bigg) v_\infty^4  \notag \\
	& = \pi \bigg(
	\frac{96731149}{26880} - \frac{10812865}{32768}\pi^2  + \frac{201}{2}\log\left(\frac{v_\infty}{2}\right) + \frac{2817}{16}\zeta(3)+{\color{cyan}\frac{630661}{8960}}
	\bigg) v_\infty^4	\label{eq:chi6nu2const}\,
	 \end{align}
where the total (local) memory-like corrections are highlighted in {\color{cyan}cyan}. The full Feynman computation would also require the inclusion of the  $\tilde\chi_{j \rm (M\text{-}like)PBnloc}^{(n, \nu^2)\rm cons}$ corrections ($n=5,6$), accounting for nonlocal-in-time memory-like contributions. However, the latter are expected to combine with a corresponding nonlocal term in the dissipative sector to yield the local-in-time result shown in \eqref{chi4tot1}-\eqref{chi6tot1}. We will discuss momentarily how to build on the local part of the scattering angle displayed above in order to introduce a complete local representation of the conservative sector.\vskip 4pt

The alert reader will immediately notice that, upon translating between different expansion parameters, the result for $\tilde\chi_{j(\rm P+T)}^{(5, \nu^2)\rm cons}+\tilde\chi_{j \rm(T\text{-}like)}^{(5, \nu^2)\rm cons}$ displayed in \eqref{eq:chi5nu2const} [in black] is in perfect agreement with the associated value in \cite{Driesse:2026qiz}.\footnote{\label{footconv}The results of~\cite{Driesse:2026qiz} are expressed in terms of the velocity variable 
$v \equiv \frac{\sqrt{\gamma^2 - 1}}{\gamma}$, whereas the value for $\tilde\chi_{j(\rm P+T)}^{(5, \nu^2)\rm cons}$ obtained in \cite{Bini:2021gat,Bini:2022enm} is formulated 
using $v_\infty \equiv \gamma v$. Although this distinction (as well as the factors of $1/\Gamma$ in \eqref{chirel}) are immaterial at the 
level of contributions to $\chi$ already entering at 5PN order, it does modify the explicit form of the 
expression in \cite{Driesse:2026qiz} (in terms of $v_\infty$) through the propagation of lower-order potential and tail corrections.} Moreover, it is also straightforward to see that the combined  
$\tilde\chi_{j \rm(M\text{-}like)PBloc}^{(5, \nu^2)\rm cons}$ [quoted in {\color{cyan} cyan}] is exactly {\it opposite} to the value obtained in~\cite{Driesse:2026qiz} under the $\gamma\text{-}3$ prescription (that sets $c_{\rm M}=1$ in their final expression). We return to the origin of this discrepancy in \S\ref{alternative}.

\section{Local-in-time isotropic decomposition}\label{isotropic}

The identification of scattering observables with (gauge-dependent) dynamical quantities is often achieved by working in an isotropic-like coordinate system, in which for instance a Hamiltonian can be arranged to depend only on $r$ and $\bp^2$~\cite{Kalin:2019rwq,Kalin:2019inp}. This construction admits a natural extension to the dissipative sector, where matching computations to either an energy flux~\cite{Cho:2021arx} or an effective force~\cite{Manohar:2022dea} provide the appropriate bridge between observables and dynamical variables. In what follows, we build on these ideas to construct isotropic-like representations for the various nonlinear gravitational radiation-reaction effects identified in this work. For reasons that will become clear momentarily we decompose the full force in isotropic-like coordinates as follows: 
\begin{align}
	\ba_{(\rm iso)} = -\frac{G M}{r^2}\bn + \sum_{n}\frac{G^n }{r^{n+1}}\bigg[
	\left(\alpha_n(\bv^2)+ \delta_n(\bv^2)(\bv \cdot \bn)\right)\bn
	+\left(\beta_n( \bv^2) (\bv\cdot \bn)+ \gamma_n (\bv^2) \right)\bv\label{aiso}
	 \bigg]\; .
\end{align}
As we shall see, the various terms appearing in this decomposition play distinct physical roles. Some naturally account for the irreversible losses of energy and angular momentum associated with dissipative dynamics, whereas others encode contributions that exhibit a conservative-like character. The latter, however, should not be regarded as genuinely conservative in a strict Hamiltonian sense, since in general they do not admit an underlying Hamiltonian representation. This distinction highlights the subtle interplay between conservative and dissipative structures beyond the conventional Hamiltonian framework. Furthermore, the representation introduced above provides a particularly natural setting in which to formulate a local-in-time, on-shell prescription for separating the conservative and dissipative sectors. Importantly, this construction achieves the desired decomposition directly at the level of the local dynamics, thus avoiding any explicit reference to intermediate nonlocal-in-time effects.

\subsection{Dissipative}
In the general parametrization of~\eqref{aiso}, the components responsible for energy and angular-momentum loss are encoded in the $\{\delta_n(\bv^2),\gamma_n(\bv^2)\}$ coefficients~\cite{Manohar:2022dea}. For instance, at leading order in radiation-reaction, we have
\begin{align}
	\ba_{\rm (RR, iso)} = 	
	\frac{G^2 M^2 \nu}{r^{3}}\bigg[
	\bar\gamma_2\, \bv^2\, \bv
	+ \bar\delta_2\,\bv^2(\bv \cdot \bn)\bn \bigg]+ \frac{G^3 M^3 \nu}{r^{4}}\bigg[
	\bar\gamma_3\, \bv
	+ \bar\delta_3(\bv \cdot \bn)\bn
	\bigg] + \cdots\,,
\end{align} 
which leads to
\begin{align}
	\Delta^{(3)} E_{\rm (RR, iso)} & = \bigg(
	\frac{\bar{\delta}_3+4\bar{\gamma}_3+6\bar{\gamma}_2}{8}
	\bigg)\pi\frac{G^3 M^4\nu^2}{b^3}v_\infty \; , \\
	\Delta^{(4)} E_{\rm (RR, iso)} & = 4\bigg(
	\frac{\bar{\delta}_3+5\bar{\gamma}_3+7\bar{\gamma}_2}{3}
	\bigg)\frac{G^4 M^5\nu^2}{b^4 v_\infty} \; , \\
	\Delta^{(2)} \bL^i_{\rm (RR, iso)} & = \big(
	2\bar{\gamma_2}
	\big)\frac{G^2 M^3\nu^2}{b^2}v_\infty\varepsilon^{ijk}b^j v_\infty^k \; ,
\end{align}
where $\{\bar\delta_n,\bar\gamma_n\}$ are numerical coefficients and $\bar{\delta}_2 = -3\bar{\gamma}_2$ so that the energy emission is zero at $O(G^2)$. Matching the energy and angular-momentum losses, together with the value of the impulse in \eqref{prr1}-\eqref{prr3}, we find
\begin{align}
	\ba^{\rm 2.5PN}_{\rm (RR, iso)} = 	
	-\frac{8}{5}\frac{G^2 M^2 \nu}{r^{3}}\bigg[
	 \bv^2\, \bv
	-3\,\bv^2(\bv \cdot \bn)\bn\bigg] - \frac{8G^3 M^3 \nu}{5r^{4}}\bigg[
	3 \bv
	-\frac{17}{3}(\bv \cdot \bn)\bn
	\bigg]
	 \; ,\label{eq:BTisoLO}
\end{align}
at leading 2.5PN order. It is straightforward to show that the above expression is equivalent to the original Burke-Thorne force in \eqref{eq:BTaLO}, via the coordinate transformation
\beq
\delta \br_{\rm iso \to BT}  = \frac{8}{3} \frac{G^2M^2\nu}{r} \left(\bv - (\bv\cdot \bn)\bn\right)\,.\label{eq:shift} 
\eeq
Moreover, it is also in agreement in the overlap with the PM derivation in \cite{Manohar:2022dea}.\vskip 4pt

Amusingly, the acceleration in \eqref{eq:BTisoLO} not only reproduces the linear losses of energy and angular momentum, but also captures the second-order contributions arising from RR--RR effects. It does not, however, yield the correct RR--RR impulse at 5PN order [cf.~\eqref{prr21}--\eqref{prr22}]. This discrepancy can only be resolved by the presence of additional contributions to the coordinate transformation in \eqref{eq:shift}. Moreover, such extra terms can affect only the $\{\alpha_n,\beta_n\}$ sector: the exact agreement in the energy and angular-momentum losses leaves no room for modifications elsewhere. This observation underpins the incorporation of RR--RR effects into an extended conservative sector that, furthermore, can be embeded into an (isotropic) Hamiltonian framework.

\subsection{Conservative}

 From the perspective of an isotropic frame, a conservative-like force can be entirely captured by the $\{\alpha_n(\bp^2),\beta_n(\bp^2)\}$ coefficients appearing in the parametrization~\eqref{aiso}. Indeed, upon restriction to this sector we find the existence of the following conserved quantities, %
\begin{align}
M\nu \left(\ba_{\rm (iso)} \cdot \bv\right)_{\{\alpha_n,\beta_n\}}&= 0 -\frac{\dd E_{\rm (iso)}}{\dd t}  \; , \quad
 2M\nu \left(\br^{[i}\ba^{j]}_{\rm (iso)}\right)_{\{\alpha_n,\beta_n\}}= 0 - \frac{\dd L^{ij}_{\rm (iso)}}{\dd t}  \; .
	\end{align}
Namely, once added to the Newtonian counterparts, the energy and angular-momentum (see \eqref{eq1} and \eqref{eq2} below) remain constant throughout the flow dictated by the equations of motion. In particular, to 5PN order (with $ \{\bar{\alpha}_i,\bar{\beta}_i\}$ numerical coefficients)
\begin{align}
	\ba_{\rm (iso)}\Big|_{\{\alpha_n,\beta_n\}}= 
	& \bar\alpha_6\frac{ G^6 M^6\nu^2  }{r^7} \bn
	+\frac{ G^5 M^5\nu^2  }{r^6}\bigg[
	\bar\alpha_5 \bv^2 \bn + \bar\beta_5(\bv\cdot \bn)\bv
	\bigg]
	\notag \\&+\frac{ G^4 M^4\nu^2  }{r^5}\bigg[
	\bar\alpha_4 \bv^4 \bn + \bar\beta_4 \bv^2(\bv\cdot \bn)\bv
	\bigg]
	 +\frac{ G^3 M^3\nu^2  }{r^4}\bigg[
	 \bar\beta_3\bv^4(\bv\cdot \bn)\bv
	\bigg] +\cdots \; ,\label{aisoX}
\end{align}
with the corresponding conserved charges associated to energy and angular-momentum given by
\begin{align}
	\label{eq1} E^{\rm 5PN}_{\rm (iso)} = 
	& +\bigg(\frac{\bar\beta_3}{3}\bigg)\frac{G^3 M^4\nu^3}{r^3}\bv^6
	+\bigg(\frac{\bar\alpha_4+\bar\beta_4-2\bar\beta_3}{4}\bigg)\frac{G^4 M^5\nu^3}{r^4}\bv^4  \\
	& +\bigg(\frac{\bar\alpha_5+\bar\beta_5-\bar\alpha_4-\bar\beta_4+2\bar\beta_3}{5}\bigg)\frac{G^5 M^6\nu^3}{r^5}\bv^2 \notag \\
	& +\bigg(\frac{5\bar\alpha_6-2\bar\alpha_5-2\bar\beta_5+2\bar\alpha_4+2\bar\beta_4-4\bar\beta_3}{30}\bigg)\frac{G^6 M^7\nu^3}{r^6} \notag \,\\
\label{eq2} \bL^{i,\rm 5PN}_{\rm (iso)} & =\varepsilon^{ijk}\br^j \bv^k\bigg[
	+\frac{G^3 M^4\nu^3}{r^3}\bigg(\frac{\bar{\beta}_3}{3}\bigg) \bv^4
	+\frac{G^4 M^5 \nu^3}{r^4}\bigg(\frac{3\bar{\beta}_4 - 4\bar{\beta}_3}{12}\bigg) \bv^2 \notag \\
	&\qquad\qquad 
	+\frac{G^5 M^6 \nu^3}{r^5}\bigg(\frac{6\bar{\beta}_5 - 3 \bar{\beta}_4 + 4\bar{\beta}_3}{30}\bigg) 
	\bigg] \; ,
\end{align}
to be added to the Newtonian values. These conservation laws do not, however, automatically imply the existence of an underlying Hamiltonian (or Lagrangian) description. Indeed, as we shall see, while the full (in-in) FT, M, and RR$^{2}$ forces are conservative-like in the above sense, they cannot be described by a Hamiltonian (or Lagrangian).

\subsubsection{Non-Hamiltonian}
After including the $\{\alpha_n,\beta_n\}$-shifts to the radiation-reaction acceleration induced by matching the (second-order) RR--RR impulse, the final form of the nonlinear 5PN in-in accelerations, $\ba_{\rm X}$ with $\rm X = \big\{ FT,M,RR^2, 
(RR\text{--}RR)_{\{\alpha,\beta\}}\big\}$, can be mapped into the isotropic parameterization in~\eqref{aisoX}. The matching can be performed via comparison with the corresponding observables at infinity. Since the $\bF_{\rm X}$ forces do not admit a Hamiltonian description, the matching requires not only the impulse/angle but also an additional (independent) observable, which we take to be the scattering time delay, defined as the first moment of the acceleration in the direction of the incoming velocity.\footnote{As is well known, the Newtonian trajectory induces an infrared divergence. Therefore, we introduce a {\it renormalized} value upon subtracting the divergent contribution. (See App.~\ref{app:delay}.)} After including the (already isotropic) Newtonian force,  we obtain the following (gauge-invariant) values:\footnote{Notice that for this type of conservative-like forces we automatically have $\Delta \bp_v = 2 \frac{G}{v_\infty^2} \Delta \bp_b$, such that the impulse and scattering angle carry the same amount of information.} %
\begin{align}
	\Delta\ord{4}\bp_{(\rm iso)} & =
	\frac{G^4 M^5 \nu^3}{b^4}\bigg[
	\bigg(
	\frac{\pi(3\bar\alpha_4-2\bar\beta_3)}{8}
	\bigg)\frac{\bb}{b}\bigg]v_\infty^3 \; , \\
	\Delta\ord{5}\bp_{(\rm iso)} &=
	\frac{G^5 M^6 \nu^3}{b^5}\bigg[
	\bigg(
	\frac{4(31 \bar\alpha_4+4\bar\alpha_5-2\bar\beta_4-18\bar\beta_3)}{15}	\bigg)\frac{\bb}{b}
	+\bigg(
	\frac{\pi(3\bar\alpha_4-2\bar\beta_3)}{4} 
	\bigg)\frac{\bv_\infty}{v_\infty}
	\bigg]v_\infty \;  \\
	\Delta\ord{6}\bp_{(\rm iso)}&=
	\frac{G^6 M^7 \nu^3}{b^6}\bigg[
	\bigg(
	\frac{\pi(134 \bar\alpha_4+34 \bar\alpha_5+5 \bar\alpha_6-2\bar\beta_5-16 \bar\beta_4-68\bar\beta_3)}{16} \bigg)\frac{\bb}{b}
\notag \\
	&\hspace{1.5cm}
	+\bigg(
	\frac{8(31 \bar\alpha_4+4\bar\alpha_5-2\bar\beta_4-18\bar\beta_3)}{15}
	\bigg)\frac{\bv_\infty}{v_\infty}
	\bigg]\frac{1}{v_\infty} \; .
\end{align}
Analogously, we find the following results for the (renormalized) time delay
\begin{align}
	\tau^{(3)}_{\rm (iso)}  & = 
	\bigg(\frac{2}{3}\bar{\beta}_3\bigg)\frac{G^3 M^3 \nu^2}{b^2}v^3_\infty  \; , \\
	\tau^{(4)}_{\rm (iso)}  & = \bigg(\frac{\bar{\alpha}_4+6 \bar{\beta}_3+\bar{\beta}_4}{8}\pi\bigg)\frac{G^4 M^4 \nu^2}{b^3}v_\infty  \; , \\
	\tau^{(5)}_{\rm (iso)}  & =  \bigg(\frac{61 \bar{\alpha}_4+4 \bar{\alpha}_5+28 \bar{\beta}_3+21 \bar{\beta}_4+4 \bar{\beta}_5}{15}\bigg)\frac{G^5 M^5 \nu^2}{b^4 v_\infty}  \; , \\
	\tau^{(6)}_{\rm (iso)}  & = \bigg(\frac{738 \bar{\alpha}_4+82 \bar{\alpha}_5+5 \bar{\alpha}_6-356 \bar{\beta}_3+18 \bar{\beta}_4+22 \bar{\beta}_5}{80}\pi
	\bigg)\frac{G^6 M^6 \nu^2}{b^5 v^3_\infty} 
\end{align}
Matching these values to those from the original $\bF_X$ forces---derived within the background harmonic gauge \cite{Porto:2024cwd}---we obtain the isotropic conservative-like accelerations, 
\begin{align}
\label{aftiso}	\ba_{\rm (FT,iso)} = 
	& \frac{5}{6}\frac{ G^6 M^6\nu^2  \bn}{r^7}
	+\frac{ G^5 M^5\nu^2  }{r^6}\bigg[
	-\frac{49}{20} \bv^2 \bn + \frac{289}{30}(\bv\cdot \bn)\bv
	\bigg]  \\
	& \qquad+\frac{ G^4 M^4\nu^2 }{r^5}\bigg[
	-\frac{23}{3} \bv^4 \bn + \frac{184}{15}\bv^2(\bv\cdot \bn)\bv
	\bigg] \nn \; , \\
	\ba_{\rm (M,iso)} = 
	& -\frac{67}{63}\frac{ G^6 M^6\nu^2  \bn}{r^7}
	+\frac{ G^5 M^5\nu^2  }{r^6}\bigg[
	\frac{1951}{630} \bv^2 \bn  -\frac{5476}{315}(\bv\cdot \bn)\bv
	\bigg]  \\
	&+\frac{ G^4 M^4\nu^2  }{r^5}\bigg[
	\frac{1058}{105} \bv^4 \bn  -\frac{5612}{315}\bv^2(\bv\cdot \bn)\bv
	\bigg] \; , \nn \\
\label{a2rriso}		\ba_{{\rm (2RR,iso)}} = 
	& \bigg(
	\frac{1994}{225} +{\color{red}\frac{228}{175}}
	\bigg)\frac{ G^6 M^6\nu^2  }{r^7}\bn \\
	&+\frac{ G^5 M^5\nu^2  }{r^6}\bigg[
	\bigg(\frac{1861}{225}-{\color{red}\frac{6}{7}}\bigg) \bv^2 \bn 
	+ \bigg(\frac{3382}{45}+{\color{red}\frac{100}{7}}\bigg)(\bv\cdot \bn)\bv
	\bigg] \notag \\
	&
	+\frac{ G^4 M^4\nu^2  }{r^5}\bigg[
	\bigg(-\frac{388}{75}-{\color{red}\frac{728}{15}}\bigg) \bv^4 \bn + \bigg(\frac{3344}{45}+{\color{red}\frac{704}{15}}\bigg) \bv^2(\bv\cdot \bn)\bv
	\bigg] \; \nn .
\end{align}
where we have combined all 2RR effects and highlighted [in {\color{red}{red}}] the RR--RR contributions.\vskip 4pt 

We have checked that none of the forces derive directly from a Hamiltonian/Lagrangian valid for generic orbits. However, it is straightforward to derive the conserved quantities,
\begin{align}
\label{Einin}
\delta \hat E_{\rm (iso)} &= \frac{1}{M\nu}\left( E^{\rm 5PN}_{(\rm  FT, iso)}+E^{\rm 5PN}_{(\rm M, iso)}+E^{\rm 5PN}_{(\rm  2RR, iso)}\right)\,, \\
\label{Linin}
\delta \hat \bL^i_{ \rm (iso)} &= \frac{1}{M\nu} \left(\bL^{i ,\rm 5PN}_{(\rm  FT, iso)}+\bL^{i ,\rm 5PN}_{(\rm M, iso)}+\bL^{i ,\rm 5PN}_{(\rm  2RR, iso)}\right)
\end{align}

with 
\begin{align}
	E^{\rm 5PN}_{\rm (FT,iso)} = 
	&- \frac{1}{30}\frac{G^6 M^7\nu^3}{r^6}
	+\frac{31}{60}\frac{G^5 M^6\nu^3}{r^5}\bv^2
	+\frac{23}{20}\frac{G^4 M^5\nu^3}{r^4}\bv^4 \; , \\
	E^{\rm 5PN}_{\rm (M,iso)} = 
	& +\frac{7}{27}\frac{G^6 M^7\nu^3}{r^6}
	-\frac{55}{42}\frac{G^5 M^6\nu^3}{r^5}\bv^2
	-\frac{1219}{630}\frac{G^4 M^5\nu^3}{r^4}\bv^4 \; , \\
	E^{\rm 5PN}_{{\rm (2RR, iso)} } = 
	&+\bigg(\frac{118}{225}-{\color{red}\frac{412}{525}}\bigg)\frac{G^6 M^7\nu^3}{r^6}
	+\bigg(\frac{643}{225}+{\color{red}\frac{526}{175}}\bigg)\frac{G^5 M^6\nu^3}{r^5}\bv^2  \\
	& +\bigg(\frac{3889}{225}-{\color{red}\frac{2}{5}}\bigg)\frac{G^4 M^5\nu^3}{r^4}\bv^4 \; , \notag\\
 \bL^{\rm 5PN}_{i \rm (FT,iso)} = & \varepsilon_{ijk}\br_j \bv_k\bigg[
	\frac{7}{10}\frac{G^5 M^6 \nu^3}{r^5} 
	+\frac{46}{15}\frac{G^4 M^5 \nu^3}{r^4} \bv^2
	\bigg]  \; , \\
	 \bL^{\rm 5PN}_{i\rm (M,iso)} = 
	&  \varepsilon_{ijk}\br_j \bv_k\bigg[
	-\frac{178}{105}\frac{G^5 M^6 \nu^3}{r^5} 
	-\frac{1403}{315}\frac{G^4 M^5 \nu^3}{r^4} \bv^2
	\bigg] \; , \\
	 \bL^{ \rm 5PN}_{i \rm (2RR, iso) } =  
	&  \varepsilon_{ijk}\br_j \bv_k\bigg[
	\bigg(
	\frac{38}{5}
	-{\color{red}\frac{964}{525}}
	\bigg)\frac{G^5 M^6 \nu^3}{r^5} 
	+\bigg(\frac{836}{45}
	+{\color{red}\frac{176}{15}}
	\bigg)\frac{G^4 M^5 \nu^3}{r^4} \bv^2
	\bigg] \; . 
\end{align}	
 We have verified that the above accelerations can be mapped to the original ones in \cite{Porto:2024cwd} (including the Burke-Thorne force) by adding to~\eqref{eq:shift} the transformation,
\begin{align}
\label{risoFT}	\delta_{\rm (FT)}\br & = -\frac{1}{30}\frac{G^5 M^5\nu^2}{r^4}\bn
	+\frac{G^4 M^4\nu^2}{r^3}
	\bigg[
	\bigg(
	\frac{7}{12}\bv^2
	+\frac{53}{15}(\bv\cdot\bn)^2
	\bigg)\bn
	-\frac{9}{2}(\bv\cdot\bn)\bv
	\bigg] \notag \\
	& \quad+\frac{G^3 M^3\nu^2}{r^2}
	\bigg[
	\bigg(
	-\frac{76}{15}\bv^4
	+\frac{52}{5}\bv^2(\bv\cdot\bn)^2
	+4(\bv\cdot\bn)^4
	\bigg)\bn \notag \\
	& \hspace{2cm} +\bigg(\frac{20}{3}\bv^2
	-16(\bv\cdot\bn)^2
	\bigg)(\bv\cdot\bn)\bv
	\bigg] \; , \\
	\delta_{\rm (M)}\br & = -\frac{71}{105}\frac{G^5 M^5\nu^2}{r^4}\bn
	+\frac{G^4 M^4\nu^2}{r^3}
	\bigg[
	\bigg(
	-\frac{1679}{630}\bv^2
	-\frac{136}{35}(\bv\cdot\bn)^2
	\bigg)\bn
	+\frac{128}{105}(\bv\cdot\bn)\bv
	\bigg] \notag \\
	& \quad+\frac{G^3 M^3\nu^2}{r^2}
	\bigg[
	\bigg(
	\frac{44}{5}\bv^4
	-\frac{2022}{35}\bv^2(\bv\cdot\bn)^2
	+\frac{358}{7}(\bv\cdot\bn)^4
	\bigg)\bn \notag \\
	& \hspace{2cm}
	+\bigg(\frac{566}{35}\bv^2
	-\frac{706}{35}(\bv\cdot\bn)^2
	\bigg)(\bv\cdot\bn)\bv
	\bigg] \; , \\
\label{risoRR2}	 \delta_{\rm (2RR)}\br & = \bigg(\frac{502}{225}-{\color{red}\frac{412}{525}}\bigg)\frac{G^5 M^5\nu^2}{r^4}\bn \notag \\
	& \quad +\frac{G^4 M^4\nu^2}{r^3}
	\bigg[
	\bigg(
	\bigg(\frac{67}{9}-{\color{red}\frac{334}{45}}\bigg)\bv^2
	+\bigg(\frac{1084}{225}+{\color{red}\frac{712}{45}}\bigg)(\bv\cdot\bn)^2
	\bigg)\bn
	+\bigg(-\frac{314}{45}-{\color{red}\frac{124}{15}}\bigg)(\bv\cdot\bn)\bv
	\bigg] \notag \\
	& \quad+\frac{G^3 M^3\nu^2}{r^2}
	\bigg[
	\bigg(
	-\frac{2816}{75}\bv^4
	+\frac{5312}{25}\bv^2(\bv\cdot\bn)^2
	-176(\bv\cdot\bn)^4
	\bigg)\bn \notag \\
	& \hspace{2cm}
	+\bigg(-\frac{256}{3}\bv^2
	+\frac{432}{5}(\bv\cdot\bn)^2
	\bigg)(\bv\cdot\bn)\bv
	\bigg] \; .
\end{align}

\subsubsection{Hamiltonian}

The existence of an isotropic form of a conservative-like force does not, in general, guarantee the existence of a (local) Hamiltonian. In such an isotropic coordinate system, the latter takes the form,
\begin{align}
	H(r, \bp^2) = \frac{\bp^2}{2M\nu} 
	+ \sum_n c_n(\bp^2)\left(\frac{G}{r}\right)^n \; ,
\end{align}
such that the equations of motion read
\begin{align}
	\dot{\br}   = \mathcal{A}(r, \bp^2)\bp \; ,	\qquad \dot{\bp}   = \mathcal{B}(r, \bp^2)\bn \,,
	\end{align}
	with
	\begin{align}
	  \mathcal{A}(r, \bp^2) \equiv 2\frac{\partial H}{\partial\bp^2} =\frac{1}{M\nu}+2\sum_n G^n \frac{c^\prime_n(\bp^2)}{r^n} \; , \quad 
	 \mathcal{B}(r, \bp^2)  \equiv  -\frac{\partial H}{\partial r} = \sum_n nG^n \frac{c_n(\bp^2)}{r^{n+1}} \; .
\end{align}

Using that $\bp^2$ is a function of  $\bv^2$, and vice versa, we immediately conclude that the acceleration derived from an isotropic Hamiltonian has the same form as in \eqref{aiso}, 
\begin{align}
	\ba &  = \alpha^{\rm H}(r, \bv^2) \bn 
	+(\bv\cdot \bn)\beta^{\rm H}(r, \bv^2) \bv  \, ,
\end{align}
with vanishing $\{\delta_n,\gamma_n\}$, and obeys the conditions 
\begin{align}
	\alpha^{\rm H}(r, \bv^2) = \mathcal{A}(r, \bp^2)\mathcal{B}(r, \bp^2) \; ,
	& & 
	\beta^{\rm H}(r, \bv^2) = \frac{\partial \mathcal{A}(r, \bp^2)}{\partial r}
	+2\frac{\partial \mathcal{A}(r, \bp^2)}{\partial \bp^2}\frac{\mathcal{B}(r, \bp^2)}{\mathcal{A}(r, \bp^2)} \; .
	\label{eq:alphabetadef}
\end{align}

Moreover, we can show that
\begin{align}
	\alpha^{\rm H}_n  = \frac{n c_n}{M\nu} +\sum_{k=1}^{n-1}2(n-k)c^\prime_{k}c_{n-k}\; , \quad
	\beta^{\rm H}_n = -2nc_n^\prime +\sum_{k=1}^{n-1}4c^{\prime\prime}_{k}d_{n-k}\; ,
\end{align}
such that the $d_n$ coefficients, defined through the relation
\begin{align}
	\frac{\mathcal{B}}{\mathcal{A}} & = -\frac{G M^3\nu^2}{r^2} + \sum_n G^n\frac{d_n(\bp^2)}{r^{n+1}} \; ,
\end{align}
 obey the following recurrence relation
\begin{equation}
	d_n = M \nu \bigg( n c_n 
	-\sum_{k=1}^{n-1}2c_k^\prime d_{n-k}\bigg)\,.
\end{equation}

Despite the existence of a conserved energy [cf.~\eqref{Einin}], it is straightforward to show that the full in-in forces do not satisfy these conditions. Nevertheless, a local-in-time description of the conservative sector may be naturally introduced at the level of the isotropic gauge by means of retaining the associated {\it local-only} contributions to the scattering angle. A Hamiltonian, or \textit{impetus} formula, can then be  reconstructed through the standard B2B manipulations described in \cite{Kalin:2019rwq,Kalin:2019inp}. Equivalently, the same information may be repackaged into a local-in-time EOB potential, as we will return to describe shortly.\vskip 4pt In order to compute the contribution to the Hamiltonian at 5PN order we require the value for the three coefficients $\{c_{4,2},c_{5,1},c_{6,0}\}$, defined through the PN expansion
\bea
c_4(\bp^2) &=& M^5\nu \left(\cdots + c_{4,2}\, \hat \bp^4 +\cdots\right) \,, \\
c_5 (\bp^2) &=& M^6\nu \left( \cdots + c_{5,1} \, \hat \bp^2 + \cdots\right) \,,\\
c_6 (\bp^2) &=& M^7\nu \left(c_{6,0} + \cdots \right) \,,
\eea
with $\hat \bp \equiv \bp/M\nu$. Following the B2B map~\cite{Kalin:2019rwq,Kalin:2019inp}, we find the following relationship at 5PN order\footnote{Since $\Gamma = \sqrt{1+2\nu(\gamma-1)}$, and $\gamma-1 
\simeq {\cal O}(v^2)$, we can set $\Gamma = 1 + {\cal O}(v^2)$ at 5PN order.}    
 \begin{equation}
\begin{aligned}
c_{4,2} &= -\frac{4}{3\pi v_\infty^6}\,\tilde\chi_{j}^{(4)}\,, \quad c_{5,1} =- \frac{3}{8 v_{\infty}^5}\,\tilde \chi_{j}^{(5)}
+\frac{28}{3\pi v_\infty^6}\,\tilde\chi_{j}^{(4)}\,,\\
c_{6,0} &= -\frac{16 }{15\pi v_\infty^4}\,\tilde\chi_{j}^{(6)}
+\frac{9}{4 v_{\infty}^5}\,\tilde\chi_{j }^{(5)}
-\frac{72}{3 \pi v_\infty^6}\,\tilde\chi_{j}^{(4)}.
\end{aligned}
\end{equation}
Combined with \eqref{eq:chi4nu2const}-\eqref{eq:chi6nu2const} we arrive at the following contribution to the Hamiltonian from FT, M and 2RR tail-like [in {\color{ForestGreen} green}] and memory-like [in {\color{cyan}cyan}] effects, with $\hat H \equiv H/(M\nu)$, 
\begin{align}
 \frac{\delta \hat H_{(\rm iso)} }{\nu^2}&=  -{\color{ForestGreen} \frac{2973}{350}}\, \hat\bp^4 \left(\frac{GM}{r}\right)^4 - \left({\color{ForestGreen}{\frac{617}{150}}}+{\color{cyan}{\frac{608}{45}}}\right) \hat \bp^2\left(\frac{GM}{r}\right)^5 \nn \\
&-\left({\color{ForestGreen}{\frac{104833}{25200}}} - {\color{cyan} \frac{50299}{8400}}\right)  \left(\frac{GM}{r}\right)^6\label{hatHloc} \,.
\end{align}
From here we obtain the associated conservative accelerations, with the {\it dissipative} (yet `energy'-conserving) counterparts given by subtracting from the total value in \eqref{aftiso}-\eqref{a2rriso}. By construction, this split remains local in time and entirely consistent with the full dynamics. 

\subsection{Local EOB coefficients}\label{loceob}
Using the results in \cite{Bini:2020wpo,Bini:2021gat,Bini:2025vuk}, 
\bea
\tilde\chi_{j(\rm tot)}^{(5, \nu^2)\rm cons, TF} &=& \left(  \frac{1408}{45}\log(2v_\infty) -\frac{365555}{6048} - \frac{4}{15} 
\bar d_{5\rm loc}^{{\rm rat} \, \nu^2}-\frac{224057}{1440}\pi^2\right)v_\infty^5 \\
\tilde\chi_{j(\rm tot)}^{(6, \nu^2)\rm cons, TF} &=& \pi \left( \frac{2817}{16}\zeta(3)+\frac{30161}{192}+\frac{201}{2}\log\left(\frac{v_\infty}{2}\right) - \frac{10812865}{32768}\pi^2 \right. \\
&& \left. - \frac{15}{32} \left( \bar d_{5\rm loc}^{{\rm rat} \, \nu^2}+a_{6\rm loc}^{{\rm rat} \, \nu^2}\right) \right) v_\infty^4\,,\nn
\eea
and matching the WEFT and Tutti-Frutti deflection angles, we can then also determine the rational part of the missing EOB coefficients, which become\footnote{Although tail-like terms appear to introduce an additional $\pi^2$ correction at ${\cal O}(G^6)$ [in {\color{ForestGreen}green}], we notice this is neatly canceled by memory-like contributions [in {\color{RoyalBlue}blue}].}
\bea
\label{eob1}
 \bar d_{5\rm loc}^{{\rm rat} \, \nu^2}&=& -\frac{394747}{63} - {\color{cyan}\frac{1216}{9}} = -\frac{403259}{63} \,,\\
\label{eob2} a_{6\rm loc}^{{\rm rat} \, \nu^2} &=& -\frac{13559209}{12600} -{\color{cyan}\frac{189583}{12600}} = -\frac{1718599}{1575}\,.
\eea
Notice that the contribution arising from the memory-like region [in {\color{cyan}cyan}] is at most of the order of $2\%$ compared to the sum of potential and tail-like terms. The latter have now been independently established at ${\cal O}(G^5)$ within both the PN and PM frameworks, which provides a solid foundation for the corresponding 5PN contribution at ${\cal O}(G^6)$. 

\section{Alternative Memories}\label{alternative} 

Building on the manifest locality of the associated in-in forces in the isotropic-like representation, in the previous section we constructed a separation between conservative and dissipative effects without the need to invoke nonlocal-in-time structures (beyond what is already inherent from standard tail terms, see \eqref{tails}). In what follows we explore other possible options. 

\subsection{Origin of nonlocality} 

The nonlocal-in-time structure of the action can be traced directly to the appearance of a product of absolute values, namely $|\omega_1|\,|\omega_2|$, in the (real) Feynman effective action. This feature is not optional: it is enforced by the choice of propagator that renders the symmetry $\omega\to-\omega$ manifest at the level of the frequency integrand. Indeed, the reader will immediately recognize that $|\omega|$ is essentially the unique parity-even function that is linear in $\omega$ in magnitude (any putative odd-in-$\omega$ linear term would be eliminated by the $\omega\to-\omega$ symmetry). The ensuing time nonlocality is therefore nothing but the Fourier-space imprint of this parity-symmetrized linearity, amplified by the specific convolution of Green's functions that arises in nonlinear gravitational contributions to the effective action.\vskip 4pt 
 
As discussed already in \cite{Porto:2024cwd}, we may isolate a tail-like region within the class of propagator convolutions in which the two frequencies effectively collapse onto a single one through an intermediate \textit{static} propagator. As we demonstrated, this prescription yields perfect agreement with the corresponding results across the existing PM literature~\cite{Dlapa:2021npj,Dlapa:2021vgp,Driesse:2024xad,Driesse:2026qiz}. Moreover, the resulting contribution to the scattering angle can be shown to follow from a manifestly local action [cf.~\eqref{7.2n} in {\color{ForestGreen} green}]. However, in contrast to genuine tail (and failed-tail) terms---where the multipoles couple through a conserved quantity---the tail-like sector is characterized by two crucial features: {\it i)} one must keep one of the three multipoles \textit{unperturbed}; and {\it ii)} the $\omega\to-\omega$ symmetry, implemented through absolute values, discards some of the contributions. The latter mechanism is essential for obtaining the one-to-one match with the PM results. As a result, the tail-like action can only be understood in a PM-expanded sense, in which we remove certain otherwise admissible contributions (see \eqref{7.2} and Eq. (7.2) in \cite{Porto:2024cwd}).\vskip 4pt

Another crucial aspect of the derivation is that Feynman's prescription does more than drop some of the corrections associated with \textit{static} multipoles, it also eliminates from the tail-like region an extra term arising from the product of three quadrupole moments in the action. Concretely, the piece proportional to the frequency structure [cf.\ \eqref{Smem1}]
\beq
\sg(\omega_1)\sg(\omega_3)\omega_1^2\,\omega_3^2\,\omega_2^4\,\delta(\omega_1+\omega_2+\omega_3)\,,\label{123}
\eeq
which, for the tail-like region, is identically zero once the conditions in \eqref{w3I}-\eqref{w2sign} are imposed. This constitutes another essential ingredient in securing agreement with both 4PM \cite{Dlapa:2021npj,Dlapa:2021vgp} and 5PM \cite{Driesse:2024xad,Driesse:2026qiz} results. The extra term contributes, starting at ${\cal O}(G^5)$, yielding to  the nonlocal-in-time part of the memory-like effective action \cite{Porto:2024cwd}, and becomes a key ingredient of the full conservative sector [cf. \eqref{eq:chi5nu2const}-\eqref{eq:chi6nu2const} in {\color{violet}{violet}}]. These observations underscore the various subtle aspects of Feynman's prescription. 

\subsection{Feynman \& Wightman}

As shown in \cite{Kalin:2022hph}, the use of Feynman propagators as a vehicle to construct conservative quantities follows directly from the in-in construction, notably in the original $(1,2)$ basis, combined with the relationship between Feynman and retarded/advanced propagators
\begin{align}
\label{retF}
\Delta_{\rm ret/adv} (x) &= \Delta_F(x) + \Delta_{\mp} (x)\,,\quad
\Delta_{\rm ret/adv} (x) =  \int_k \frac{e^{ik \cdot x}}{(k^0\pm i0^+)^2-\bk^2}\,,\\
\Delta_F(x) &=  \int_k \frac{e^{ik \cdot x}}{k^2+i0^+} \,, \quad \Delta_{\pm} (x) = (2\pi i) \int_k e^{ik \cdot x}\delta(k^2) \theta(\pm k^0)\,.\nn
\end{align}
In the present work, the relevant Green's functions that enter in the EFT computations involving time-dependent multipole moments (which absorb the dependence on the spatial coordinates \cite{Goldberger:2009qd}) are obtained after
integrating over $\dd^3\bk$, yielding (with the center-of-mass located at $\bx=0$)
\begin{equation}
	\begin{aligned}
	\Delta_{\rm ret/adv}(\omega) 
	&=\mp\frac{i}{4\pi}\omega \; , \quad
	\Delta_{F}(\omega)  	= -\frac{i}{4\pi}\sg(\omega)\omega\; , 
	&\quad \Delta_{\pm} (\omega) = \pm\frac{i}{2\pi} \theta(\pm \omega)\omega \; ,
	\end{aligned}
\end{equation}
From here it follows that the relation in \eqref{retF} is nothing but a reflection of the identity $\sg(\omega) = 1-2\theta(-\omega)$, which the reader will recall is also behind the identification of a local part through the PB prescription. The apparent nonlocality in the memory-like integral is therefore paired with a corresponding nonlocal \textit{dissipative} contribution that arises from products of Wightman correlators. For illustrative purposes, let us return to the memory contribution in \eqref{eq:fin}. Upon a few manipulations, it can be rewritten as%
\begin{align}
	S_{\rm (M)} & = -\frac{16 \pi^2 G^2}{70} \int \frac{\dd \omega_1 \dd\omega_2 \dd\omega_3}{(2\pi)^3} \mathcal{F}(\omega_1, \omega_2, \omega_3)
	 \bigg[
	I^{ij}_-(\omega_1) I^{jk}_+(\omega_2) I^{ki}_+(\omega_3)  \Delta_{\rm adv}(\omega_1)\Delta_{\rm ret}(\omega_3)
	 \notag \\ 
		&+ I^{ij}_-(\omega_3) I^{jk}_+(\omega_1) I^{ki}_+(\omega_2)  \Delta_{\rm adv}(\omega_3)\Delta_{\rm ret}(\omega_1)
	+ I^{ij}_-(\omega_2) I^{jk}_+(\omega_3) I^{ki}_+(\omega_1)  \Delta_{\rm ret}(\omega_1)\Delta_{\rm ret}(\omega_3)
	 \bigg] \, ,
\end{align}
with 
\begin{equation}
	\mathcal{F}(\omega_1, \omega_2, \omega_3) =
	\left(7 \omega_1 ^3 \omega_3^3-2 \omega_1 ^2 \omega_3^2 \omega_2^2+2 \omega_1 \omega_3 \omega_2^4\right)\ddl(\omega_1+\omega_2+\omega_3) \; .
\end{equation}
From here we obtain for the acceleration,
\begin{equation}
	\ba^{\ell}_{\rm (M)}(t) = -\frac{16\pi^2 G^2}{35}\delta^{\ell\langle i}\br^{j \rangle}(t) F^{ij}_{\rm (M)}(t) \; ,
\end{equation}
where $F^{ij}_{\rm (M)}(t)$ is the Fourier transform of
\begin{align}
	F^{ij}_{\rm (M)}(\omega) & = \int \frac{\dd\omega_2 \dd\omega_3}{(2\pi)^2} \mathcal{F}(\omega, \omega_2, \omega_3)I^{jk}(\omega_2) I^{ki}(\omega_3)  \Delta_{\rm adv}(\omega)\Delta_{\rm ret}(\omega_3)  \label{eq:FullininExp} \\
	& 
	+\int \frac{\dd\omega_1 \dd\omega_2}{(2\pi)^2} \mathcal{F}(\omega_1, \omega_2, \omega)I^{jk}(\omega_1) I^{ki}(\omega_2)  \Delta_{\rm adv}(\omega)\Delta_{\rm ret}(\omega_1) \notag \\
	& 
	+\int \frac{\dd\omega_1 \dd\omega_3}{(2\pi)^2} \mathcal{F}(\omega_1, \omega, \omega_3)I^{jk}(\omega_3) I^{ki}(\omega_1)  \Delta_{\rm ret}(\omega_1)\Delta_{\rm ret}(\omega_3)\nn \, .
\end{align}
The conservative part is controlled instead by
\begin{align}
\label{sconsM2}
	S^{\rm cons}_{\text{(M)}} & = -\frac{16\pi^2G^2}{70} \int \frac{\dd \omega_1 \dd \omega_2\dd \omega_3}{(2\pi)^3}
\ddl(\omega_1+\omega_2+\omega_3) \Delta_F(\omega_1)\Delta_F(\omega_3)  \\
	&\hspace{3cm}  
	\times I_{ij}(\omega_1)I_{jk}(\omega_3)I_{ki}(\omega_2) \mathcal{F}(\omega_1, \omega_2, \omega_3) \notag  \, .
\end{align}
from which we find
\begin{equation}
	\ba^{{\rm cons},\ell}_{\rm (M)} = -\frac{16\pi^2 G^2}{35}\delta^{\ell\langle i}\br^{j \rangle}(t) F^{ij\, \rm cons}_{\rm (M)}(t) \; ,
\end{equation}
where
\begin{align}
 F^{ij\, \rm cons}_{\rm (M)}(\omega) & = \int \frac{\dd\omega_2 \dd\omega_3}{(2\pi)^2} \mathcal{F}(\omega, \omega_2, \omega_3)I^{jk}(\omega_2) I^{ki}(\omega_3)  \Delta_{F}(\omega)\Delta_{F}(\omega_3)   \\
	& 
	+\int \frac{\dd\omega_1 \dd\omega_2}{(2\pi)^2} \mathcal{F}(\omega_1, \omega_2, \omega)I^{jk}(\omega_1) I^{ki}(\omega_2)  \Delta_{F}(\omega)\Delta_{F}(\omega_1) \notag \\
	& 
	+\int \frac{\dd\omega_1 \dd\omega_3}{(2\pi)^2} \mathcal{F}(\omega_1, \omega, \omega_3)I^{jk}(\omega_3) I^{ki}(\omega_1)  \Delta_{F}(\omega_1)\Delta_{F}(\omega_3) \, .\nn
\end{align}
It is straightforward to show that $F^{ij\, \rm cons}_{\rm (M)}$ follows directly from the total $F^{ij}_{\rm (M)}$ upon replacing  $\Delta_{\rm ret/adv}\to\Delta_{F}$, with the remaining dissipative part, 
$F^{ij\, \rm diss}_{\rm (M)}(\omega) \equiv F^{ij}_{\rm (M)}-F^{ij\, \rm cons}_{\rm (M)}$, involving the Wightman Green's functions.

\subsubsection*{Off-shell}

The above separation into conservative and dissipative terms is {\it independent} of the trajectories. This allows us, in principle, to implement a decomposition of the force into local and nonlocal contributions at the level of the {\it off-shell} action, as we demonstrate below.\vskip 4pt Using the following relationships,
\begin{align}
	\Delta_{\rm adv}(\omega)\Delta_{\rm ret}(\omega_3)-\Delta_F(\omega)\Delta_F(\omega_3) & = 
	\frac{\omega \omega_3}{8\pi^2}\bigg(
	\theta(-\omega)\theta(-\omega_3)+\theta(\omega)\theta(\omega_3)
	\bigg)\\
	\Delta_{\rm ret}(\omega)\Delta_{\rm ret}(\omega_3)-\Delta_F(\omega)\Delta_F(\omega_3) & = 
	-\frac{\omega \omega_3}{8\pi^2}\bigg(
	\theta(\omega)\theta(-\omega_3)+\theta(-\omega)\theta(\omega_3)	\bigg) \;,
\\
	& = -\frac{\omega \omega_3}{8\pi^2}\bigg(1-
\theta(\omega)\theta(\omega_3)-\theta(-\omega)\theta(-\omega_3)
	\bigg) \; , \nn
\end{align}
the dissipative force can be written as (for simplicity we drop the (M)-label from here on)

\begin{align}
	F^{ij\, \rm diss (off)}_{{\rm loc }}(\omega) 
	& = -\int \frac{\dd\omega_2 \dd\omega_3}{(2\pi)^2} \mathcal{F}(\omega_1, \omega, \omega_3)I^{jk}(\omega_1) I^{ki}(\omega_3)  \frac{\omega_1 \omega_3}{8\pi^2}   \\
	F^{ij\, \rm diss (off)}_{{\rm nloc }}(\omega) 	& = \int \frac{\dd\omega_2 \dd\omega_3}{(2\pi)^2} \mathcal{F}(\omega, \omega_2, \omega_3)I^{jk}(\omega_2) I^{ki}(\omega_3)  \frac{\omega \omega_3}{8\pi^2}\Big(
	\theta(\omega)\theta(\omega_3)+\theta(-\omega)\theta(-\omega_3)
	 \Big)  \notag \\
	& 
	+\int \frac{\dd\omega_1 \dd\omega_2}{(2\pi)^2} \mathcal{F}(\omega_1, \omega_2, \omega)I^{jk}(\omega_1) I^{ki}(\omega_2)  \frac{\omega \omega_1}{8\pi^2}\Big(
	\theta(\omega)\theta(\omega_1)+\theta(-\omega)\theta(-\omega_1)
	 \Big) \notag \\
	& 
	+\int \frac{\dd\omega_1 \dd\omega_3}{(2\pi)^2} \mathcal{F}(\omega_1, \omega, \omega_3)I^{jk}(\omega_3) I^{ki}(\omega_1)  \frac{\omega_1 \omega_3}{8\pi^2}\Big(
	\theta(\omega_1)\theta(\omega_3)+\theta(-\omega_1)\theta(-\omega_3)
	 \Big) \, .
\end{align}
Likewise, following similar manipulations we can decompose the conservative force as
\begin{align}
	F^{ij\, \rm cons (off)}_{{\rm loc}}(\omega) 
	& = \int \frac{\dd\omega_2 \dd\omega_3}{(2\pi)^2} \mathcal{F}(\omega, \omega_2, \omega_3)I^{jk}(\omega_2) I^{ki}(\omega_3)  \frac{\omega \omega_3}{16\pi^2}  \label{loconsoff}\\
	& 
	+\int \frac{\dd\omega_1 \dd\omega_2}{(2\pi)^2} \mathcal{F}(\omega_1, \omega_2, \omega)I^{jk}(\omega_1) I^{ki}(\omega_2)  \frac{\omega \omega_1}{16\pi^2} \notag \\
	& 
	+\int \frac{\dd\omega_1 \dd\omega_3}{(2\pi)^2} \mathcal{F}(\omega_1, \omega, \omega_3)I^{jk}(\omega_3) I^{ki}(\omega_1)  \frac{\omega_1 \omega_3}{16\pi^2}\, ,\nn \\
	 F^{ij\, \rm cons (off)}_{{\rm nloc}}(\omega) 	 
		& = -\int \frac{\dd\omega_2 \dd\omega_3}{(2\pi)^2} \mathcal{F}(\omega, \omega_2, \omega_3)I^{jk}(\omega_2) I^{ki}(\omega_3)  \frac{\omega \omega_3}{8\pi^2}\Big(
	\theta(\omega)\theta(\omega_3)+\theta(-\omega)\theta(-\omega_3)
	 \Big)  \notag \\
	& 
	-\int \frac{\dd\omega_1 \dd\omega_2}{(2\pi)^2} \mathcal{F}(\omega_1, \omega_2, \omega)I^{jk}(\omega_1) I^{ki}(\omega_2)  \frac{\omega \omega_1}{8\pi^2}\Big(
	\theta(\omega)\theta(\omega_1)+\theta(-\omega)\theta(-\omega_1)
	 \Big) \notag \\
	& 
	-\int \frac{\dd\omega_1 \dd\omega_3}{(2\pi)^2} \mathcal{F}(\omega_1, \omega, \omega_3)I^{jk}(\omega_3) I^{ki}(\omega_1)  \frac{\omega_1 \omega_3}{8\pi^2}\Big(
	\theta(\omega_1)\theta(\omega_3)+\theta(-\omega_1)\theta(-\omega_3)
	 \Big) \, .\\
	 &= - F^{ij\, \rm diss (off)}_{{\rm nloc}}(\omega)\label{nloconsoff}  \nn\,.
\end{align}
The sum of local conservative and dissipative terms then reproduces the total (in-in) force.\vskip 4pt 

From the above analysis we conclude that time nonlocality is therefore just a direct consequence of how the Feynman prescription reorganizes the support of the frequency integrals. These manipulations imply that, within an off-shell prescription, we may simply retain the local contributions from both dissipative and conservative parts without ever referring to the intermediate nonlocal~terms.  Although suggestive, the PM components of the (relative) conservative scattering angle under such a prescription depart from our previous derivation already at~${\cal O}(G^4)$. We will comment on this point in~\S\ref{outlook}. 	

\subsubsection*{On-shell}

The above procedure should be contrasted with our earlier decomposition of the effective action, where the {\it on-shell}~local conservative part is isolated only \emph{after} substituting the solution to the equations of motion as an expansion in $G$, see~\eqref{w3I}. The two key distinctions are: {\it i)}  the on-shell approach discards (consistently with PM derivations) certain terms (via \eqref{w2sign}) that are instead retained in \eqref{loconsoff}; and moreover {\it ii)} it allows us to naturally incorporate 2RR effects into the conservative sector. These differences imply that, to reproduce the on-shell impulse one must include contributions which, from the off-shell perspective, are encoded in a subset of the (a priori) nonlocal part of the force (although involving $I^{(2)ij}_{(0)}$ pieces).\vskip 4pt

An essential point of the on-shell prescription is that only the full Feynman action admits a well-defined variational principle. Once we substitute the solution to the equations of motion and subsequently split the result into local and nonlocal pieces, this decomposition no longer corresponds to a separation at the level of a well-defined action principle. Yet, owing to the local-in-time nature of the resulting in-in force, in \S\ref{isotropic} we reconstructed a local action that reproduces the local part of the (relative) impulse and associated scattering angle to the desired PN order [cf. \eqref{eq:chi4nu2const}-\eqref{eq:chi6nu2const}]. The dissipative counterpart is then unambiguously defined as the (local) remainder. Time locality is preserved throughout and the in-in dynamics is recovered by construction. We return to this point in~\S\ref{outlook}. 

\subsection{Symmetrizing \& (Re)routing}

As we mentioned earlier, the natural way to isolate the conservative sector of the in-in effective theory is to undo the Keldysh representation, expressed in terms of the $\bx_\pm$ variables, and return to the original $\bx_{1,2}$ basis. As discussed in \cite{Galley:2012hx,Kalin:2022hph}, rewriting the action as
\beq
S_{\rm inin} = \int \dd t\big[ L(\bx_{a,1},\bv_{a,1}) - L(\bx_{a,2},\bv_{a,2}) + K(\bx_{a,1},\bv_{a,1},\bx_{a,2},\bv_{a,2})\big]\,,
\eeq
leads to equations of motion that naturally decompose into conservative and dissipative sectors: the former derives from an ordinary Lagrangian, $L(\bx_a,\bv_a)$, while the latter is encoded in the nonconservative kernel $K$. As shown in \cite{Kalin:2022hph}, Feynman's prescription furnishes precisely such a decomposition. It is therefore not surprising that, when this procedure is implemented directly at the level of the final off-shell action, written in terms of retarded propagators, one recovers a conservative contribution that is fully equivalent to the one obtained from the off-shell Feynman prescription, in which the nonlocal pieces cancel among themselves. In~contrast, if one performs the symmetrization at the level of the decomposition in \eqref{retF}, while keeping $\Delta_F$ throughout, one is naturally led back to the same sequence of steps that gave rise to the PB prescription.\vskip 4pt 

There are in principle other ways to reorganize the full in-in computation. For instance, let us return to the memory contribution to the force in \eqref{eq:FullininExp}, which can be written as 
\begin{align}
	F^{ij}_{\rm (M)}(\omega) & = \int \frac{\dd\omega_2 \dd\omega_3}{(2\pi)^2} \mathcal{F}(\omega, \omega_2, \omega_3)I^{jk}(\omega_2) I^{ki}(\omega_3)  \frac{\omega\omega_3}{16\pi^2}\label{eq:FullininExp2} \\
	& 
	+\int \frac{\dd\omega_1 \dd\omega_2}{(2\pi)^2} \mathcal{F}(\omega_1, \omega_2, \omega)I^{jk}(\omega_1) I^{ki}(\omega_2) \frac{\omega\omega_1}{16\pi^2} \notag \\
	& 
	{\color{PineGreen}-\int \frac{\dd\omega_1 \dd\omega_3}{(2\pi)^2} \mathcal{F}(\omega_1, \omega, \omega_3)I^{jk}(\omega_3) I^{ki}(\omega_1)  \frac{\omega_1\omega_3}{16\pi^2}}\nn \, .
\end{align}
We have highlighted [in {\color{PineGreen} green}] the term that involves both retarded propagators. We immediately noticed that {\it rerouting} this term in the Feynman diagrams, namely replacing 
\beq \Delta_{\rm ret} (\omega_1)\Delta_{\rm ret}(\omega_3) \to \Delta_{\rm ret} (\omega_1)\Delta_{\rm ret}(-\omega_3) =\Delta_{\rm ret} (\omega_1)\Delta_{\rm adv}(\omega_3)\,,\label{relocalPB}
\eeq
 would flip the sign of the respective contribution, yielding a symmetrized total version (with three $\Delta_{\rm ret}\Delta_{\rm adv}$ combinations). The resulting force is therefore conservative by construction and furthermore it coincides with the value given in \eqref{loconsoff}. Unfortunately, similarly to the off-shell prescription, when applied across all regions (without taking into account Feynman's symmetries) the associated conservative sector is already modified at ${\cal O}(G^4)$. Nonetheless, when restricted to the memory-like region, the above rerouting produces the same structure as the local  version obtained through the PB prescription. This has important implications, in particular when combined with the tail-like contributions, as we discuss next. 
 
 \subsection{$\gamma\text{-}3$ prescription}

An alternative conservative-like construction that could in principle incorporate all the relevant features at 4PM and 5PM orders was introduced in~\cite{Driesse:2026qiz}, and dubbed the ``$\gamma\text{-}3$" prescription. This {\it region-dependent} framework lies somewhere in between an on-shell prescription and a conservative-like rerouting. In particular, all tail-like contributions are derived using~Feynman's $i0$'s, whereas memory-like effects are computed by choosing all Green's functions to be {\it incoming}  retarded propagators. In practice, the prescription becomes
\beq
\Delta_F(\omega_1)\Delta_F(\omega_3)\delta(\omega_1+\omega_2+\omega_3) \to \Delta_{\rm ret}(\omega_1)  \Delta_{\rm ret}(\omega_3)\delta(\omega_1+\omega_2+\omega_3)\,.
\eeq
For instance, the $\gamma\text{-}3$ prediction for the memory-like region of the action in \eqref{sconsM2} yields
\begin{align}
	 & S^{\rm cons (\gamma\text{-}3)}_{\text{(M-like)}}  = -\frac{16\pi^2G^2}{70} \int \frac{\dd \omega_1 \dd \omega_2\dd \omega_3}{(2\pi)^3}
\ddl(\omega_1+\omega_2+\omega_3) \Delta_{\rm ret}(\omega_1)  \Delta_{\rm ret}(\omega_3)  \\
	&\hspace{3cm}  
	\times I^{ij}_{\rm (N)}(\omega_1)I^{jk}_{\rm (N)}(\omega_3)I^{ki}_{\rm (N)}(\omega_2) \mathcal{F}(\omega_1, \omega_2, \omega_3)  \, .\nn
\end{align}
 It is straightforward to demonstrate that the resulting force is precisely the \emph{opposite} of that obtained from the rerouting in \eqref{relocalPB}. The latter, in turn, once restricted to the memory-like region, is the one equivalent to the PB prescription.\vskip 4pt
 It is instructive to track the origin of the sign mismatch by comparing the contributions from the tail- and memory-like regions of the conservative action \eqref{sconsM2}, rewritten as
 \begin{align}
	S^{\rm cons} & = -\frac{16\pi^2G^2}{70} \int \frac{\dd \omega_1 \dd \omega_2\dd \omega_3}{(2\pi)^3}
\ddl(\omega_1+\omega_2+\omega_3) \sg(\omega_1)\sg(\omega_3)  \\
	&\hspace{3cm}  
	\times \Delta_{\rm ret}(\omega_1)\Delta_{\rm ret}(\omega_3)I_{ij}(\omega_1)I_{jk}(\omega_3)I_{ki}(\omega_2)  \mathcal{F}(\omega_1, \omega_2, \omega_3) \notag  \, .
\end{align}
Because of the distributional identity in \eqref{w2sign}, the tail-like region becomes (see \eqref{Tlike})
 \begin{align}
	S^{\rm cons}_{\text{(T-like)}} &= -\frac{16\pi^2G^2}{70} \int \frac{\dd \omega_1 \dd\omega_2 \dd \omega_3}{(2\pi)^3}
\ddl(\omega_1+\omega_2+\omega_3) \sg(\omega_1)\sg(\omega_3) \\
&\qquad \qquad \times \Delta_{\rm ret}(\omega_1)\Delta_{\rm ret}(\omega_3) \mathcal{F}(\omega_1,\omega_2, \omega_3) I^{ik}_{(0)}(\omega_2) I^{ij}_{\rm (N)}(\omega_1)I^{jk}_{\rm (N)}(\omega_3) \nn \\
&= +\frac{16\pi^2G^2}{70} \int \frac{\dd \omega_1 \dd \omega_2\dd \omega_3}{(2\pi)^3}
\ddl(\omega_1+\omega_2+\omega_3) \Delta_{\rm ret}(\omega_1)\Delta_{\rm ret}(\omega_3)\nn  \\
	&\hspace{3cm}  
	\times I^{ij}_{(\rm N)}(\omega_1)I^{jk}_{(\rm N)}(\omega_3)I^{ki}_{(0)}(\omega_2)  \mathcal{F}(\omega_1, \omega_2, \omega_3)\nn\,,
\end{align}
where we used that $I_{(0)}^{ik}(\omega_2)$ is proportional to (up to two) derivatives of $\delta(\omega_2)$ [cf.~\eqref{i02d}].\vskip 4pt

Crucially, there is a sign flip due to a factor of $\sg(\omega_1)\sg(-\omega_1) = -1$. Since, by construction, the $\gamma\text{-}3$ prescription preserves the tail-like region, we find \begin{align}
\label{gm3cons}
&\hspace{2.5cm} S^{\rm cons}_{(\gamma\text{-}3)}= S^{\rm cons(\gamma\text{-}3)}_{(\rm T\text{-}like)} + S^{\rm cons(\gamma\text{-}3)}_{(\rm M
\text{-}like)} = \\ & \frac{16\pi^2G^2}{70} \int \frac{\dd \omega_1 \dd \omega_2\dd \omega_3}{(2\pi)^3}
\ddl(\omega_1+\omega_2+\omega_3) \Delta_{\rm ret}(\omega_1)\Delta_{\rm ret}(\omega_3)\nn  \\
	&\hspace{3cm}  
	\times I^{ij}_{(\rm N)}(\omega_1)I^{jk}_{(\rm N)}(\omega_3)\left[I^{ki}_{(0)}(\omega_2)-I_{(\rm N)}^{ki}(\omega_2) \right] \mathcal{F}(\omega_1, \omega_2, \omega_3)\,.\nn
\end{align}
Hence, adopting the $\gamma\text{-}3$ routing in all of the memory-like contributions at 5PN in our WEFT, while keeping the same potential-only and tail-like terms, we obtain 
\beq	\label{gm35pm} \tilde\chi_{j (\rm tot)(\gamma\text{-}3)}^{(5,\nu^2)\rm cons} = \bigg(
	\frac{9522061}{6048} - \frac{224057}{1440}\pi^2 + \frac{1408}{45}\log(2v_\infty) 
	\bigg) v_\infty^5  \; ,
	\eeq
	\beq
\label{gm36pm}\tilde\chi_{j (\rm tot)(\gamma\text{-}3)}^{(6,\nu^2)\rm cons}   =  \pi\bigg(
	\frac{47419583}{13440}
	- \frac{10812865}{32768}\pi^2
	{\color{blue}
	+\frac{2721}{2560}\pi^2
	}
	 + \frac{201}{2}\log\left(\frac{v_\infty}{2}\right) + \frac{2817}{16}\zeta(3) 
	\bigg) v_\infty^4  \;.
\eeq%
The alert reader will immediately notice that, upon translating between expansion parameters and conventions (see footnote \eqref{footconv}), the value of the scattering angle in \eqref{gm35pm} is in perfect agreement with the choice $c_M=1$ in \cite{Driesse:2026qiz}. This confirms that both the PN and PM computations are formally producing the same integrand in the overlapping regime of validity.\vskip 4pt In contrast, when choosing Feynman's propagators, the local PB prescription leads instead to the combination
\begin{align}
&\hspace{2.5cm} S^{\rm cons}_{\rm PBloc}= S^{\rm cons}_{(\rm T\text{-}like)} + S^{\rm cons}_{(\rm M
\text{-}like)loc} = \\ & \frac{16\pi^2G^2}{70} \int \frac{\dd \omega_1 \dd \omega_2\dd \omega_3}{(2\pi)^3}
\ddl(\omega_1+\omega_2+\omega_3) \Delta_{\rm ret}(\omega_1)\Delta_{\rm ret}(\omega_3)\nn  \\
	&\hspace{3cm}  
	\times I^{ij}_{(\rm N)}(\omega_1)I^{jk}_{(\rm N)}(\omega_3)\left[ I^{ki}_{(0)}(\omega_2) + I_{(\rm N)}^{ki}(\omega_2)\right] \mathcal{F}(\omega_1, \omega_2, \omega_3)\,,\nn
\end{align}
which adopts the same conservative routing for both tail and memory effects. This ultimately yields the values for the scattering angle reported in~\eqref{eq:chi4nu2const}-\eqref{eq:chi6nu2const}, and the sign mismatch for the memory-like contributions starting at ${\cal O}(G^5)$. 

\section{Summary \& Outlook} \label{outlook}

In this paper we have completed the knowledge of the total even-in-velocity relative impulse and scattering angle at 5PN order [cf.~\eqref{chi4tot1}--\eqref{chi6tot1}]. We have shown how this information, combined with the time delay and losses of energy and angular momentum, allows us to introduce an isotropic-like description that uniquely determines the coefficients of the full gravitational forces without  ambiguities. Hence, modulo the identification of nonlocal-in-time tail-type effects [cf.~\eqref{tails}] \cite{Bini:2020hmy,Dlapa:2024cje,Bini:2024tft,Dlapa:2025biy}, this framework provides a direct link to the complete gravitational dynamics through a matching procedure involving only observable quantities defined at infinity---very much in the spirit of the B2B correspondence \cite{Kalin:2019rwq,Kalin:2019inp,Cho:2021arx}.\vskip 4pt  For the particular case of failed-tail, memory, and second-order radiation-reaction effects, the isotropic representation allows us to decompose the associated forces naturally into dissipative [cf.~\eqref{eq:BTisoLO}] and {\it conservative-like} interactions [cf.~\eqref{aftiso}--\eqref{a2rriso}], although the latter do not derive from an autonomous Hamiltonian. Nevertheless, we have found that the extended sector may be characterized by conserved {\it energy} and (mechanical) {\it angular-momentum} charges [cf.~\eqref{Einin}-\eqref{Linin}] 
\begin{align}
\label{Einin2}
 \delta \hat E_{\rm (iso)}
&=
-\frac{13\nu^2}{378}\,\left(\frac{GM}{r}\right)^6
+\frac{6389\nu^2}{1260}\,\left(\frac{GM}{r}\right)^5\,\bv^2
+\frac{33809\nu^2}{2100}\left(\frac{GM}{r}\right)^4\,\bv^4 \,,\\
\delta \hat \bL^i_{(\rm iso)} & =
	\varepsilon^{ijk}	\br^j \bv^k\bigg[
	\frac{1669}{350}\frac{G^5 M^5 \nu^2}{r^5} 
	+\frac{3037}{105}\frac{G^4 M^4 \nu^2}{r^4} \bv^2
	\bigg] \;.
\end{align}
Combined with the associated nonlinear acceleration (such that, e.g., $-\tfrac{d}{dt}\delta \hat E=\bv\cdot \delta \ba$ using Newton's equations), 
\begin{align}
\label{eq:baiso}
\delta \ba_{(\rm iso)}
&=
\frac{6259}{630}\frac{G^6M^6\nu^2}{r^7}\,\bn
+
\frac{G^5M^5\nu^2}{r^6}
\left[
\frac{50783}{6300}\,\bv^2\bn
+
\frac{3431}{42}\, (\bv\cdot\bn)\bv
\right] \nn\\
&+
\frac{G^4M^4\nu^2}{r^5}
\left[
-\frac{8977}{175}\,\bv^4\bn
+
\frac{12148}{105}\bv^2(\bv\cdot \bn)\bv
\right],
\end{align}
this provides a starting point for constructing an alternative effective description which may naturally incorporate an enlarged conservative sector that is not necessarily described in terms of canonical variables. For instance, in the particular case of circular orbits, it would be straightforward to add \eqref{Einin2} into the full $E(\Omega)$, and use a balance-type law, $\dot E =-{\cal F}$, to provide a more accurate description of the evolution of the orbital frequency.~This highlights an important feature of nonlinear gravitational interactions, namely the emergence of {\it Schott-like} terms that do not vanish upon orbital averaging, which can ultimately have a significant impact on the evolution of the system.\vskip 4pt 

Following the analysis in \cite{Porto:2024cwd}, we used the Feynman $i0^+$ prescription to isolate a conservative sector that can still be described in terms of a canonical structure. This provides a framework in which part of the dynamics may be organized in terms of standard phase-space variables, such as action-angle variables. Although such a split may ultimately retain a degree of arbitrariness, the construction of an effective framework built around a symplectic structure, such as the EOB approach, can nevertheless provide a systematic way to incorporate the impact of additional contributions, including nonperturbative corrections.\vskip 4pt

The Hamiltonian sector contains tail-like and memory-like components, the latter depending on a double Principal-Value integral. The induced nonlocality is a byproduct of imposing a time-symmetric prescription designed to isolate the conservative contribution independently of each integration region, while consistently accounting for the correct frequency scaling of all the terms entering the convolution of Green's functions generated by nonlinear gravitational interactions. Building on Feynman's boundary conditions, yet exploiting the locality of the full in-in effective theory, we have formulated a PB-inspired prescription that cleanly separates local-in-time conservative and dissipative sectors while preserving the full dynamics. Concretely, this is accomplished by employing the B2B dictionary to reconstruct an isotropic local Hamiltonian [cf.~\eqref{hatHloc}] from the on-shell scattering angle [cf.~\eqref{eq:chi4nu2const}--\eqref{eq:chi6nu2const}], 
\beq
\delta \hat H_{\rm (iso)}=  -\frac{\nu^2}{50}\left(\frac{2973}{7}\, \hat\bp^4 \left(\frac{GM}{r}\right)^4 + \frac{7931}{9} \, \hat \bp^2\left(\frac{GM}{r}\right)^5-\frac{5758}{63}\left(\frac{GM}{r}\right)^6 \right)\,,\label{Hiso2}
\eeq
to be added to the potential and (standard) tail contributions \cite{Galley:2015kus,Blumlein:2020pyo,Almeida:2021xwn,Blumlein:2021txe,Almeida:2023yia} (or alternatively the memory-like components [cf. \eqref{hatHloc} in {\color{cyan}cyan}] may be added to the tail-like Hamiltonians in~\cite{Khalil:2022ylj,Dlapa:2024cje,Bini:2024tft,Dlapa:2025biy}). The associated dissipative sector is then uniquely determined as the local remainder, upon subtracting from~\eqref{eq:baiso} the Hamiltonian equations of motion. \vskip 4pt 
 
A~central aspect of our construction is that the variational principle is controlled by the full Feynman action. Accordingly, after imposing the solution to the equations of motion, the ensuing split into local and nonlocal terms should be regarded as an organizational device, not as evidence that each term separately defines an autonomous Hamiltonian description. Nevertheless, because the associated in-in forces remain local in time, one can reconstruct an action that reproduces the local contribution to the relative impulse and corresponding scattering angle to the required PN order. In~practice, these manipulations amount to adding and subtracting compensating pieces so as to enforce the existence of a local Hamiltonian description that reproduces the associated value of the scattering angle.\footnote{The procedure is reminiscent of the \textit{flexibility} function introduced in the Tutti-Frutti approach~\cite{Bini:2020wpo}.}  One of the virtues of such approach is that the associated Hamiltonian (or impetus formula) is tailored to reproduce on-shell scattering observables, and therefore it can be subsequently extrapolated to all-orders-in-velocity expressions using PM input. Moreover, this construction automatically preserves the mass scaling inherited from the scattering impulse, which ensures the absence of an ${\cal O}(G^{4}\nu^{2})$ contribution to the scattering angle.\vskip 4pt The on-shell procedure thus provides a systematic characterization of a universal contribution from failed-tail, memory and second-order radiation-reaction effects to the two-body Hamiltonian [cf.~\eqref{Hiso2}], while preserving time locality without sacrificing the full dynamics. As~we~have shown, this also fixes the (rational) EOB coefficients~[cf.~\eqref{eob1}-\eqref{eob2}] in a manner that is consistent with the Tutti-Frutti formalism and $\pi^2$-conjecture in~\cite{Bini:2025vuk}, yielding
\beq
\bar d_5^{\rm loc} =  \left(\frac{331054}{175}-\frac{63707}{512}\pi^2\right)\nu + \left(-\frac{403259}{63}+ \frac{306545}{512}\pi^2\right) \nu^2 +\left(\frac{1069}{3}-\frac{205}{16}\pi^2\right)\nu^3 \,,
\eeq
\beq
a_6^{\rm loc} = \left(-\frac{1026301}{1575}+\frac{246367}{3072}\pi^2\right)\nu + \left(-\frac{1718599}{1575}
+\frac{25911}{256}\pi^2\right)\nu^2 + 4\nu^3\,.
\eeq

For completeness, we have examined alternative prescriptions that define a local conservative sector. For instance, exploiting explicit cancellations, we have shown that an off-shell Feynman prescription may be implemented by effectively disregarding the Principal-Value contributions. This prescription gives rise to a well-defined action principle for the relative dynamics. However, it is straightforward to show that it leads to a non-vanishing contribution to $\tilde \chi_{j(\rm tot)}^{(4,\nu^2)\rm cons}$. This, in turn, implies that---although the relative dynamics is described by a Hamiltonian---a violation of the mass-scaling rule at ${\cal O}(G^4)$ would necessarily entail non-vanishing recoil effects (from the point of view of the incoming center-of-mass frame). While this does not by itself lead to an immediate inconsistency, such contributions would need to be incorporated in a self-consistent description of the dynamics.\vskip 4pt 

We have also analysed the $\gamma\text{-}3$ prescription introduced in the recent scattering computation at ${\cal O}(G^5\nu^2)$ in \cite{Driesse:2026qiz}. We find that, upon implementing the same routing of Green's functions within our WEFT formalism, we reproduce exact agreement with the results reported in \cite{Driesse:2026qiz} over the common domain of validity [cf.~\eqref{gm35pm}]. This provides a nontrivial check of the formal equivalence of the integrands in the PN and PM approaches. However, implementing instead the PB prescription yields a local memory-like contribution with the exact {\it opposite} sign [cf.~\eqref{eq:chi5nu2const}]. A relative sign, to some extent, might be expected: the $\gamma\text{-}3$ prescription was specifically devised to generate a memory-like correction that cancels a divergent contribution from the other regions in the $\gamma \to 3$ limit \cite{Driesse:2026qiz}. By contrast, the PB prescription preserves, by construction, the same routing in both the tail-like and memory-like regions. It is precisely this sign mismatch that manifests itself not only in the discrepancy at 5PM order, but also in the absence of a cancellation among the $\pi^2$ contributions from tail and memory terms at ${\cal O}(G^6\nu^2)$[cf.~\eqref{gm36pm}]. For instance, extended throughout the 5PN regime, the $\gamma\text{-}3$ prescription applied to our WEFT leads to the EOB values,
\bea
\label{eob1n}\bar d_{5{\rm loc}}^{\nu^2(\gamma\text{-}3)} &=&-\frac{42915}{7} + \frac{306545}{512}\pi^2\,, \\ \label{eob2n} a_{6{\rm loc}}^{\nu^2(\gamma\text{-}3)}&=&-\frac{742757}{700}+\frac{25911}{256}\pi^2 {\color{blue}-\frac{907}{400}\pi^2},\eea 
whereas the PB prescription preserves the conjectured $\pi^2$ structure.\footnote{We should stress, however, that the $\pi^2$ coefficient of the total even-in-velocity scattering angle is not uniquely fixed by the potential region (see~\eqref{chi62dis}). Although rather unnatural (and somewhat arbitrary), one could in principle add and subtract appropriate $\pi^2$ terms on both conservative and dissipative sides of the memory-like contribution in order to enforce agreement with the $\gamma\text{-}3$ prescription.}\vskip 4pt

Given that the same routing is obtained for both tail and memory terms, the discrepancy discussed above suggests that Feynman's boundary conditions---either in their fully nonlocal form or in the local version---most likely do not generate a memory-like contribution that cancels the $\gamma=3$ divergence arising from the other regions. In this sense, the issue appears not to be merely one of assigning an alternative value to the coefficient $c_{\rm M}$ introduced in~\cite{Driesse:2026qiz}, but rather of identifying a more fundamental difference in the structure of the relevant contributions. In fact, we find that the PB prescription applied to Feynman's boundary conditions would imply the boundary values
\beq
I_{1\rm (PB)}^{\rm M} = -\frac{1}{15(8\pi)^4\epsilon}+{\cal O}(\epsilon^0)\,, \quad I_{2\rm  (PB)}^{\rm M} = +\frac{5}{6(8\pi)^8\epsilon^2}+{\cal O}(\epsilon^{-1})\,,\label{Fi1i2}
\eeq 
with both integrals changing their overall signs. Pending a more complete understanding of Feynman's calculation---including, in particular, the proper treatment of infrared divergences in the scattering problem---the above values, while still retaining the $\gamma=3$ singular behavior of the conservative result, would nevertheless be compatible with a finite limit as $\epsilon \to 0$.\vskip 4pt In light of these findings, one may wonder whether the cancelation of $\gamma=3$ singularities should be regarded as a guiding principle for defining the conservative sector.
 The expectation is that, indeed, once conservative and dissipative terms are combined, the full result remains finite in the $\gamma \to 3$ limit. However, even in such a case, enforcing independent cancellations may not provide a compelling rationale for the conservative part. An indication can be seen from the fact that the $\gamma\text{-}3$ prescription reverses the sign of the $\Delta_{\rm ret}\Delta_{\rm adv}$ term in the memory force [cf.~\eqref{eq:FullininExp}], allegedly to remove the $\gamma=3$ singularity. However, since the complete result must reproduce the original routing, the sign would then have to be recovered through the associated dissipative terms. This could, in principle, restore the divergence back into the full answer. On the other hand, it is possible that the cancelation occurs against the $\Delta_{\rm ret}\Delta_{\rm ret}$ term instead, while the  $\Delta_{\rm ret}\Delta_{\rm adv}$ pieces remain finite. However, this term is unmodified by the $\gamma\text{-}3$ prescription and, in turn, is the one reversed by Feynman. In such a scenario, the cancelation observed in the $\gamma\text{-}3$ routing would then be nothing but a direct manifestation of the cancelation displayed by the full in-in result, rather than a vehicle for defining a natural conservative part.\footnote{In this respect, it would also be instructive to investigate the behavior of the conservative sector implied by the $\gamma\text{-}3$ prescription in the $\gamma \to \infty$ limit.} \vskip 4pt  We may also argue that, unlike in the PN framework---where the role of Feynman boundary conditions (and associated origin of $\pi^2$ terms) emerges in a comparatively natural and internally consistent manner---the extrapolation to PN of a prescription solely based on the removal of singular behavior at $\gamma=3$, that moreover reverses the sign of the force depending on the region [cf. \eqref{gm3cons}], is far less compelling. That is especially the case since this point corresponds to a value of the relative velocity, $v=\tfrac{2\sqrt{2}}{3}$, far exceeding the regime in which PN theory is expected to be reliable.\vskip 4pt There is, of course, the alternative possibility that $\gamma\text{-}3$ poles persist. While somewhat puzzling---given that it is not tied to a kinematical configuration---the singularity at $\gamma=3$ may nevertheless still be an artifact of a truncation of an expansion in Newton's constant.
In this respect, an independent computation of the scattering angle within the gravitational self-force programme \cite{Bini:2024icd,Barack:2026izc} (performed with both time-symmetric and retarded boundary conditions) would be especially valuable in order to shed light on these issues.\vskip 4pt 
 
These considerations underscore the importance of determining the entire ${\cal O}(G^5)$ scattering dynamics at second order in the self-force expansion. The full (even-in-velocity) PN results derived in this paper provide a key step in that direction---while opening the door to a complete characterization of the underlying two-body dynamics via scattering observables. %

\vskip 12pt
{\bf Acknowledgements}
\vskip 4pt 
We thank Christoph Dlapa and Gregor K\"alin for useful discussions, and Donato Bini for his help adapting results from the linear-response formalism. We~are grateful to the authors of \cite{Driesse:2026qiz} for discussions on the $\gamma\text{-}3$ prescription and cancelation of $\epsilon$-poles. The work presented here was partially financed by the Deutsche Forschungsgemeinschaft (DFG) under Germany's Excellence Strategy (EXC 2121) ``Quantum Universe" (390833306). MMR research is supported by the
U.S. Department of Energy (award no. DE-SC0011941).

\appendix

\section{Trajectories}
\label{app:integration}

As we mentioned in the main text, the derivation of the impulse does not necessitate an explicit analytic expression for the trajectory, which instead can be kept in integrated form. For instance, let us consider the computation of $\Delta^{(6)} \bp_{\rm FT}$, which includes a contribution of the form
\begin{align}
	\Delta^{(6)} \bp_{\rm FT} & \supset m\nu \int \dd t\, \ba_{\rm (N)}\Big|_{\delta^{(3)}_{\rm FT}} = -G^6 m^2 \nu\int \dd t\bigg[\frac{\delta\ord{3} \br_{\rm FT}}{r_0^3} -3 \frac{ (\br_0 \cdot \delta\ord{3} \br_{\rm FT}) \br_0}{r_0^5}\bigg] \notag \\
	&  = -G^6 m^2 \nu \bigg[\int \dd t \frac{1}{r_0^3}\int_{-\infty}^t \dd t^\prime \,   \delta\ord{3} \bv_{\rm FT}
	-3\int \dd t \frac{\br_0}{r_0^5}\int_{-\infty}^t \dd t^\prime \, \br_0\cdot  \delta\ord{3} \bv_{\rm FT}\,,
	\bigg] \, .
	\label{eq:ImpDeltaInt}
\end{align}
with $\delta\ord{3} \br_{\rm FT}$ entering at ${\cal O}(G^5)$. Using the following identities,
\begin{align}
	\frac{1}{r_0^3(t)} & = \frac{\dd}{\dd t}\bigg( \frac{t}{b^2 r_0(t)}\bigg) \,, \\
	\frac{\br^i_0(t) \br_0^k(t)}{r_0^5(t)} & = \frac{\dd}{\dd t}\bigg(\frac{t(3b^2+2v_\infty^2 t^2)}{3b^4 r_0^3(t)}\bb^i \bb^k -\frac{2\bb^{(i}\bv_\infty^{k)}}{3 v_\infty^2 r_0^3(t)} + \frac{t^3}{3b^2 r_0^3(t)} \bv_\infty^{i}\bv_\infty^{k}\bigg) \,,
\end{align}
we can substitute these expressions in~\eqref{eq:ImpDeltaInt} and, upon integrating by parts, we find
\begin{align}
	\Delta \bp_{\rm FT} &\supset -G^6 m^2 \nu \bigg[
	\Delta \bp_{\rm Boundary} + \Delta \bp_{\rm extra}
	\bigg] \, ,
\end{align}
where
\begin{align}
	\Delta p^i_{\rm Boundary} & = \bigg[ 
	\frac{t}{b^2 r_0(t)}\int_{-\infty}^t \dd t^\prime \,   \delta\ord{3} \bv^i_{\rm FT} \notag \\
	& 
	-3\bigg(\frac{t(3b^2+2v_\infty^2 t^2)}{3b^4 r_0^3(t)}\bb^i \bb^k -\frac{2b^{(i}\bv_\infty^{k)}}{3 v_\infty^2 r_0^3(t)} + \frac{t^3}{3b^2 r_0^3(t)} \bv_\infty^{i}\bv_\infty^{k}\bigg)\int_{-\infty}^t \dd t^\prime \,   \delta\ord{3} v^k_{\rm FT}
	\bigg]_{-\infty}^{+\infty} \notag \\
	& = \frac{1}{b^2 v_\infty} \int_{-\infty}^{+\infty} \dd t \,   \delta\ord{3} \bv^i_{\rm FT}
	-\frac{1}{b^2 v_\infty}\bigg(\frac{\bv_\infty^i \bv_\infty^k}{v_\infty^2}+\frac{2\bb^i \bb^k}{b^2}\bigg)\int_{-\infty}^{+\infty} \dd t \,   \delta\ord{3} \bv^k_{\rm FT} \notag \\
	& = -\frac{1}{b^2 v_\infty}\frac{\bb^i}{b^2} \, \bb\cdot \int_{-\infty}^{+\infty} \dd t \,   \delta\ord{3} \bv_{\rm FT} \, , \label{eq:pBoundaryInt}\\
	\Delta p^i_{\rm extra} & = 
	\int_{-\infty}^{+\infty} \dd t \bigg[
	\, \frac{t}{b^2 r_0(t)}   \delta\ord{3} \bv^i_{\rm FT} \notag \\
	& 
	\qquad\qquad -\bigg(\frac{t(3b^2+2v_\infty^2 t^2)}{b^4 r_0^3(t)}\bb^i \bb^k -\frac{2\bb^{(i}\bv_\infty^{k)}}{v_\infty^2 r_0^3(t)} + \frac{t^3}{b^2 r_0^3(t)} \bv_\infty^{i}\bv_\infty^{k}\bigg)    \delta\ord{3} \bv^k_{\rm FT}
	\bigg] \, .
\end{align}
where in the last step of \eqref{eq:pBoundaryInt} we decomposed the trajectory into the impact parameter and velocity directions, \beq \delta\ord{3} \bv_{\rm FT} = (\bb\cdot \delta\ord{3} \bv_{\rm FT}) \bb/b^2 + (\bv_\infty\cdot \delta\ord{3} \bv_{\rm FT}) \bv_\infty/v_\infty^2\,.\eeq At this point we notice that, since $\delta\ord{3} \bv_{\rm FT}$ can be obtained analytically, the integrals can be explicitly performed. Moreover, the integral in \eqref{eq:pBoundaryInt} is proportional to the impulse in the $\bb$-direction at ${\cal O}(G^5)$, which is already computed. Similar steps can be carried out for all other computations involving high-order deflections.
\section{Poincar\'{e}--Bertrand}
\label{app:PBFourier}

The PB theorem states \cite{Davies:1996gee, king2009hilbert}
\begin{equation}
  \int \dd x \frac{\PV}{u-x} \int \dd y \frac{\PV}{x-y} f(x,y)
  = \int \dd y \int \dd x \frac{\PV}{(u-x)(x-y)} f(x,y) - \pi^2 f(u, u) \, .\label{pbth}
\end{equation}
Let us introduce the notation
\begin{align}
  \text{LHS}(u) & \equiv \int \dd x \frac{\PV}{u-x} \int \dd y \frac{\PV}{x-y} f(x,y) \, ,\\
  \text{RHS}_1(u) & \equiv \int \dd y \int \dd x \frac{\PV}{(u-x)(x-y)} f(x,y) \, ,\quad
  \text{RHS}_2(u)  \equiv -\pi^2 f(u, u) \,,
\end{align}
such that LHS = RHS$_1 +$ RHS$_2$. Our goal is to illustrate how the relation in \eqref{pbth}, combined with the distributional identity  
\begin{equation}
	\int \dd x \frac{\PV}{x-a} e^{i \omega x} 
	 = i \pi \sg(\omega)e^{i\omega a} \, ,
	\label{eq:PVfinal}
\end{equation}
becomes a simple condition in Fourier space.\vskip 4pt  

Let us start with the LHS, which using \eqref{eq:PVfinal} can be written as
\begin{align}
  \mathscr{F}[\text{LHS}(u)](\omega) 
  & =
  \int \frac{\dd \omega_1 \dd \omega_2}{(2\pi)^2}f(\omega_1, \omega_2)\,
  \ddl(\omega - \omega_1 - \omega_2)
  \bigg[ - \pi^2 \sg(\omega_2) \sg( \omega_1 +\omega_2) \bigg] \, .
\end{align}
Let us move onto the RHS. The Fourier transform of the last term, RHS$_2$, is trivial.  For RHS$_1$ we use the following identity (see Eq.~(2.38) of \cite{Davies:1996gee})
\begin{equation}
  \frac{\PV}{(x-u)(y-x)} = \frac{\PV}{y-u} \bigg[\frac{\PV}{x-u} - \frac{\PV}{x-y} \bigg] \, ,
\end{equation}
such that
\begin{equation}
  \text{RHS}_1(u) =
  \int \dd y \frac{\PV}{y-u}
  \bigg[\int \dd x \frac{\PV}{x-u} - \int \dd x \frac{\PV}{x-y} \bigg] f(x, y) \, ,
\end{equation}
and therefore
\begin{align}
  \mathscr{F}[\text{RHS}_1(u)](\omega) 
  =
  \int & \frac{\dd \omega_1 \dd \omega_2}{(2\pi)^2}f(\omega_1, \omega_2)\,
  \ddl(\omega - \omega_1 - \omega_2)  \notag \\
  & \times\bigg[ - \pi^2 \sg(\omega_1) \bigg( \sg(\omega_2) -  \sg(\omega_1 + \omega_2) \bigg)\bigg] \, .
\end{align}
The relation in \eqref{pbth} then becomes, 
after performing the integral over $\omega_2$, %
\begin{align}
  \int \frac{\dd \omega_1}{2\pi} f(\omega_1, \omega -\omega_1)\,
  \bigg[ \sg(\omega) \sg( \omega -\omega_1) \bigg]
  =
  & \int \frac{\dd \omega_1}{2\pi} f(\omega_1, \omega -\omega_1)\notag \\
  & \times
  \bigg[ \sg(\omega_1) \bigg( \sg(\omega-\omega_1) -  \sg(\omega) \bigg)  + 1\bigg] \,.
\end{align}
It is now straightforward to show that the PB theorem in Fourier space is therefore a consequence the simple relation
\begin{equation}
  \sg(\omega) \sg( \omega -\omega_1) - \sg(\omega_1) \bigg( \sg(\omega-\omega_1) -  \sg(\omega) \bigg) = 1 \,,
  \label{eq:PBF}
\end{equation}
which can be explicitly checked on a case-by-case basis. It is instructive to rewrite the  above relation as follows,
\begin{align}
  \sg(\omega_1) \sg(\omega) - \sg(\omega_1)  \sg(\omega-\omega_1)
  =
  2\theta(\omega)\theta(\omega_1-\omega) + 2\theta(-\omega)\theta(\omega-\omega_1) \,,
\end{align}
such that the PB theorem turns into the relation,%
\begin{equation}
  \sg(\omega) \sg( \omega_1)
  = 1 -2\theta(\omega)\theta(-\omega_1) - 2\theta(-\omega)\theta(\omega_1) \,.
  \label{eq:PBf22}
\end{equation}
This condition can then be systematically implemented to extract a local-support contribution from the convolution of Principal-Value integrals, as explained in the main text.  

\section{Time delay} \label{app:delay}

\subsection*{General structure}
The solution for the trajectory may be obtained by integrating the equations of motion, yielding
\begin{equation}
	\label{delay1} \br(t) = \bb + \bv_\infty t +
	\int_{-\infty}^t \dd t' (t-t') \ba(t') \; ,
\end{equation}
with $\bb\cdot \bv_\infty = 0$.  From here we can define a scattering {\it time delay} ($\tau$) as follows. We start by considering a large sphere, of radius $R$, and define the time endpoints as the value of the trajectory at which $r(T_\pm) = R$. For instance, for a free theory we have 
\begin{equation}
	\tau_{\rm free} = T_+ - T_- = \frac{2\sqrt{R^2-b^2}}{v_\infty}\simeq \frac{2R}{v_\infty}\, \qquad  (R \gg b) \; .
\end{equation}
For an interacting theory, and provided $\ba$ vanishes at infinity sufficiently fast (see below),  the system may be taken initially on free motion, such that 
\begin{equation}
	T_- = -\frac{\sqrt{R^2-b^2}}{v_\infty}\simeq -\frac{R}{v_\infty} \; .
\end{equation}
 The final trajectory, at $T_+$, can also be taken as straight motion, such that
\begin{equation}
	\br_+(t) \simeq \bb_+ + \bv_+ t \; ,
\end{equation}
where explicitly
\begin{equation}
	\bb_+ = \bb -\int_{-\infty}^{+\infty} \dd t\, t \, \ba(t) \; , \qquad
	\bv_+ = \bv_\infty +\Delta \bv \; , \qquad 
	\Delta \bv\equiv \int_{-\infty}^{+\infty} \dd t \,  \ba(t) \; .
\end{equation}
and the value of $T_+$ can be found by solving:
\begin{equation}
	b_+^2 + v_+^2 T_+^2  + 2\bb_+\cdot \bv_+ T_+ = R^2 \; ,
\end{equation}
yielding 
\begin{equation}
	T_+ = -\frac{\bv_+\cdot \bb_+}{v_+^2}+\frac{1}{v_+}\sqrt{R^2-b_+^2+ \bigg(\frac{\bb_+\cdot \bv_+}{v_+^2}\bigg)^2}\simeq\frac{R}{v_+}-\frac{\bv_+\cdot \bb_+}{v_+^2}  \; .
\end{equation}
From here we find %
\begin{equation}
	\tau_{\rm int} = T_+ -T_-
	= \frac{R}{v_+}+\frac{R}{v_\infty} -\frac{\bv_+\cdot \bb_+}{v_+^2} \; .
\end{equation}
After subtracting the free part, we can then introduce the {\it time delay} for the interacting theory as
\begin{equation}
	\tau_{\rm int} = -\frac{\bv_+\cdot \bb_+}{v_+^2} =-\frac{\bb \cdot \Delta \bv}{v_+^2} + \int \dd t\frac{t (\bv_+\cdot \ba(t))}{v_+^2}  \; .
	\label{eq:Deltatdef}
\end{equation}
Notice that, for conservative scattering we have $|\bv_+| = |\bv_\infty|$, such that
\begin{equation}
	\tau^{\rm cons}_{\rm int} 
	= -\frac{\bv_+\cdot \bb_+}{v_\infty^2} = -\bb\cdot \frac{\Delta \bv}{v_\infty^2} +  \bv_\infty \cdot \int \dd t \, \frac{t \, \ba(t)}{v_\infty^2} +{\cal O}(\ba^2) \; .
\end{equation}

Since the total change in the velocity, $\Delta\bv$, is a gauge-invariant observable related to the impulse, the only new information encoded in \eqref{eq:Deltatdef} is given by the second term, which we will therefore retain as the quantity that will allow us to match the conservative-like forces between generic and isotropic gauges.\vskip 4pt  
It is instructive to notice that the above definition in \eqref{eq:Deltatdef} is also consistent with the traditional choice, \beq \Delta \label{delt2} T^{\rm cons}_{\rm int} = 2\int^{\infty}_{r_{\rm min}} \frac{\dd r}{\dot r(r,\hat E,\hat J)}-2\int_{r_{\rm min}}^{\infty} \frac{\dd r}{\dot r_{0}(r,\hat E,\hat J)}\,,\eeq at leading order in the acceleration. Here $r_{\rm min}$ is given by 
\begin{equation}
	r(t_{\rm p})=r_{\min} \;,\qquad \dot r(t_{\rm p})=0 \; ,
\end{equation}
and 
\beq
\dot r(r,\hat E,\hat J) \equiv \sqrt{2(\hat E-\hat V(r))-\hat J^2/ r^2}
\eeq
with $\hat E,\hat J$ the (normalized) energy and angular momentum, $\hat V(r)$ the associated potential, and $r_0$ the free motion (with $\hat V(r)=0$). The connection can be shown directly upon expanding the radial motion at leading order,  such that
\beq
\tau^{\rm cons}_{\rm int} \simeq  - 2\int_{r_{\rm min}}^{\infty}  \frac{\delta \dot r}{\dot r_0^2} dr_0 + {\cal O}(\ba^2)\,.
\eeq
Hence, using $dr_0 = \dot r_0 \dd t$ together with
\beq
\begin{aligned}
\frac{\delta \dot r}{\dot r_0} =  \frac{\hat \br_0}{\dot r_0} \cdot \delta \bv + \frac{\dot{\hat \br}_0}{\dot r_0}\cdot \delta \br &= \frac{\bv_\infty\cdot\delta \bv}{v_\infty^2}  +  \frac{\bb\cdot \delta \bv}{t \, v_\infty^2} + \frac{\bv_\infty\cdot \delta \br}{t \, v_\infty^2} - \frac{\br_0\cdot \delta \br}{r_0^2} + {\cal O}(\ba^2)
\end{aligned}
\eeq
combined with \eqref{delay1},  
\beq
\tau^{\rm cons}_{\rm int}  \simeq  -\bb\cdot \frac{\Delta \bv}{v_\infty^2} +  \lim_{\Lambda \to \infty} \left(\bv_\infty \cdot \int_{-\infty}^{\Lambda} \dd t \, \frac{t \, \ba(t)}{v_\infty^2} +{\cal O}(\ba^2) - \Lambda\frac{\bv_\infty\cdot \Delta \bv}{v_\infty^2}\right)  + {\cal O}(\ba^2)\,.
\eeq
where $\Lambda$ is a time cutoff. At first glance, the reader may be puzzled by the emergence of a divergent contribution. Its origin, however, is straightforward. Although the time delay introduced in \eqref{delt2} is itself perfectly well defined, expanding it at linear order generates spurious divergences that are artifacts of the approximation and are removed once higher-order terms are consistently included. Indeed, for a conservative scattering process the asymptotic speed is preserved, $|\bv_\infty+\Delta \bv| = |\bv_\infty|$,
which immediately implies
\beq
\bv_\infty\!\cdot\!\Delta \bv = -\tfrac{1}{2} |\Delta \bv|^2 \sim {\cal O}(\ba^2)\, .
\eeq
Hence the seemingly linear contribution is in fact quadratic in the deflection. Assuming---as expected---that the higher-order terms supply the compensating pieces which cancel this spurious divergence, the remaining finite part is well defined and agrees with \eqref{eq:Deltatdef}.

\vskip 4pt

For the sake of notation, in what follows (as well as in the main text) we drop the `cons' and `int' labels on $\tau$.

\subsection*{Renormalized contribution}

In general, for a Newtonian potential scaling as $V_{(\rm N)} \simeq 1/r$, it is well known that the force does not decouple at infinity. In our calculations, notably for scattering events, this is remediated by dim. reg., such that $V^{(d)}_{(\rm N)}\simeq 1/r^{d-2}$. Nonetheless, for the computation of the time delay, the Newtonian force induces a logarithmically divergent contribution,
\begin{equation}
	\tau^{\rm (N)}  
	= -\frac{G M}{v_\infty^2}\int \dd t\frac{t (\bv_\infty\cdot \bn(t))}{r^2(t)}
	\label{eq:DeltatN}   \; ,
\end{equation}
which at leading PM order becomes (in $d=3$)
\begin{equation}
	\tau^{\rm (N)} 
	= -G M\int \dd t \frac{ t^2}{(b^2+v_\infty^2 t^2)^{3/2}}   \; ,
\end{equation}
leading to a $\int \dd t/t$ divergence. To deal with this feature we will introduce a renormalized time delay in which we subtract the divergent piece from the computation. The main culprit is the contribution from the Newtonian acceleration evaluated on the deflected trajectory, namely
\begin{equation}
	\tau^{\rm (N)}_{{\rm (X)}} 
	= \int \dd t\frac{t (\bv_\infty\cdot \ba_{\rm (N)}(t))}{v_\infty^2} \bigg|_{\delta_{\rm(X)}\br(t)}  \; .
\end{equation}
The divergent part comes from the $t \to +\infty$ limit of integration, for which we have the scaling 
\begin{equation}
	\frac{t (\bv_\infty\cdot \ba_{\rm (N)}(t))}{v_\infty^2}\bigg|_{\delta_{\rm(X)}\br(t)} = -\frac{G M}{v_\infty^2}\frac{\bv_\infty \cdot \bv_+ }{v_\infty^3 }\frac{1}{t} \; .
\end{equation}
Rather than using dim. reg., it is simpler in this context to introduce a time cutoff, such that
we may directly subtract the problematic term from the definition, 
\beq	\tau_{\rm (X)} =  \int \dd t\frac{t (\bv_\infty\cdot \ba_{\rm (X)}(t))}{v_\infty^2}  
	+\lim_{\Lambda \to\infty}\bigg\{\int_{-\infty}^\Lambda \dd t
	\frac{t (\bv_\infty\cdot \ba_{\rm (N)}(t))}{v_\infty^2} \bigg|_{\delta_{\rm(X)}\br(t)} 
	+\frac{G M}{v_\infty^2}\frac{\bv_\infty \cdot \bv_+ }{v_\infty^3 }\log(\Lambda)
	\bigg\} \; .
	\label{eq:TdelayX}
\eeq%
As expected, the first term involving the radiation-reaction forces decays as $1/r^{n}$ with $n>2$, and therefore does not lead to divergent integrals, whereas the second term has been renormalized by explicitly subtracting the $\log (\Lambda)$ divergence. (It is straightforward to show that the divergence is actually invariant under coordinate transformations, and therefore it also cancels out in the matching.) \vskip 4pt 

Following the above steps we find the explicit results
\begin{align}
	\tau^{(3)}_{\rm (FT)}  & = 0  \; , & & 
	& \tau^{(4)}_{\rm (FT)}  & = \bigg(\frac{23}{40}\pi\bigg)\frac{G^4 M^4 \nu^2}{b^3}v_\infty  \; , \\
	\tau^{(5)}_{\rm (FT)}  & =  \bigg(-\frac{544}{45}\bigg)\frac{G^5 M^5 \nu^2}{b^4 v_\infty}  \; , & &
	&\tau^{(6)}_{\rm (FT)}  & = \bigg(-\frac{2711}{40}\pi
	\bigg)\frac{G^6 M^6 \nu^2}{b^5 v^3_\infty}  \; , \\
	\tau^{(3)}_{\rm (M)}  & = 0  \; , & & 
	& \tau^{(4)}_{\rm (M)}  & = \bigg(-\frac{1219 }{1260}\pi\bigg)\frac{G^4 M^4 \nu^2}{b^3}v_\infty  \; , \\
	\tau^{(5)}_{\rm (M)}  & =  \bigg(\frac{11552}{945}\bigg)\frac{G^5 M^5 \nu^2}{b^4 v_\infty}  \; , & &
	&\tau^{(6)}_{\rm (M)}  & = \bigg(\frac{6109}{70}\pi
	\bigg)\frac{G^6 M^6 \nu^2}{b^5 v^3_\infty} \; , \\
	\tau^{(3)}_{\rm (2RR)}  & = 0  \; , & & 
	& \tau^{(4)}_{\rm (2RR)}  & = \bigg(\frac{3889}{450}\pi +{\color{red} \frac{19}{25}\pi}\bigg)\frac{G^4 M^4 \nu^2}{b^3}v_\infty  \; , \\
	\tau^{(5)}_{\rm (2RR)}  & =  \bigg(\frac{4736}{45} -{\color{red} \frac{196352}{1575}}\bigg)\frac{G^5 M^5 \nu^2}{b^4 v_\infty}  \; , & &
	&\tau^{(6)}_{\rm (2RR)}  & = \bigg(-\frac{587}{450}\pi
	-{\color{red}\frac{2989999}{6300}\pi}
	\bigg)\frac{G^6 M^6 \nu^2}{b^5 v^3_\infty} \; ,
\end{align}
with RR--RR effects highlighted in {\color{red}red}.

\section{Conservative 2RR effects}\label{app:split}

We elaborate here on the separation between second-order radiation-reaction effects in the effective action, where \begin{equation}
S[\br] = S_{(\rm N)}[\br] + \varepsilon S_{(\rm RR)}[\br],\label{stot1}
\end{equation}
with $S_{\rm (N)}$ denotes the conservative Newtonian part,
\beq
S_{\rm (N)} = M\nu \int \dd t  \left(\frac{1}{2} \bv^2 - \frac{GM}{r}\right)\,,
\eeq
and $S_{\rm RR}$ is the radiation-reaction correction in \eqref{eq:srr2}. We consider the {\it background} solution to be the (exact) Newtonian trajectory \beq \bar \br \equiv \br_0 + \br_{(\rm N)}\,,\eeq 
obeying
\begin{equation}
\label{statN}
\frac{\delta S_{\rm (N)}}{\delta \br}\Big|_{\bar \br} = 0\,
\end{equation}
and expand the trajectory as
\begin{equation}
\br_\eps(t) = \bar \br (t)+ \varepsilon  \br_1(t) + \varepsilon^2  \br_2(t) + {\cal O}(\eps^3)\,,\label{brtot2}
\end{equation}
with boundary conditions, 
\begin{align}
	\br_1(t)\to 0 \; , \qquad
	\dot{\br}_1(t) \to 0 \; , 
	\qquad \text{for } t\to -\infty \; .
\end{align}

\subsection*{Radial action}

Our goal is to compute the on-shell radial action on the physical trajectory to ${\cal O}(\varepsilon^2)$. It is convenient to resort to a Hamiltonian formulation, where the radial action is given by\footnote{Strictly speaking, when the Lagrangian contains higher derivatives the radial action cannot be read off directly from \eqref{eq:Ir-def}. Instead, we must either eliminate the accelerations via  a field redefinition, or formulate the system in an enlarged phase space. However, the quantity relevant for our purposes is the on-shell action evaluated on the physical trajectory, which is invariant under redefinitions that vanish at infinity.}
\begin{equation}
\label{eq:Ir-def}
{\cal I}_r(E,J;\varepsilon)
=
2\int_{r_{\min}(\varepsilon)}^{\infty} p_r(r;E,J,\varepsilon)\,\dd r,
\end{equation}
where $p_r(r;E,J,\varepsilon)$ is the radial momentum, determined by
\begin{equation}
\label{eq:H-constraint}
H\!\left(r,p_r(r;E,J,\varepsilon);E,J,\varepsilon\right)=E\,,
\end{equation}
with $H$ the Hamiltonian, and  $r_{\min}(\varepsilon)$ the turning point, obeying
\begin{equation}
\label{eq:turning-point}
p_r(r_{\min}(\varepsilon);E,J,\varepsilon)=0.
\end{equation}
We consider a family of Hamiltonians, 
\begin{equation}
\label{eq:H-expansion}
H(z_\varepsilon)=\bar H+\varepsilon H_1 + \varepsilon^2 H_2 +{\cal O}(\eps^3) \,,
\end{equation}
where $\bz \equiv (\br,\bp)$ denotes collectively the relevant phase-space variables, with
\begin{equation}
\label{eq:traj-expansion}
\bz_\varepsilon=\bar \bz+\varepsilon \bz_1+\varepsilon^2 \bz_2 + {\cal O}(\eps^3) \,. 
\end{equation}

We start by differentiating \eqref{eq:Ir-def} with respect to $\eps$, obtaining  
\begin{equation}
\frac{d{\cal I}_r}{d\varepsilon}
=
2\int_{r_{\min}}^{\infty}\partial_\varepsilon p_r(r;\varepsilon)\,dr -2\,p_r(r_{\min};\varepsilon)\frac{dr_{\min}}{d\varepsilon}.
\end{equation}
Since $r_{\min}$ obeys $p_r(r_{\min};\varepsilon)=0$, one gets
\begin{equation}
\label{eq:dIr-deps-1}
\frac{d{\cal I}_r}{d\varepsilon}
=
2\int_{r_{\min}}^{\infty}\partial_\varepsilon p_r(r;\varepsilon)\,\dd r.
\end{equation}

Now differentiate the Hamiltonian constraint \eqref{eq:H-constraint} w.r.t. $\eps$ (at fixed $(r,E,J)$)
\begin{equation}
\partial_{p_r}H\,\partial_\varepsilon p_r+\partial_\varepsilon H=0 \to \partial_\varepsilon p_r
=
-\frac{\partial_\varepsilon H}{\partial_{p_r}H}.
\end{equation}
Substituting into \eqref{eq:dIr-deps-1} gives us the relation
\begin{equation}
\label{eq:dIr-deps-2}
\frac{d{\cal I}_r}{d\varepsilon}
=
-2\int_{r_{\min}}^{\infty}\dd r\,
\frac{\partial_\varepsilon H}{\partial_{p_r}H}\,,
\end{equation}
which,  along the radial trajectory,
\begin{equation}
\dot r=\partial_{p_r}H,
\qquad
\dd t=\frac{\dd r}{\partial_{p_r}H},
\end{equation}
yields (over the full orbit) 
\begin{equation}
\label{eq:dIr-full}
\frac{d{\cal I}_r}{d\varepsilon}
=
-\int_{-\infty}^\infty \dd t\,\partial_\varepsilon H.
\end{equation}
Inserting the expansion of the Hamiltonian, 
\begin{equation}
\partial_\varepsilon H = H_1 + 2 \eps H_2 + {\cal O}(\eps^3) \,,
\end{equation}
and the perturbed orbit,
\begin{equation}
H_1(\bz_\varepsilon)= H_1(\bar \bz)+\varepsilon \bz_1 \cdot \nabla H_1 (\bar \bz) +\cdots \,, \quad H_2(z_\varepsilon)= H_2(\bar \bz)+\cdots\,,
\end{equation}
such that, matching to the expansion of the radial action, 
\begin{equation}
{\cal I}_r=I_r^{(0)}+\varepsilon {\cal I}_r^{(\eps)}+\varepsilon^2 {\cal I}_r^{(\eps^2)}+{\cal O}(\varepsilon^3),
\quad
\frac{d{\cal I}_r}{d\varepsilon}=I_r^{(\eps)}+2\varepsilon {\cal I}_r^{(\eps^2)}+{\cal O}(\varepsilon^2).
\end{equation}
we arrive at
\begin{equation}
{\cal I}_r^{(\eps)}
= -\int \dd t\,H_1(\bar\bz)\,, \quad
{\cal I}_r^{(\eps^2)}
=-\int \dd t\,H_2(\bar\bz) -\frac12\int \dd t\,\bz_1\cdot \nabla H_1(\bar \bz).\label{eq:Ir2-H}
\end{equation}

\subsection*{Lagrangian form}

It is instructive to re-derive the above relationships in terms of a Lagrangian involving a linear perturbation. In our case, we have
\begin{equation}
\label{eq:L-pert}
L(\br,\bv;\varepsilon)=L_{(\rm N)}(\br,\bv)-\varepsilon V_{\rm (RR)}(\br,\bv),
\qquad
L_{(\rm N)}=\frac12 M\nu \bv^2-V_{(\rm N)}(\br)\,,
\end{equation}
 with the corresponding momentum given by
\begin{equation}
\label{eq:p-from-L}
\bp=\frac{\partial L}{\partial \bv}
=
M\nu \bv-\varepsilon\,V_{{(\rm RR)}\,,\bv} 
\end{equation} 
such that
\begin{equation}
\begin{aligned}
H
=\bp\cdot \bv - L =
\bp\cdot \bv
-\left(\frac12 M\nu \bv^2-V_{\rm (N)} (\br)-\varepsilon V_{\rm (RR)}(\br,\bv)\right).
\end{aligned}
\end{equation}
Substituting
\beq
M\nu \bv=\bp+\varepsilon\,V_{{(\rm RR)}\,,\bv}+{\cal O}(\varepsilon^2)\,,
\eeq
we find the expected form 
\begin{equation}
\label{eq:H-expanded-from-V}
H =
\bar H+\varepsilon H_1+\varepsilon^2 H_2+{\cal O}(\varepsilon^3)\,,
\end{equation}
with
\begin{equation}
\label{eq:H1H2-from-V}
H_{(\rm N)}=\frac{\bp^2}{2M\nu}+V_{(\rm N)},\qquad
H_1=V_{(\rm RR)},\qquad
H_2=\frac12\big|V_{(\rm RR)\,, \bv}\big|^2\,.
\end{equation}

Concentrating on second order effects, from \eqref{eq:Ir2-H} we have
\begin{equation}
\label{eq:Ir2-H-mid}
{\cal I}_r^{(\eps^2)}=
-\frac12\int \dd t\,\big|V_{(\rm RR)\,, \bv}\big|_{\bar \br,\bar \bv}^2
-\frac12 \int  \dd t\,
\big[
\br_1\cdot (V_{(\rm RR),\br})_{\bar \br,\bar \bv}
+
\bp_1\cdot (V_{(\rm RR),\bv})_{\bar \br,\bar \bv}
\big].
\end{equation}
Plugging the expression for the momentum, we then arrive at
\begin{equation}
\begin{aligned}
\label{eq:Ir2-V}
{\cal I}_r^{(2)}
&=
-\frac12\int \dd t\,
\left[
\br_1\cdot \frac{\partial V_{(\rm RR)}}{\partial \br}(\bar\br,\bar\bv)
+
\dot{\br}_1\cdot \frac{\partial V_{(\rm RR)}}{\partial \bv}(\bar\br,\bar\bv)
\right] =\frac12\,\delta S_{(\rm RR)}\Big|_{\bar \br;\,\delta \br=\br_1}\,,
\end{aligned}
\end{equation}
where the explicit $H_2$ term cancels against the correction from $\bp_1$, yielding---up to a crucial factor of $1/2$---the variation of the action, as expected.\footnote{Notice that, in principle, the expression in \eqref{eq:Ir2-V} is not invariant under a shift $L \to L + \frac{\dd F(\br,t)}{dt}$. This is particularly important whenever $F(\br,t)$ does not vanish at infinity. In such cases, the expression in \eqref{eq:Ir2-V} picks up a boundary term, such that
\begin{align}
\label{eq:extrav}
\mathcal I_r^{(2)}
&= \frac12\,\delta S_{(\rm RR)}\Big|_{\bar \br;\,\delta \br=\br_1}-\frac{1}{2}\lim_{T\to \infty}  V_{{(\rm RR)}\,,\bv} \big(\bar\br(T),\bar \bv(T)\big)\cdot \br_1(T) = \frac12\, \int \dd t \, \bF_{(\rm RR)}\cdot \br_1\,,
\end{align}
which is not affected by total derivatives.} 

\subsection*{On-shell action}

The previous results are very suggestive. Indeed, we can derive it directly from the computation of  the action in \eqref{stot1} on the on-shell trajectory in \eqref{brtot2}.  A straightforward Taylor expansion gives (schematically)
\begin{equation}
\begin{aligned}
S[\br_\varepsilon;\varepsilon]
&=
S[\bar \br]
+\varepsilon\,\delta S_{(\rm N)}[\bar \br](\br_1)+\varepsilon^2
\big(
\delta S_{(\rm N)}[\bar \br](\br_2) \\
&+\frac12\,\delta^2 S_{(\rm N)}[\bar \br](\br_1,\br_1)
+\delta  S_{(\rm RR)}[\bar \br](\br_1)\big)
+{\cal O}(\varepsilon^3).
\end{aligned}
\end{equation}
At the same time, the first-order correction obeys the linearized equation
\begin{equation}
\delta^2 S_{(\rm N)}[\bar \br](\br_1,-)+\delta S_{(\rm RR)}[\bar \br](-)=0\,,
\end{equation}
and substituting back we get
\begin{equation}
\label{eq:S2-half}
S_{\rm on-shell}^{(\eps^2)}= \delta S_{\rm (N)}[\bar \br](\br_2)
+\frac12\,\delta  S_{(\rm RR)}[\bar \br](\br_1)\,.
\end{equation}

At linear order in the perturbation, the variation of the on-shell action, $\delta S_{\rm (N)}[\bar \br](-)$, is only given by boundary terms. It is useful to perform the variation in first order form and polar coordinates, for which
\begin{equation}
S_{\rm (N)}=\int \dd t\,\big(p_r\dot r+J\dot\phi-H_{\rm (N)}\big),
\qquad H_{(\rm N)}=E \ \text{on shell},
\end{equation}
and the first variation is therefore given by
\begin{equation}
\delta S_{\rm (N)}=
\Big[p_r\,\delta r+J\,\delta\phi-E\,\delta t\Big]_{t_i}^{t_f}\,,
\end{equation}
hence, keeping only the contribution to the real part,\footnote{For instance, in the direct computation the boundary term includes (see App.~\ref{app:delay})
\begin{align}
 \bv_\infty \cdot  (\bb_+-\bb)_2   =  -\bv_\infty \cdot \int_{-\infty}^{+\infty} \dd t \, t\, \ddot \br_2  
= -M\nu \, v_\infty^2 (t_f-t_i)  - p_\infty \frac{\bb\cdot \Delta_2 \bv}{v_\infty}\,,
\end{align}
which combined with the leading kinetic term, $\int K \dd t = \tfrac{M\nu}{2} v_\infty^2 (t_f-t_i)$, and using the relationship
\beq 
2\sin \frac{(\phi_f-\phi_i)}{2} \simeq (\phi_f-\phi_i) + \cdots = 
-\frac{\hat \bb\cdot \Delta \bv}{v_\infty} +\cdots \,,
\eeq
it provides the factor of $-E(t_f-t_i) + J (\phi_f-\phi_i)$, as expected.} 
\begin{equation}
\label{eq:B2}
 \delta S_{\rm (N)}[\bar \br](\br_2) 
=
\Big[p_{r0}\, r_2+J\,\phi_2-E\,t_2\Big]_{t_i}^{t_f} = p_{r0}\, r_2 + J\Delta\phi_2 - E \tau_2\,,
\end{equation}
such that, using the known relationship between on-shell and radial actions through a Legendre transformation \cite{Kim:2025sey,Kim:2025gis}, we find 
\begin{equation}
\label{eq:S2f}
{\cal I}_{r(\rm RR)}^{(\eps^2)}= S_{\rm on\text{-}shell}^{(\eps^2)} - J\Delta\phi_2 + E \tau_2= +\frac12\,\delta  S_{(\rm RR)}[\bar \br](\br_1) \,,
\end{equation}
after removing the divergent piece proportional to the radial coordinate going to infinity.\vskip 4pt

\subsection*{RR$^2$ vs RR--RR}

Inputing the explicit form for our case, we find
\beq
	{\cal I}^{(\eps^2)}_{r (\rm RR)} =
	-\frac{2\pi G}{5}\int\frac{\dd \omega_1\dd\omega_2}{(2\pi)^2} I_1^{ij}(\omega_1)\bar{I}^{ij}(-\omega_1)\omega_1^4\Delta_F(\omega_1)\,,
\eeq	
where the first-order perturbation of the quadrupole moment may be written~as 
\beq
	 I_1^{ij}(\omega_1) = 2M \nu \int \frac{\dd \omega_3}{2\pi}\br_1^{\langle i}(\omega_3)\bar{\br}^{j\rangle}(\omega_1-\omega_3)\,.
\eeq
In order to compute the relevant terms we must solve for $\br_1$, which can be done using the form of the accelerations, given by
\begin{align}
	\ba_{\rm (N)}(t) = -\frac{G M}{ r^3(t)}\br(t) \; , \quad	\ba^\ell_{\rm (RR)}(\omega) = -\frac{8\pi G}{5}\int\frac{\dd \omega^\prime}{2\pi}\br^j(\omega-\omega^\prime) \omega^{\prime 4}\Delta_F(\omega^\prime)I^{j\ell}(\omega^\prime) \;,
\end{align}
yielding, at first order in the perturbation,
\begin{align}
	\ddot{\br}_1 = ( \br_1 \cdot \nabla)\bar\ba_{(\rm N)} + \bar{\ba}_{\rm (RR)}\;.
\end{align}
The solution with the correct boundary conditions can then be found via 
\begin{align}
	\br^i_1(t) & = 
	-\frac{8\pi G}{5}\int \dd t_2 \int \dd t^\prime  \, 
 \partial_{t^\prime}^2\Delta_F(t^\prime-t_2) G^{ik}_{\rm ret}(t,t^\prime)\bar{\br}^j(t^\prime) \bar{I}^{(2), k j}(t_2)\label{eqr1d} \; ,
\end{align}
 where
  \begin{equation}
	\bigg(
	\delta^{ij}\frac{\dd^2}{\dd t^2}
	-\mathcal{E}^{ij}(t)
	\bigg)G_{\rm ret}^{jk}(t,t^\prime) = \delta^{ik}\delta (t-t^\prime) \; ,
	\qquad 
	G^{jk}_{\rm ret}(t,t^\prime) = 0 \text{ for } t<t^\prime \; ,
\end{equation}
or directly in Fourier space,
\begin{equation}
	\br^i_1(\omega) = \int \frac{\dd \omega_3}{2\pi}G^{ij}_{\rm ret}(\omega, -\omega_3) \bar{\ba}_{\rm RR}^j(\omega_3)\,,
\end{equation}
with\footnote{Notice that the appearance of retarded propagators on the worldlines, i.e. $(\omega+i0)^{-1}$, resembles the same type of contributions in the WEFT approach for the PM domain \cite{Kalin:2020mvi}.}  
\begin{align}	
	G^{ij}_{\rm ret}(\omega, -\omega_3) & \equiv -\frac{\delta^{ij}\ddl(\omega-\omega_3)}{(\omega+i0^+)^2}
	+ \frac{\mathcal{E}^{ij}(\omega-\omega_3)}{(\omega+i0^+)^2(\omega_3+i0^+)^2}
	\notag \\
	& \qquad 
	-\int\frac{\dd \omega_4}{2\pi} \frac{\mathcal{E}^{i k}(\omega-\omega_4)\mathcal{E}^{ k j}(\omega_4-\omega_3)}{(\omega+i0^+)^2(\omega_3+i0^+)^2(\omega_4+i0^+)^2} + \dots
\end{align}
with the Newtonian Hessian given by
\begin{equation}
	\mathcal{E}^{ij}(t) \equiv -G M\bigg(
 \frac{\delta^{ij}}{\bar{r}^3(t)}
 -3\frac{\bar{\br}^i(t)\bar{\br}^j(t)}{\bar{r}^5(t)}
 \bigg)\,,\quad {\cal E}_{ij}(\omega)= \int \dd t e^{i\omega t}{\cal E}_{ij}(t)\,.
\end{equation}

Since on the unperturbed quadrupole,
\begin{equation}
	\int \dd t_1 \Delta_F(t-t_1) I^{(2)}_{(0)ij}(t_1) \propto \frac{i}{\pi}\int \dd t_1  \frac{\PV}{(t-t_1)^2 }\bv_\infty^i \bv^j_\infty \to  0 \; ,
\end{equation}
it is straightforward to see from \eqref{eqr1d} that $\br_1$ is nonvanishing only when the background quadrupole $\bar{I}^{(2)kj}$ is evaluated along the Newtonian trajectory. Therefore, effects at second-order in the radiation-reaction force may be suggestively written in the on-shell (radial) action~as,
\begin{align}
\label{finalrr}
{\cal I}_{r (\rm 2RR)}  &= \frac{M\nu}{2} \int\frac{\dd \omega_1\dd\omega_2}{(2\pi)^2} \bar\ba^i_{\rm (RR)}  (-\omega_1) G^{ij}_{\rm ret}(\omega_1,-\omega_2) \bar \ba^j_{\rm (RR)}(\omega_2)\\
&=-\frac{32\pi^2 G^2 M \nu}{25}\int\frac{\dd \omega_1\dd\omega_2}{(2\pi)^2}  \omega_1^4\omega_2^4\Delta_F(\omega_1)\Delta_F(\omega_2)I_{\rm (N)}^{ij}(-\omega_1) I_{\rm (N)}^{k \ell}(\omega_2){\cal O}^{k\ell;ij}(\omega_1,-\omega_2) \; \nn
 \end{align}
with
\begin{align}
\label{eq:Ojk}
{\cal O}^{k\ell;ij}(\omega_1,-\omega_2) \equiv
	\int \frac{\dd \omega_3\dd \omega_4}{(2\pi)^2}\bar{\br}^k(\omega_1-\omega_3) G^{\ell i}_{\rm ret}(\omega_3, -\omega_4)\bar{\br}^j(\omega_4-\omega_2) \; .
\end{align}	

The computation of scattering observables, such as the scattering angle, then follows upon evaluating the above expression on the physical trajectories to the given PM order. This is achieved by inputing into the expression in \eqref{eq:Ojk} the background values, 
\bea
\bar \br(\omega) &=& \br_0(\omega) + \br_{\rm (N)}(\omega) =  \ddl (\omega)\bb + i \bv_\infty {\ddl}^\prime (\omega) + \br_{\rm (N)}(\omega)\\
\bar \bv(\omega) &=& \bv_\infty \ddl(\omega)+ \bv_{\rm (N)}(\omega)
\eea
  as well as the corrections to the Green's function,
\begin{align}
G^{jk}_{\rm ret}(\omega_1, -\omega_2) &= \delta^{jk}\ddl(\omega_1-\omega_2)G_{(0)}(\omega_1) 
 	+ G_{(1)}^{jk}(\omega_1, -\omega_2)
 	+ G_{(2)}^{jk}(\omega_1, -\omega_2) \; ,\label{Gexp}
 \end{align}
 with 
 \begin{align}
 \label{G0}	G_{(0)}(\omega_1) &= -(\omega_1+i0^+)^{-2}\,,\quad
 	G_{(1)}^{jk}(\omega_1, -\omega_2)  =
	G_{(0)}(\omega_1)
	\mathcal{E}^{j k }(\omega_1-\omega_2)G_{(0)}(\omega_2) \; ,  \\ 
	G_{(2)}^{jk}(\omega_1, -\omega_2) & =
	\int \frac{\dd\tilde{\omega}}{2\pi}G_{(0)}(\omega_1)
	\mathcal{E}^{j n}(\omega_1-\tilde{\omega})G_{(0)}(\tilde{\omega})\mathcal{E}^{ n k}(\tilde{\omega} - \omega_2)G_{(0)}(\omega_2) \nn\;,
 \end{align}
yielding the corresponding expansion,
\begin{align}
\label{Oexp}
	\mathcal{O}^{k\ell;ij}(\omega_1, -\omega_2) = \mathcal{O}_{(0)}^{k\ell;ij}(\omega_1, -\omega_2) 
	+ \mathcal{O}_{(1)}^{k\ell;ij}(\omega_1, -\omega_2) 
	+ \mathcal{O}_{(2)}^{k\ell;ij}(\omega_1, -\omega_2)  \;.
\end{align}
Returning to \eqref{finalrr} we can therefore naturally identify two types of contributions, in particular
\beq
{\cal I}^{(0)}_{r (\rm 2RR)}  = -\frac{M\nu}{2} \int\frac{\dd \omega}{(2\pi)} \frac{\bar\ba_{\rm (RR)}  (-\omega) \cdot \bar \ba_{\rm (RR)}(\omega)}{(\omega+i0)^2}\,,
\eeq
which contains all of the RR$^2$ effects, whereas the remaining corrections involve the Hessian-dependent factors in \eqref{Gexp}. 
\subsection*{Tail-like}
It is straightforward to show that all the pieces involving the Hessian vanish in the tail region. Therefore, the latter is dominated by ${\cal I}^{(0)}_{r (\rm 2RR)}$ and furthermore arises from  
plugging the unperturbed trajectories for the background quantities. Recalling the explicit expression for the Feynman propagator,
\begin{equation}
	\Delta_F(\omega) = -\frac{i}{4\pi}|\omega| = -\frac{i}{4\pi}\omega\sg(\omega) \; ,
\end{equation}
and entering the relevant expressions, we find 
\begin{align}
{\cal I}_{r (\rm 2RR,\, T\text{-}like)}& = {\cal I}^{(0)}_{r (\rm 2RR,\, T\text{-}like)}  =  \frac{2 G^2M\nu}{25}\int \frac{\dd \omega_1 \dd \omega_2}{(2\pi)^2} \frac{\ddl(\omega_1-\omega_2)}{(\omega_1+i0^+)^2}\sg(\omega_1)\sg(\omega_2) \notag \\
& \qquad \times(\bb^j+i\bv^j_\infty\partial_{\omega_1})\Big[ \omega_1^5 I^{ij}_{\rm (N)}(-\omega_1)\Big]
	  (\bb^k-i\bv^k_\infty\partial_{\omega_2})\Big[\omega_2^5 I^{ik}_{\rm (N)}(\omega_2)\big]\,,
\end{align}
where we have discarded factors involving derivatives of $\sg(\omega)$ leading to $\omega^n \delta(\omega)$ terms with $n \geq 3$.\footnote{More generally, throughout our computations we use the distributional identity  $(-\omega^2-i0)^{\tfrac{1}{2}-\epsilon} \delta^\prime (\omega) \to 0$ in dim. reg., with $\epsilon <0$. This also provides a natural regulator for the derivation of the memory-like region of the radial action (see below). \label{foot}} Hence, using the identity $\sg(\omega_1)\sg(\omega_2)\delta(\omega_1-\omega_2) \to \delta(\omega_1-\omega_2)$, we arrive at 
\begin{align}
 	{\cal I}_{r (\rm 2RR,\, T\text{-}like)} & = \frac{2G^2 M \nu}{25}\int\frac{\dd \omega}{2\pi}\left(
 	(\omega^4 \bb^i
 	-5 \omega^3i \bv_\infty^i
 	-i\omega^4 \bv_\infty^i\partial_{\omega})\Big[ I_{\rm (N)}^{ij}(\omega) \Big] \times \text{c.c.}\right) \; .
 \end{align}
 Transforming back into direct space, and using  
\begin{align}
\big(\omega^4 \bb^j
	-i5\omega^{3}\bv^j_\infty
	-i\omega^{4}\bv^j_\infty\partial_{\omega}\big)[I_{\rm (N)}^{ij}(\omega)
	\big]  & =
	i \int \dd t \Big(
	\br^j_{(0)}(t) I^{(4), ij}_{{\rm (N)}}(t) 
	- \bv_\infty^j I^{(3), ij}_{\rm (N)}(t)
\Big) e^{i\omega t} \; ,
\end{align}
we get
\begin{align}
 	 	{\cal I}_{r (2\rm RR,\, T\text{-}like)} & =
 	\frac{2G^2 M \nu}{25}\int \dd t\bigg[
 	2  \br_0^{i} \bv_\infty^{j}  I^{(2)}_{j k}  I^{(5)}_{i k}
- \br_0^{i} \br_0^{j}  I^{(3)}_{i k} I^{(5)}_{j k}
 	\bigg]  \;.
\end{align}
In terms of the variables introduced in \cite{Porto:2024cwd},
 \begin{equation}
 	Q_{(0)}^{ij} = M\nu \br_0^i(t) \br_0^j(t) \; ,
 	\qquad 
 	L_{ij} = 2M\nu \br_0^{[i}(t) \bv_{\infty}^{j]} \; ,
 \end{equation}
 we have 
\begin{align}
 		{\cal I}_{r (\rm 2RR,\, T\text{-}like)} & =
 	\frac{2G^2}{25}\int \dd t\bigg[
 	-L_{ki}I^{(4)}_{k j }I^{(3)}_{i j }
 	+Q_{(0),ki}I^{(4)}_{k j }I^{(4)}_{i j }
 	+Q^{(2)}_{(0),ki}I^{(3)}_{k j }I^{(3)}_{i j }
 	\bigg]  = S^{\rm cons}_{(\rm RR^2,\, T\text{-}like)} \; ,
\end{align}
leading to the expression in \eqref{7.2}, and in agreement with Eq. 7.2 in \cite{Porto:2024cwd}. 
\subsection*{Memory-like}

Next we input the Newtonian deflections. For simplicity we will study the combined 2RR contributions. We start with $\mathcal{O}_{(1)}^{jk;i\ell}(\omega_1, -\omega_2)$. After several manipulations it may be written as,
\begin{align}
	\mathcal{O}_{(1)}^{k\ell;ij}(\omega_1, -\omega_2) & = 
	-\delta^{\ell i}(\bb^k -i\bv_\infty^k \partial_{\omega_1})\bigg[\frac{ \br_{\rm (N)}^{j}(\omega_1-\omega_2)}{(\omega_1+i0^+)^2}\bigg]
	-\delta^{\ell i}(\bb^j +i\bv_\infty^j \partial_{\omega_2})\bigg[\frac{\br_{\rm (N)}^{k}(\omega_1-\omega_2)}{(\omega_2+i0^+)^2}\bigg]\ \notag \\
	& 
	+ (\bb^k -i\bv_\infty^k \partial_{\omega_1})(\bb^j +i\bv_\infty^j \partial_{\omega_2})\bigg[\frac{\mathcal{E}^{\ell i}(\omega_1-\omega_2)}{(\omega_1+i0^+)^2 (\omega_2+i0^+)^2}\bigg] \; .
\end{align}
Inputting this expression into the radial action we find\footnote{In arriving here we have used dim. reg. to remove a boundary term that could in principle arise through derivatives of the $\sg(\omega)$ function. See footnote \ref{foot}.}
\begin{align}
	{\cal I}^{(1)}_{r (\rm 2RR,\, M\text{-}like)} & = \frac{2 G^2M\nu}{25}\int \frac{\dd \omega_1 \dd \omega_2}{(2\pi)^2}\sg(\omega_1)\sg(\omega_2)\notag \\
	&\times 
	\bigg\{
	\frac{\br_{\rm (N)}^j(\omega_1-\omega_2)}{(\omega_1+i0^+)^2}
	\omega_2^5 I_{\rm (N)}^{i j}(\omega_2)\omega_1^4(\omega_1 \bb^k +5 i\bv_\infty +i\omega_1 \bv_\infty^k\partial_{\omega_1})\big[I_{\rm (N)}^{ki}(-\omega_1)
	\big]\notag \\
	& 
	+ \frac{\br_{\rm (N)}^k(\omega_1-\omega_2)}{(\omega_2+i0^+)^2}
	\omega_1^5I_{\rm (N)}^{ki}(-\omega_1)\omega_2^4
	(\omega_2 \bb^j -5i\bv_\infty^j-i\omega_2 \bv_\infty^j \partial_{\omega_2})\big[I_{\rm (N)}^{ji}(\omega_2)
	\big] \notag \\
	&  
	- \frac{\mathcal{E}^{\ell i}(\omega_1-\omega_2)}{(\omega_1+i0^+)^2 (\omega_2+i0^+)^2}\omega_1^4 \omega_2^4
	(\omega_1 \bb^k +5 i\bv^k_\infty +i\omega_1 \bv_\infty^k\partial_{\omega_1})\big[I_{\rm (N)}^{ki}(-\omega_1)
	\big]\notag \\
	& \qquad\qquad \times
(\omega_2 \bb^j -5i\bv_\infty^j-i\omega_2 \bv_\infty^j \partial_{\omega_2})\big[I_{\rm (N)}^{ji}(\omega_2)
	\big]\bigg\}\; .
\end{align}

Although somewhat intricate, the above expression has the expected memory-like structure from which we can extract a local contribution via the PB prescription, which becomes
 \begin{align}
 \label{2rrg5}
 	{\cal I}^{(1)\rm PBloc}_{r (2\rm RR,\, M\text{-}like)} & = \frac{2 G^2}{25}\int \dd t\bigg\{
 	- M \nu\Big(
 	\br^i_{(0)}  I^{(3), i j}_{\rm (N)}-2\bv_{(0)}^i I^{(2), ij}_{\rm (N)}\Big) \mathcal{E}^{jk}\Big(
	\br^\ell_{(0)}  I^{(3), \ell k}_{{\rm (N)}}  
	- 2\bv_{(0)}^\ell I^{(2), \ell k}_{\rm (N)} 
\Big) \notag \\
& \hspace{3cm}
- M \nu\left(3{\ba}_{\rm (N)}^i\br_{(0)}^jI^{(3), jk}_{\rm (N)} I^{(3), ki}_{\rm (N)} 
-4{\ba}_{\rm (N)}^i \bv_{(0)}^jI^{(2), jk}_{\rm (N)} I^{(3), ki}_{\rm (N)} \right)

 	\notag \\
&\hspace{3cm}
 	 +Q^{ij}_{\rm(N)}I^{(4), jk}_{\rm (N)} I^{(4), ki}_{\rm (N)}
	+ Q^{(2),ij}_{\rm(N)}I^{(3), jk}_{\rm (N)} I^{(3), ki}_{\rm (N)}
 	\bigg\} \; ,
 \end{align}
where we identified (at this order)
 \begin{equation}
 	Q^{ij}_{\rm (N)} = 2M\nu \br_{(0)}^{(i}\br_{\rm (N)}^{j)} \; .
 \end{equation}
The expression in \eqref{2rrg5} is sufficient to derive all the 2RR effects entering at 5PM order. \vskip 4pt  

The situation becomes slightly more subtle at next order in the expansion. Although---upon including $\mathcal{E}^{ij}_{(1)}$ and increasing the order in $G$ of the $\br_{\rm (N)}$ and $\bv_{\rm (N)}$ deflections (or multipole moments)---many of the previous manipulations leading to \eqref{2rrg5} generalize straightforwardly to 6PM, there are novel contributions. In particular, we find terms of the form (schematically)
\begin{align}
{\cal I}^{(G^6)}_{r (2\rm RR,\, M\text{-}like)} \supset \int \dd t_1 \dd t_2 \dd t_3 \dd t_4 \bigg(-\frac{i}{\pi}\frac{\PV}{t_1-t_2}\bigg)
	\bigg(-\frac{i}{\pi}\frac{\PV}{t_3-t_4} \bigg) A_{(\rm N)}(t_1)B_{\rm (N)}(t_3) {\cal F}(t_2,t_4) \,,
\end{align}
which originate either from taking $\bar{\br}_{\rm (N)}$ on both entries in \eqref{eq:Ojk}, or from ${\cal O}({\cal E}^2)$ terms in the Hessian expansion. These contributions are, in principle, genuinely nonlocal in character and, moreover, cannot be straightforwardly reduced using the PB identity. In this context, our strategy to isolate a local-in-time component is to universally extend the PB procedure through the systematic replacement $\sg(\omega)\sg(\omega_1) \to 1$ in the Fourier-space representation (see App.~\ref{app:PBFourier}). Since the associated $A_{(\rm N)}, B_{(\rm N)}$ are evaluated on the Newtonian solution, no spurious tail-like corrections are introduced in this fashion. Moreover, this prescription consistently captures local contributions supported on ${\cal F}(t_2, t_3) \propto \delta(t_2 - t_3)$.\vskip 4pt

 Following this procedure, and after carefully disentangling all the relevant pieces, we find the following structure for the local-in-time radial action, 
\beq
{\cal I}^{\rm PBloc}_{r (2\rm RR)} = {\cal I}^{\rm PBloc}_{r (\rm RR^2)}+{\cal I}^{\rm PBloc}_{r (\rm RR\text{-}RR)}\,,
\eeq 
universally valid across both tail- and memory-like regions, with
\begin{align}
\label{eq:RR2cons}
{\cal I}^{\rm PBloc}_{r (\rm RR^2)} & = -\frac{2 G^2}{25} \int \dd t \bigg(L^{k\ell}
	 I^{(4),ki} I^{(3),\ell i}   -	{Q}^{kj} I^{(4), ki}  I^{(4),ij}
	-{Q}^{(2),kj} I^{(3), ki} I^{(3),ij} \bigg)\,,
\end{align}
recovering the RR$^2$ effective action introduced in \cite{Porto:2024cwd}; whereas, for the remaining RR--RR effects,  \begin{align}
{\cal I}^{\rm PBloc}_{r (\rm RR\text{-}RR)} &=	+\frac{2 M\nu G^2 }{25} \int \dd t \bigg\{- 3{\ba}_{\rm (N)}^i\br_{(0)}^jI^{(3), jk}_{\rm (N)} I^{(3), ki}_{\rm (N)} 
+4\ba_{\rm (N)}^i \bv_{(0)}^jI^{(2), jk}_{\rm (N)} I^{(3), ki}_{\rm (N)}
	 \\
	& - \br_{\rm(N)}^k \ba_{\rm(N)}^j I^{(3),ki}_{\rm (N)}
	I^{(3),ij}_{\rm (N)} - \Big(
 	\br^i_{(0)}  I^{(3), i j}_{\rm (N)}-2\bv_{(0)}^i I^{(2), ij}_{\rm (N)}\Big) \mathcal{E}^{jk}\Big(
	\br^\ell_{(0)}  I^{(3), \ell k}_{{\rm (N)}}  
	- 2\bv_{(0)}^\ell I^{(2), \ell k}_{\rm (N)} 
\Big)\bigg\}
	\notag \\
	&
	+\frac{G M\nu}{5}\int \dd t  \bigg\{
	\ba^{k}_{\rm (N)}I^{(3), ki}_{\rm (N)} \delta_{\rm (RR)} \br^i
	- 
	\Big(\br^{k}_{\rm (N)}I^{(4), ki}_{\rm (N)}
	- \bv^{k}_{\rm (N)}I^{(3), ki}_{\rm (N)}\Big)\delta_{\rm (RR)}\bv^i	
	\notag \\
	& \qquad 
+\br^{k}_{\rm (N)}I^{(5), k\ell}_{\rm (N)} \Delta_{\cal E} \br^\ell+\Big(
	\br^k_{(0)}I_{\rm (N)}^{(3),k\ell}(t) 
	- 2\bv_{(0)}^k I^{(2),k\ell}_{{\rm (N)}}
\Big)\mathcal{E}^{\ell i} \Big(\delta_{(\rm RR)}  \br^i
	+\Delta_{\cal E} \br^i \Big)
	\bigg\}\nn
\label{eq:RRRRcons}
\end{align}
where we introduced the notation, 
\begin{align}
	\delta_{\rm (RR)}\br^i(t_1) & = -\frac{2G}{5}\int \dd t_2 \; G_0(t_1-t_2) \br^{j}_{\rm (N)}(t_2)I^{(5), ji}_{\rm (N)}(t_2) \; , \\
	 \delta_{\rm (RR)} \bv^i(t_1)  &= 	-\frac{2G}{5}\int \dd t_2 \; \dot G_0(t_1-t_2)  \ba^{j}_{\rm (N)}(t_2)I^{(3), ji}_{\rm (N)}(t_2) \; , \\
	\Delta_{\cal E}\br^\ell(t_1) & = 
	-\frac{2G}{5}\int \dd t_2 \; G_0(t_1-t_2) \mathcal{E}^{\ell i}(t_2)
	\Big(
	\br^j_{(0)}(t_2)I_{\rm (N)}^{(3),ji}(t_2) 
	- 2\bv_{(0)}^j  I^{(2), ji}_{{\rm (N)}}(t_2)
\Big) \; ,
\end{align}
 with  $\, G_0(t_1-t_2) = (t_1-t_2) \theta(t_1-t_2)$ the Fourier transform of the leading order Green's function in \eqref{G0}.
\newpage
\bibliographystyle{JHEP} 
\bibliography{refmem5PN}
\end{document}